\pdfoutput=1
\documentclass[12pt,preprint]{aastex}
\begin{document}
\tighten
\title{A Survey of Irradiated Pillars, Globules, and Jets in the Carina Nebula}


\author{
	P. Hartigan \altaffilmark{1}, 
	M. Reiter \altaffilmark{2}
	N. Smith \altaffilmark{2},
	J. Bally \altaffilmark{3},
	}

\vspace{1.0cm}

\altaffiltext{1}{Rice University, Department of Physics and Astronomy,
6100 S. Main, Houston, TX 77521-1892} 

\altaffiltext{2}{Department of Astronomy, University of Arizona, Tucson, AZ 85721}

\altaffiltext{3}{Center for Astrophysics and Space Astronomy, University of Colorado, Boulder, CO 80309}

\begin{abstract}

We present wide-field, deep narrowband H$_2$, Br$\gamma$, H$\alpha$, [S~II], [O~III], and 
broadband I and K-band images of the Carina star formation region.  The new images provide a
large-scale overview of all the H$_2$ and Br$\gamma$ emission present in over
a square degree centered on this signature star forming complex.
By comparing these images with archival HST and Spitzer
images we observe how intense UV radiation from O and B stars
affects star formation in molecular clouds. 
We use the images to locate new candidate outflows and
identify the principal shock waves and irradiated interfaces
within dozens of distinct areas of star-forming activity.
Shocked molecular gas in jets traces the parts
of the flow that are most shielded from the intense UV radiation. Combining the
H$_2$ and optical images gives a more complete view of the jets, which are sometimes only
visible in H$_2$.  The Carina region hosts several compact young clusters, and the gas
within these clusters is affected by radiation from both the cluster stars and the
massive stars nearby.
The Carina Nebula is ideal for studying the physics of young H~II regions and PDR's, as it
contains multiple examples of walls and irradiated pillars at various stages of development.
Some of the pillars have detached from their host molecular clouds to form proplyds. 
Fluorescent H$_2$ outlines the interfaces between the ionized and molecular gas, 
and after removing continuum, we detect spatial offsets between the Br$\gamma$ and H$_2$ 
emission along the irradiated interfaces.  These spatial offsets can be
used to test current models of PDRs once synthetic maps of these lines become available.


\end{abstract}

\section{Introduction} 

Most stars form within large molecular cloud complexes that are also sites of
massive star formation. Massive stars have profound effects on their
parent clouds, primarily by emitting copious amounts of extreme ultraviolet
radiation that create large bubbles of ionized gas which then expand to
produce blisters as the gas emerges from the confines of the cloud. Along
the edges of the ionized volumes, far-ultraviolet radiation impinges upon
the dark clouds, heating and dissociating the molecular gas in a photodissociation region
\citep[PDR, a.k.a. photon-dominated regions; e.g.][]{ht97}.
As the molecular cloud evaporates, the PDR boundary can develop
pillars, which later detach to create isolated globules, often with a young star located near
the apex of the globule \citep[e.g.][]{odell93,hester96}.  
Because PDR boundaries trace surfaces where massive stars deposit energy back into
their nascent clouds, following how these boundaries evolve with time is important
from the standpoint of star formation energetics. PDRs also provide examples of interfaces
between hot ionized gases and cooler, neutral gases where dynamical instabilities may develop. 

\begin{figure}  
\centering
\includegraphics[angle=0,scale=1.00]{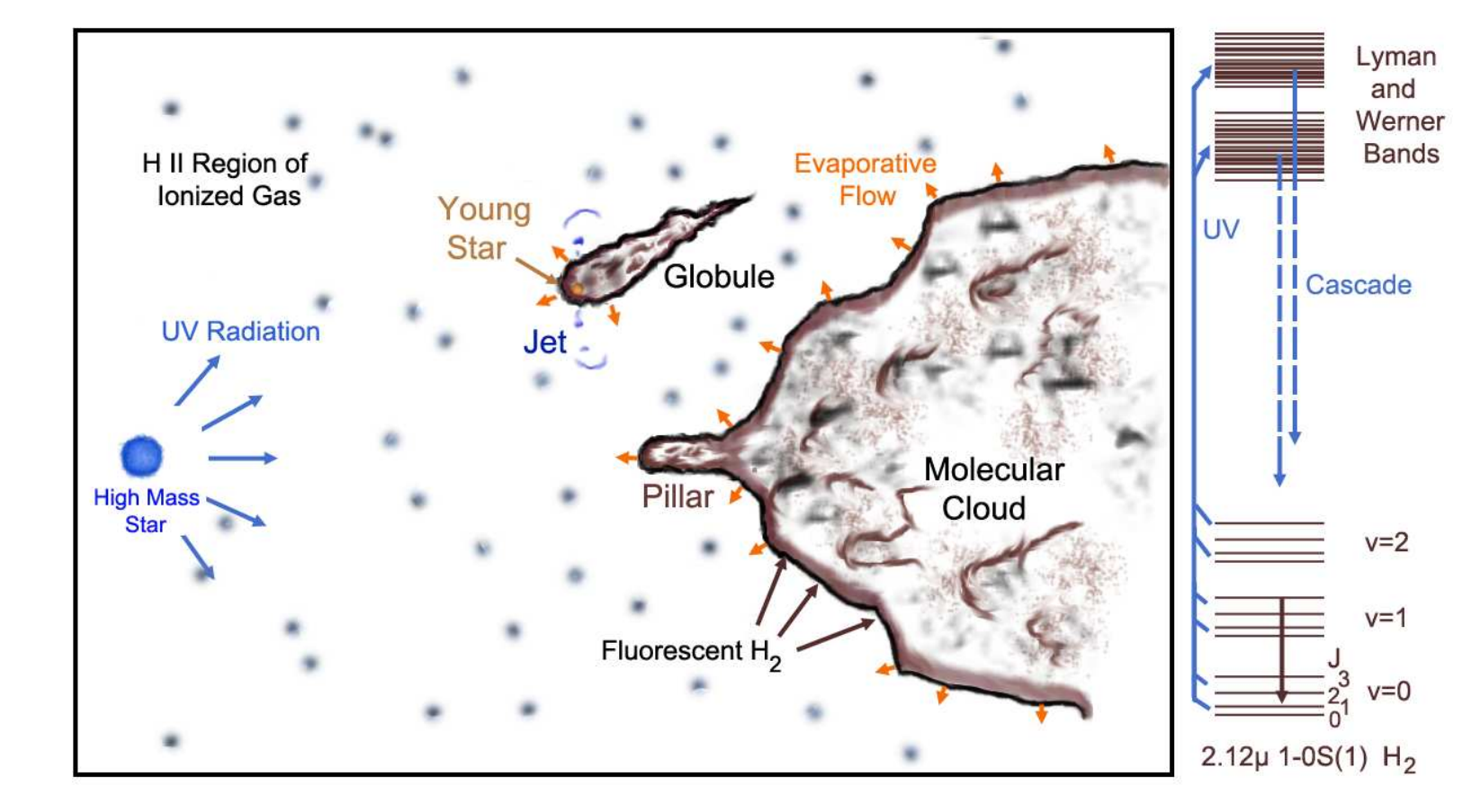}
\caption{Schematic of a photodissociation region. Ultraviolet radiation from an O-star
excites H$_2$ at the interface between an ambient molecular cloud and the H~II region,
and the molecule emits 2.12$\mu$m line radiation as it transitions to the ground state.
Recombination lines of H$\alpha$ and Br$\gamma$ form throughout the H~II region, and are
enhanced as H in the dense photoevaporative flow becomes ionized. Globules and pillars
form in response to the radiation. Jets from young stars create shock fronts that radiate
permitted and forbidden line radiation, and may also excite the 2.12$\mu$m lines of H$_2$
if molecules are present within or alongside the jet.} \label{fig:h2schematic}
\end{figure}

Because molecular clouds are shrouded in dust, extinction may
completely hide PDRs from view at optical and UV wavelengths.
Fortunately, H$_2$ fluorescence makes it possible to observe
PDR interfaces directly in the near-infrared where extinction is an order of magnitude lower. 
Figure~\ref{fig:h2schematic} summarizes the basic physics of this interaction. FUV
radiation from high-mass stars excites H$_2$ into the Lyman and Werner bands, and subsequent
cascades generate strong 2.12$\mu$m quadrupole line emission from the v = 1-0, J = 3-1 transition.
Hence, the 1-0 S(1) line defines the PDR interface, allowing us to observe precisely where the 
FUV radiation is absorbed.

Different regions of star formation possess distinct kinds of irradiated interfaces.
For example, although the Carina nebula (and to a lesser extent the Eagle Nebula)
have a rich array of pillars, globules, and walls, the more massive
but significantly older Cyg OB2 association has mostly isolated globules, while Ara OB1 shows
a single long wall \citep{hedla}. These variations arise from the combined effects of 
differing ages, number and location of massive stars, and density structures within the
parent molecular clouds. 

Located 2.3~kpc from the Sun \citep{smith06a}, Carina is an ideal target to
study firsthand how stars form in the presence of strong radiation. The region contains
$\gtrsim$ 70 O-type stars \citep{smith06b}, and as such is more representative of a typical 
H~II region we might observe in another galaxy (as opposed to regions like Orion that
have $\lesssim$ 10 O stars).  Carina is also young enough to have much
ongoing star formation, but old enough that the most massive stars have
cleared away enough material to reveal a spectacular naked-eye H~II
region replete with a dizzying array of globules and pillars.
Carina spans more than a degree on a side, but with modern imaging systems
it is now possible to acquire high-resolution large-scale images that cover
the entire star-forming region in a single observing run. Some of the most massive stars
in our galaxy are located in Carina, including spectral types as early as O2
\citep{walborn02}, WNH stars \citep{sc08}, and of course,
the famous LBV star Eta Car, a prime candidate for the next Galactic supernova.

Carina provides a snapshot of the early phases of an OB association. Its
massive O-type stars are distributed in several clusters, with the
two largest groups, Tr 14 and Tr 16, situated in the central cavity of the nebula. 
These two groups contain several of the most massive stars in the region, and
they dominate the ionizing flux of the outer Carina Nebula.  
Tr~16 consists of several sub-clusters identified from X-ray surveys
\citep{feig11,wolk11}, while Tr~14 and the nearby cluster
Tr~15 appear more compact.  Typical ages
for the young stars in Tr~14 and Tr~16 are 1$-$3 Myr \citep{ascenso07,hur12}, with some
evidence that Tr~15 is older by several million years \citep{wang11}.  
A substantial fraction of the O-type stars (about half, depending on
whether or not one counts Cr 228 as part of Tr 16; see Smith 2006a) are
scattered across 30-40 pc residing in several smaller sub-clusters
(Tr 15, Cr 228, Cr 232, Bo 10, Bo 11, and others).  The presence of multiple groups of
massive stars make it difficult to determine which source or
sources are responsible for ionizing any given feature
within the Nebula.

Historically, the Carina Nebula was considered to be an evolved H~{\sc
ii} region with little evidence for active star formation, a conclusion based on
initial far-IR surveys that targeted the inner nebula \citep{harvey79,ghosh88}.
The first wide-field IR survey that
revealed the large extent of ongoing star-formation activity was the
{\it Midcourse Space Experiment} ({\it MSX}) imaging of Carina \citep{smith00}.
These wide-field mid-IR images uncovered an extended
network of pillars, globules, bubbles, and embedded IR sources spread
across several degrees.  A number of other studies focused on signs of
star formation and bright individual PDRs in the region
\citep{megeath96,brooks00,brooks01,rathborne02,brooks03,smith04b,rathborne04,smith05}.
\citet{sb07} provide an overview of the
global multiwavelength structure and energy budget of Carina.  Later,
imaging with the {\it Spitzer Space Telescope} clarified much of the
spatial relationship between stars, ionized gas, PAHs, and dust 
and also revealed thousands of embedded protostars and disk excess sources
organized in a hierarchical network of young star clusters spread across 30-40 pc
\citep{smith10b,povich11}.  This collection of
young clusters was confirmed with wide-field X-ray images obtained
with {\it Chandra}, which documented the numerous T
Tauri stars throughout the region \citep[e.g.][]{feig11}.

Following the first ground-based discovery of a protostellar jet in
the region \citep[HH 666; ][]{smith04a}, a survey with the {\it Hubble
Space Telescope} ({\it HST}) Advanced Camera for Surveys (ACS)
revealed $\sim$ 40 jets or candidate jets (Smith et al. 2010a),
confirming the active ongoing star formation in the region.  The {\it
HST} images included a contiguous mosaic of the region surrounding the
Tr 14 and 16 clusters, plus a number of other individual fields in the
South Pillars and other regions.  Many of these protostellar jets
emerge from compact dark globules, also scattered throughout the
region, indicating that at least some of these compact globules must
contain protostars.  {\it Spitzer} also revealed a few candidate
molecular outflows based on their excess emission in the IRAC Band 2
filter (Smith et al. 2010b).  We expand the census of known
outflow sources with the wide-field images in this paper.

Several recent papers have explored the Carina star-forming region
with large surveys at both X-ray, IR, and sub-mm wavelengths.  As part of the
Chandra Carina Complex Project \citep[CCCP; ][]{townsley11,broos11}, the Carina Nebula
was targeted for deep X-ray exposures, and over 14000 point sources were discovered.
In the near-IR, wide-field JHK images of the central portion of the Carina complex
\citep{preib11a,preib11b} with VLT showed that most of the X-ray sources were likely to be
cluster members. In the sub-mm, \citet{preib11c} and \citet{pekruhl13}
surveyed an area of the Carina Nebula with 18-arcsecond resolution over
a square degree and identified all of the principal areas of
warm dust in the region, while \citet{rocca13} published far-IR maps of
the entire region using Herschel with similar spatial resolution.
In the optical, \citet{grenman14} compiled a list of dozens of small dark globules in 
Carina from archival HST H$\alpha$ images of the central portion of the Carina star formation region.
\citet{tapia11} has published H$_2$ 2.12$\mu$m images
of selected objects within Carina, and \citet{ohlendorf12}
discuss mid-IR Spitzer images of the region whose filters include lines of H$_2$. 

Carina, with its strong UV field, modest extinction, and plentiful ongoing
star formation is the premier region for studying how ionizing radiation affects jets 
\citep[e.g.][]{smith10a,ohlendorf12}.
Once jets emerge into the H~II region, they become photoionized by the ambient
ultraviolet light. Cooling from this process complicates the
interpretation of emission line images because the
observed radiation arises from both shock heating and radiative heating, and it can be
difficult to distinguish a photoionized filament from a photoionized jet.
In fact, \citet{rs13} found that most of the H$\alpha$ and [S~II] emission
from several Carina jets traces ionization fronts that arise from external UV radiation.
However, there is also evidence for shielding, in that the dense central parts
of jets appear relatively unaffected by the ambient radiation field
when they first emerge from their natal cloud cores.  In a region as complex as
Carina, searches for jets benefit from molecular tracers such as
H$_2$ to unveil embedded outflows \citep[e.g.][]{stanke00,stanke02,smith10b,lee12,lee13}
in addition to the standard optical tracers of H$\alpha$ and [S~II].

In this paper we compile the deepest H$_2$ and Br-$\gamma$ mosaics to date of the Carina 
star-forming region and compare them with new narrowband optical and broadband
NIR images. The images cover three times the area of the JHK mosaics of \citet{preib11a}, and
reveal subarcsecond detail along dozens of interfaces
where FUV radiation from massive stars is in the process of destroying ambient molecular clouds.
By subtracting appropriately-scaled continuum images from the narrowband ones, we construct
pure emission-line images of the irradiated interfaces.
The observed spatial offsets between the Br$\gamma$ and H$_2$ emission along the PDRs 
are independent of the reddening, and quantify how quickly the molecular gas flows into the
H~II region.  Shocked H$_2$ gas delineates jets while they are embedded deep within a dark cloud,
and complements the spatial structure of jets inferred from H$\alpha$ 
and [S II] emission (cooling zones of shocks and irradiated flows).
Our [O~III] images show the most highly-ionized parts
of pillars, identify the highest shock velocities, and help to quantify how ambient
UV-radiation fields affect shocked gas. 

A main focus of this paper is to identify all the observable outflows and irradiated interfaces in
Carina that are visible in our images, with the aim to provide
a snapshot of the walls, pillars, globules, clusters, jets, and
so on that are present within a typical region of ongoing massive star formation. 
Numerical simulations which include magnetic fields and radiative transfer \citep[e.g.][]{arthur11}
can use the examples shown here to compare with the types of structures present in
the models in order to better understand how star formation evolves in real systems.

\section{Observations} 

\subsection{H$_2$, Br$\gamma$ and K Mosaics}\label{ss:h2brk} 

The near-IR images of Carina in this paper were taken 11 $-$ 18 March, 2011 with the
NEWFIRM camera at the NOAO Blanco 4-m telescope at Cerro-Tololo in Chile.  NEWFIRM
is a large-format imager that consists of four 2048$\times$2048 InSb arrays \citep{newfirm}.
A single exposure covers 27.6 arcminutes with a gap of 35 arcseconds between the
arrays at a scale of 0.4 arcseconds per pixel. Images were taken in J, H, K, and through
narrowband filters centered at H$_2$ 1-0 S(1)
and Br$\gamma$ using the standard NEWFIRM filter set. The H$_2$ and Br$\gamma$ filters 
both have $\Delta\lambda$/$\lambda$ of $\sim$ 1.1\% .  All mosaics were reduced and analyzed
within the IRAF environment.

Because there are gaps between the arrays as well as numerous cosmetic defects,
the usual strategy for observing with NEWFIRM is to obtain at least several dozen dithered 
exposures of a single field and use software to combine the set of images into a single
mosaic; NOAO has devised a pipeline reduction process to perform this task. Unfortunately
the pipeline does not work well for regions such as Carina that have a large amount of 
extended emission, because the pipeline interprets the extended emission as enhanced 
background sky, and subtracts it.  As a result, artificial negative `bowls' surround
all regions of extended emission in the pipeline reductions. 

To solve this problem we observed Br$\gamma$ and H$_2$ in 6$\times$6 grids with two
small ($<$ 1 arcminute) random dithers for each grid point and employed offsets between adjacent
grid points of typically 6 arcminutes. With this strategy each point in the mosaic is observed
enough times to eliminate bad pixels, and the grid covers a wide enough area so that each array
has a substantial number of exposures that have no objects with extended emission, thereby 
eliminating sky-subtraction biases. However, this procedure demands that the sky exposures 
are always free of extended emission, so it is not possible to use
the pipeline default of a group of exposures nearest in time. 
We tested our procedure by using a more time-consuming
method of offsetting 1.5 degrees away for sky during the mosaic. Such large offsets
proved problematic as they introduced significant positional drift during the mosaic, but the
result was similar to what we were able to achieve with our 6$\times$6$\times$2 mosaic.

Exposure times at each pointing were 2 minutes, so the entire exposure time for a single
mosaic is 72 minutes.  Allowing for dithers, telescope settling, and readout, each mosaic takes
about 2 hours to complete in real time. The actual exposure time for a given location in the
mosaic is largest near the center of the mosaic, and decreases towards the edges because fewer
images contribute to the final mosaic at those locations.  Nights with the best seeing were used
for the H$_2$ and Br$\gamma$ observations. The sky was stable enough during the exposure sequence
to enable sky subtraction between any two images in a given mosaic.  Eliminating the mosaics
with the worst seeing, we were left with 4 mosaics in Br$\gamma$ and 6 mosaics in H$_2$.
Combining these produced the final mosaic for each filter.  The resulting H$_2$ and Br$\gamma$
mosaics have excellent image quality (FWHM images of 0.74 arcseconds and 0.79 arcseconds,
respectively), each span about 1.1 degrees on a side, and have central exposure times of
3.5 hours for H$_2$ and 2.4 hours for Br$\gamma$ (Table 1). Distortion corrections and
final coordinate mapping for the NEWFIRM images are determined by the pipeline process, and
make use of known positions of bright stars in the 2MASS catalog.  A check of the
coordinates for several of the brightest stars in the final mosaics against their
position in the SIMBAD database shows errors $\lesssim$ 1 pixel (0.4 arcseconds). 

We also imaged Carina in J, H, and K, with the aim of establishing a photometric
database for the region, and as an aid to subtracting ambient continuum radiation as
described below.  Large-scale nebular emission is less of a problem for these filters, so 
we could acquire them using small mosaics under less favorable weather conditions,
and later combine them to produce full mosaics.  As such, the image quality in the
J, H, and K-band mosaics, while typically 0.7 arcseconds, varies across the field of
view and approaches 2 arcseconds in a few areas.  

Interpreting narrowband images of regions like Carina can be problematic because 
reflected light from dust may be present along the same interfaces that produce
H$_2$ and Br$\gamma$ emission lines. However, because the wavelengths of H$_2$ and Br$\gamma$ are
nearly the same, the reddening is essentially identical at each point in the two images; hence,
the continuum can be removed well by simply forming a scaled difference image (H$_2$ $-$ C$\times$Br$\gamma$),
where the scale factor C takes into account the differences in the filter bandpass widths.
The NEWFIRM filter bandpasses imply C=1.19 for flat continuum sources. To verify this number
we performed photometric measurements on a sample of stars in the final mosaics and found
C = 1.19 $\pm$ 0.05. All H$_2$ $-$ Br$\gamma$ difference images shown in this paper include
this correction factor. While extended continuum subtracts out well in the H$_2$ $-$ Br$\gamma$
composites, stellar images typically leave a dark core surrounded by a light halo because the seeing
was slightly better in the final H$_2$ composite than it was in Br$\gamma$. Stars that have strong
Br$\gamma$ emission will appear white in the difference images.

The H$_2$ $-$ Br$\gamma$ difference images are very effective for showing gradients in
PDRs. However, quantifying the spatial offsets between the emission lines requires comparing
spatial profiles across the two narrowband images, and these profiles include continuum.
In principle one can remove the continuum algebraically by using the H$_2$, Br$\gamma$, and K-band
filters, scaling from their respective bandpasses and noting that any H$_2$ and Br$\gamma$
emission lines also contribute to the K-band flux (essentially three equations and three
unknowns at each point in the image). While mathematically correct, the different seeing in the
three mosaics leads to poor subtractions of point sources and distracting residuals at
the locations of the stars. Hence, we have performed this procedure only along
the edges of pillars and bright PDR interfaces where we measure the spatial offsets
between H$_2$ and Br$\gamma$ (see section \ref{ss:spatial_offsets}).

\subsection{H$\alpha$, [S II], [O III], I-band, and Spitzer IRAC Mosaics}\label{ss:sii-oiii-i-spitzer} 

We obtained narrow-band images of the Carina Nebula on 9 $-$ 11 March 2003
using the 8192 $\times$ 8192 pixel Mosaic 2 imager mounted at the prime focus of
the CTIO 4-meter Blanco telescope.  This camera has a pixel scale of 0.27 $\arcsec$
~pixel$^{-1}$ and a 35.4\arcmin\ field of view.  We used $\Delta \lambda$ = 80 \AA\
interference filters centers on the $\lambda$ = 5007 [OIII] line, the $\lambda$ =
6563 H$\alpha$ line, and the $\lambda \lambda$ = 6717 / 6731  [S~II] doublet, and
a broad-band filter transmitting the SDSS i band.  In each filter, we took several
individual exposures with slight positional offsets to correct for gaps in the CCD
array and to correct for detector artifacts.

We reduced the data in the standard
fashion with the MSCRED package in IRAF, and absolute sky coordinates were computed
with reference to USNO catalog stars. Emission-line images were flux calibrated
by comparison with images of the Orion Nebula obtained during the same run, and
the Orion images were calibrated with reference to HST data.
Distortion corrections for the Mosaic images were determined by measuring $\sim$ 100
stars in common between the optical and NEWFIRM images, resulting in alignment errors
$\sim$ 0.1 arcsecond RMS between the optical and IR images.

In this paper we also make use of existing Spitzer images of the region.
\citet{smith10b} targeted two regions in the Carina nebula with Spitzer for deep imaging
in order to search for lower-mass protostars
(1-3 M$_{\odot}$).  Based on extended mid-IR emission in lower-resolution images from
the MSX survey, those authors observed the South Pillars to further study the
complex molecular emission and evidence for current star formation amid a large
number of recently formed stars.  The western region was chosen as an edge-on
view of an ionization front and photodissociation region that also shows signs
of star formation.  These data were reduced using the GLIMPSE pipeline, and are
described in more detail in \citet{smith10b}.

In the following sections we consider the PDR interfaces as defined by H$_2$ fluorescence, and then
turn to what the NEWFIRM and MOSAIC images reveal about outflows.

\section{Observations of PDR Interfaces} 

\begin{figure} 
\centering
\includegraphics[angle=0,scale=1.00]{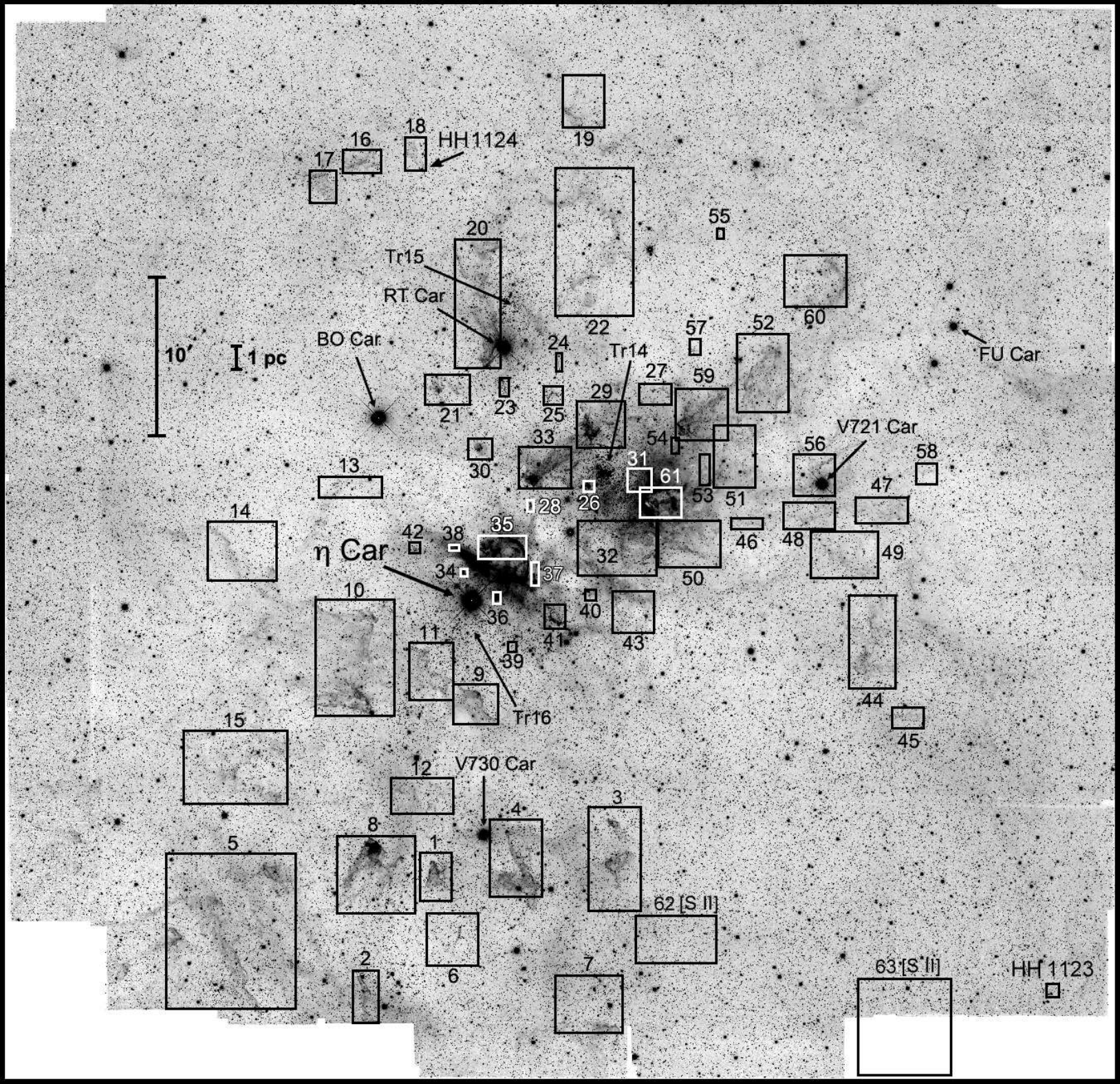}
\caption{Overview of the Carina region in H$_2$ 2.12$\mu$m.
Numbered boxes mark the regions of H$_2$ emission
highlighted in Fig. 3.  The scale bar assumes a distance of 2.3 kpc. Coordinates for
the centers of each box are compiled in Table 2. The clusters Tr 14, Tr 15, Tr 16, and the brightest
stars are marked.  The image spans $\sim$ 1.1 degrees
on a side, with a scale of 0.4 arcseconds per pixel. North is up and east to the left
in this and subsequent images.} \label{fig:h2_summary}
\end{figure}

Figure~\ref{fig:h2_summary} identifies 61 areas of interest in the H$_2$ and Br$\gamma$ mosaics
that are reproduced in Figs.~\ref{fig:spillars} to \ref{fig:western_wall}.
Each area has some some notable aspect of H$_2$ or Br$\gamma$ emission, in the form
of a spatially-resolved irradiated surface, a jet, cavity, bow shock or distinct
nebular feature.  These areas are located in one of eight larger regions in Carina:
the Southern Pillars (Figures~\ref{fig:spillars} and \ref{fig:spillars2}),
the Eastern Walls (Figures~\ref{fig:sewall}, \ref{fig:sewall2}, and \ref{fig:sewall3}),
the Northern Pillars (Figure~\ref{fig:npillars} and \ref{fig:npillars2}),
the vicinity of Tr 14 (Figure~\ref{fig:tr14}),
a region around $\eta$ Car (Figure~\ref{fig:eta}),
the Southwestern Pillars and Walls (Figure~\ref{fig:swpillar}),
the Northwestern Pillars and Walls (Figure~\ref{fig:nwpillar}),
and the Western Wall (Figure~\ref{fig:western_wall}). 
In what follows we collect the various types of objects together and refer to them 
by their area number. 

\subsection{Irradiated Walls, Pillars, and Globules}\label{ss:pillars} 

Several molecular `walls' of irradiated H$_2$ with lengths 1~pc - 5~pc exist within the boundaries
of the survey. 
As expected, pillars and waves along these walls form on the sides that face the ionizing sources
in the center of mosaic, though it is difficult to assign a definitive exciting source owing to the
large number of O stars in the region. Overplotting the 94 candidate OB stars of
\citet{povich11b} on the H$_2$ composite did not yield any insights into the identities of the
exciting sources.

The walls generally divide into three categories: (1)
{\it structured walls with few or no pillars} as exist in Areas 5, 9, 16, 46, 47, 51, and 56; (2)
{\it structured walls with `chunky' pillars that have aspect ratios 1 $\sim$ 2} in
Areas 3, 7, 10, 11, 14, 33, 44, 48, 49, 52, 53, 59, 60, and 61; and (3) 
{\it structured walls with well-defined pillars} such as those in Areas 22, 27, 29, 32, 44, and 50.
There does not seem to be any obvious relationship between the (projected) distance of the walls from 
Tr 14 and Tr 16 and the presence of pillars. For example,
pillars exist along the walls in both Areas 29 and 44,
the former situated only about 1.5 pc from Tr 14, while the latter lies at the western extreme of the
H$_2$ emission a bit more than 10 pc away in projection from both Tr 14 and Tr 16.

The pillars come in a variety of aspect ratios, ranging from those with rounded heads and 
aspect ratios $\sim$ 5 (Areas 1, 3, 8, 14, 15, 17, 18, 27, 31, 45, bottom of Area 22) to pillars
that are very long and narrow (Areas 2, 4, 21, 23, 24, 37, top-right corner of Area 10).
It is common for pillars to exhibit multiple arcs along their sides, implying a shouldered or
scalloped shape (Areas 1, 2, 4, 18, 20, 21, 27, 33, 45). Some pillars have one
one or more bright NIR continuum sources situated near their apices (Areas 2, 3, 4, 8, 18, 21,
bottom of Area 10), while others do not (Area 1, 14, 17, bottom of Area 22). There are
many background and foreground stars in the region that could by chance coincide with the apex of
any given pillar. However, in some cases (e.g. Area 4) enhanced Br-$\gamma$ emission at the location
of the NIR source associates the source with the pillar.

Isolated irradiated globules and pillars that have detached from their parent clouds
exist throughout the region, and abound in the vicinity
of Eta Car and Tr 14. Some of these show a classic head-tail `proplyd' shape
(Areas 23 and 24), while others are much more irregular (Areas 25 and 28). Several of these objects
have one or more bright NIR continuum sources near their apices
(Area 6, 21, 23, 24, 25, 40, 41, 55, upper left corner of Area 15), while others do not show any
obvious NIR continuum sources near the head of the pillar
(Area 12, 28, 32, 35, 36, 38, 39, 54, the object to right of main pillar in Area 4,
and small globules in Area 6). In many cases the NIR source at the apex may drive a jet, but these
can be difficult to detect at ground-based spatial resolution, and also because in the absence of
strong extinction, Br$\gamma$ is considerably fainter than H$\alpha$ for both shocked and irradiated
flows.  For example, \citet{smith10b} identified the star at the southern end of
the pillar in Area 24 of Fig.~\ref{fig:tr14}
as the driving source for the HH 1013 flow they discovered with HST ACS images, but these
HH objects are too small to resolve against the variable background in our ground-based images.

A collection of H$_2$ emission features in Fig.~\ref{fig:h2_summary} in the Eastern Wall region
that extends from Area~13 through Area~14 in Fig.~\ref{fig:sewall2} has some unusual
characteristics. Unlike the major pillars highlighted
above, all of which are bright in 70$\mu$m Herschel images \citep{preib12}, these regions of
H$_2$ emission are not readily-visible with Herschel. However, there is a good correlation between 
absorption in our H$\alpha$ images and these H$_2$ structures, so some dust exists in these regions.
The most likely explanation is that we are observing a low-extinction PDR that lacks sufficient
dust column density to produce observable far-IR continuum.

\subsection{Spatial Offsets between H$_2$ 1-0 S(1) and Br$\gamma$}\label{ss:spatial_offsets} 

Throughout the Carina Nebula we consistently observe spatial offsets between Br$\gamma$ and 
H$_2$ emission along irradiated interfaces in the sense that Br$\gamma$ is located closer to the
source of ionizing radiation. This spatial offset is typically $\sim$ 1$\arcsec$, which corresponds
to $\sim$ 20 times the RMS scatter between the stellar positions in the Br$\gamma$ and H$_2$ images.
This offset is independent of the reddening, and a potentially important diagnostic for PDRs.
The offset involves a rather
complex interplay between the atomic, molecular and ionized gas, the dust, and the radiation field.
Both dust and gas serve to shield the molecular cloud from the radiation, but the opacities
vary strongly with wavelength at ionization thresholds, and the situation is complicated by
photoablation that causes newly-ionized gas to flow away from the cloud. Numerical models that
include the requisite atomic and molecular radiative transfer as well as dynamics are just
becoming available \citep[e.g.][]{arthur11}, so quantifying these spatial offsets is a
timely endeavor from the observational standpoint, and will hopefully motivate numerical models
to predict offsets from their codes.

We chose the spectacular irradiated interface associated with G287.38-0.62
(\citealt{retallack83,brooks03}; the 'Western Wall' in Area~61 of Fig.~\ref{fig:western_wall};
see Fig.~\ref{fig:western_color} for a color composite) to study emission line
offsets in PDRs. This interface is very bright in all emission lines, and is large
enough to allow measurements to be made along the periphery of the cloud where the
incident ionizing radiation is more oblique, and near the apex where the radiation is more
normal to the surface.  One complication is that
irradiated interfaces emit continuum radiation from reflected starlight. 
The filter bandpasses for Br$\gamma$ and H$_2$ have
somewhat different widths, and so subtraction between the two images could potentially 
affect the spatial distribution of the emission.  We removed the continuum using the
procedure described in section~\ref{ss:h2brk}.

Intensity profiles extracted along the four line segments shown in Fig.~\ref{fig:western_color}
appear in Fig.~\ref{fig:lineout}.  We define the position of the ionization front as the location
where the normalized H$\alpha$ and [S~II] profiles intersect. In practice, this location is
simply a convenient fiducial point. In each panel there is a small offset
on the order of $10^{16}$ cm between the peaks of the H$\alpha$ and [S~II] emission, with the 
H$\alpha$ occurring, as expected, closer to the ionizing source.  The decline of the integrated
H$\alpha$ flux as the line of sight proceeds into the PDR also occurs in Br$\gamma$. The lack of 
significant spatial offsets between these two hydrogen recombination lines implies that 
differential extinction from dust does not affect these profiles significantly. The fluoresced
H$_2$ emission is offset by $\sim$ $5\times 10^{16}$ cm from the Br$\gamma$
(one arcsecond corresponds to $3.4\times 10^{16}$ cm).
The profiles look similar to one-another even though we extracted them from different locations
and projected angles relative to $\eta$ Car and Tr~16.
When interpreting these spatial offsets one must keep in mind projection effects caused by
the curvature of the globule. Ideally, numerical models would predict spatial emission
profiles integrated along the line of sight to compare with these observations.

\section{Jets, Outflows, and Nebulous Objects} 

Narrowband H$_2$ imaging provides a view of the molecular outflows that are
obscured at visual wavelengths and that may be unresolved with Spitzer's
relatively coarse angular resolution at $\lambda > 3$ \micron.  Jets
with H$_2$ emission are usually still embedded in their natal clouds
and shielded from the UV radiation that can dissociate molecules in
exposed parts of the flows.  In this section we first consider what the new
images reveal about several previously-known flows, go on to identify new molecular outflow
candidates based on their H$_2$ and [S~II] morphologies, and then discuss several
peculiar nebulous objects.  Half of the candidate molecular flows fall within the
field of view imaged with HST, whereas both new optical flows are in locations only imaged from
the ground.  Coordinates for the new candidate outflows are listed in Table~2.

\subsection{Known Outflows}\label{ss:old_flows} 

\citet{smith10a} detected 39 candidate and confirmed HH jets in the Carina nebula,
only one of which had been observed before in ground-based images
\citep[HH 666;][]{smith04a}. In hindsight, several of the HH jets
discovered with HST are also visible in earlier ground-based images, but except for
HH~666, the jet-like morphologies of the HST objects were unclear in the ground-based data.
Thirty-five of the HST jets in Carina fall within the field of view imaged with NEWFIRM (the remaining
four are in NGC 3324). While a few of these jets have H$_2$ emission,
most are too narrow or too faint to be discerned readily in the IR images. Nevertheless,
the NEWFIRM images often provide valuable insights as to how irradiated 
pillars, globules, and the driving stars relate to the jets. 
In this section we discuss what the new images reveal about several of
the most notable jets.

\begin{figure} 
\centering
\includegraphics[angle=0,scale=1.00]{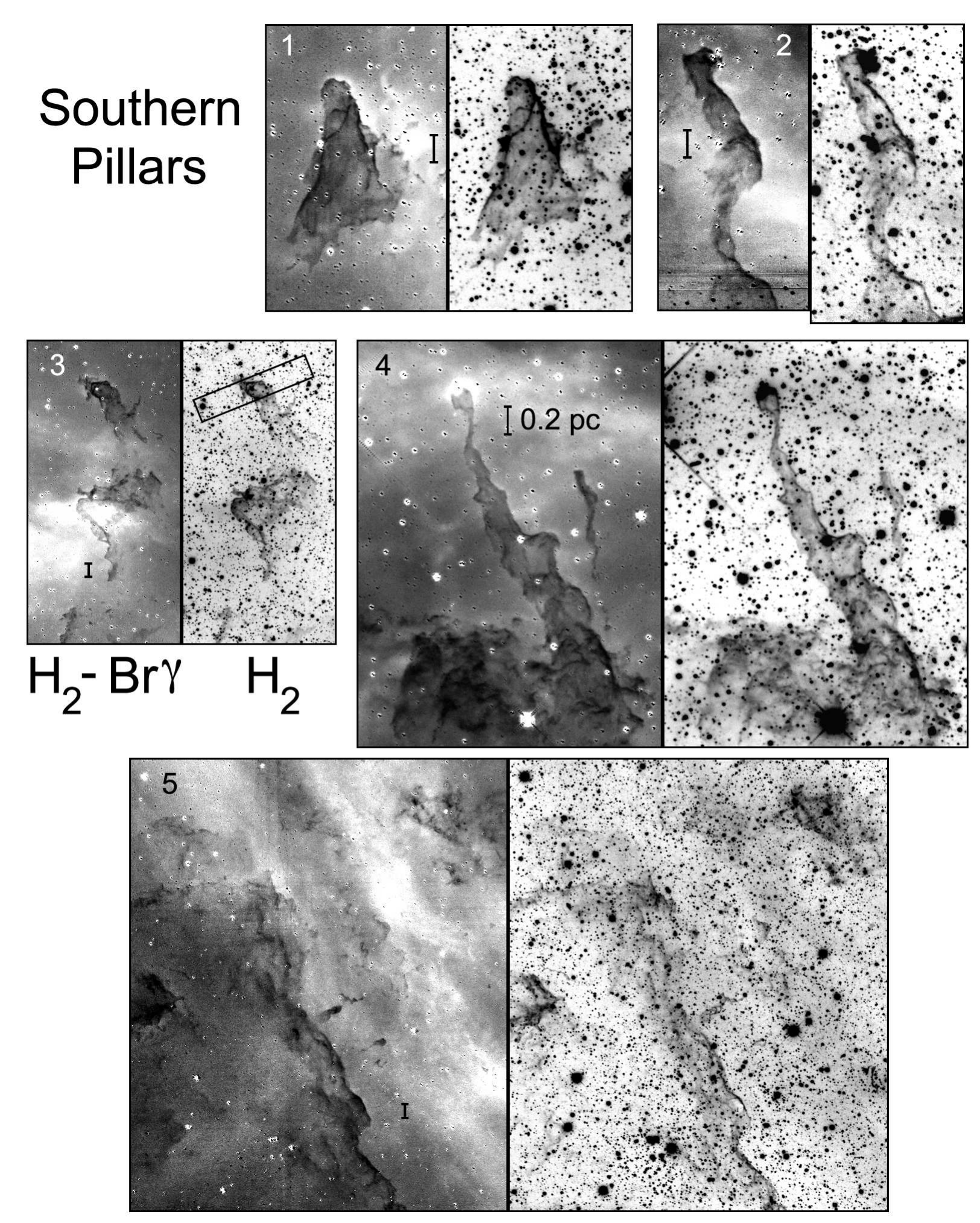}
\caption{The Southern Pillars of H$_2$ emission in Carina. Numbers refer
to the Areas 1 $-$ 5 marked in Fig~\ref{fig:h2_summary}. Right: H$_2$ 2.12$\mu$m image, Left: Difference
image of H$_2$ $-$ Br$\gamma$, with Br$\gamma$ emission relatively brighter where the
image is white, and H$_2$ relatively brighter where the image is black. The scale bar
of 0.2 pc corresponds to a distance of 2.3 kpc.  The rectangle in Area 3 denotes the region
covered in Fig.~\ref{fig:hh666}.} \label{fig:spillars}
\end{figure}

\begin{figure} 
\centering
\includegraphics[angle=0,scale=1.00]{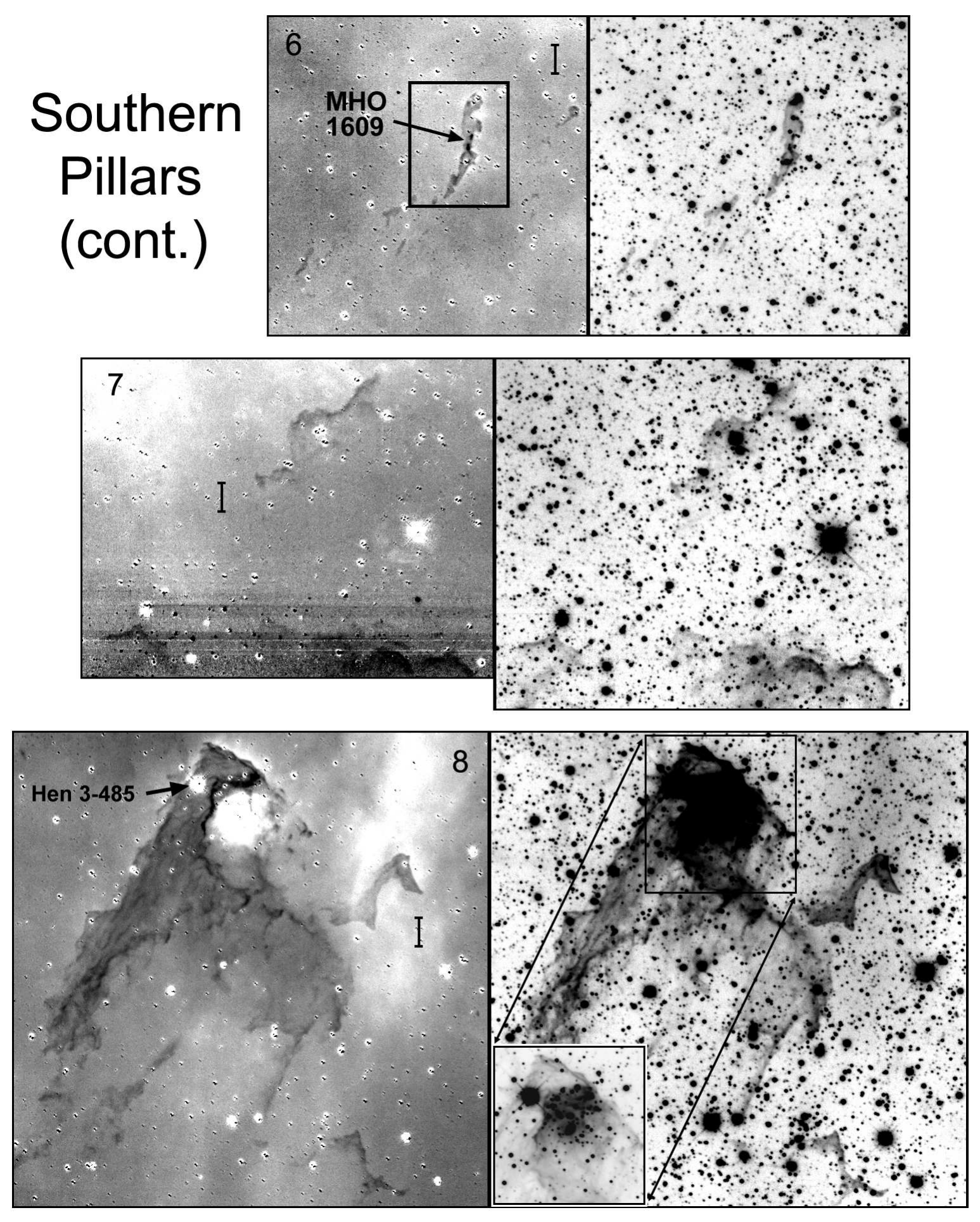}
\caption{Same as Fig.~\ref{fig:spillars} for additional Southern Pillars of H$_2$ emission in the Carina Nebula.
Numbers refer to the Area 6 $-$ 8 marked in Fig.~\ref{fig:h2_summary}. The MHO~1609 region
is expanded in Fig.~\ref{fig:MHO1609}. The inset in Area 8 is the embedded cluster known as
the Treasure Chest.} \label{fig:spillars2}
\end{figure}

\begin{figure} 
\centering
\includegraphics[angle=0,scale=1.00]{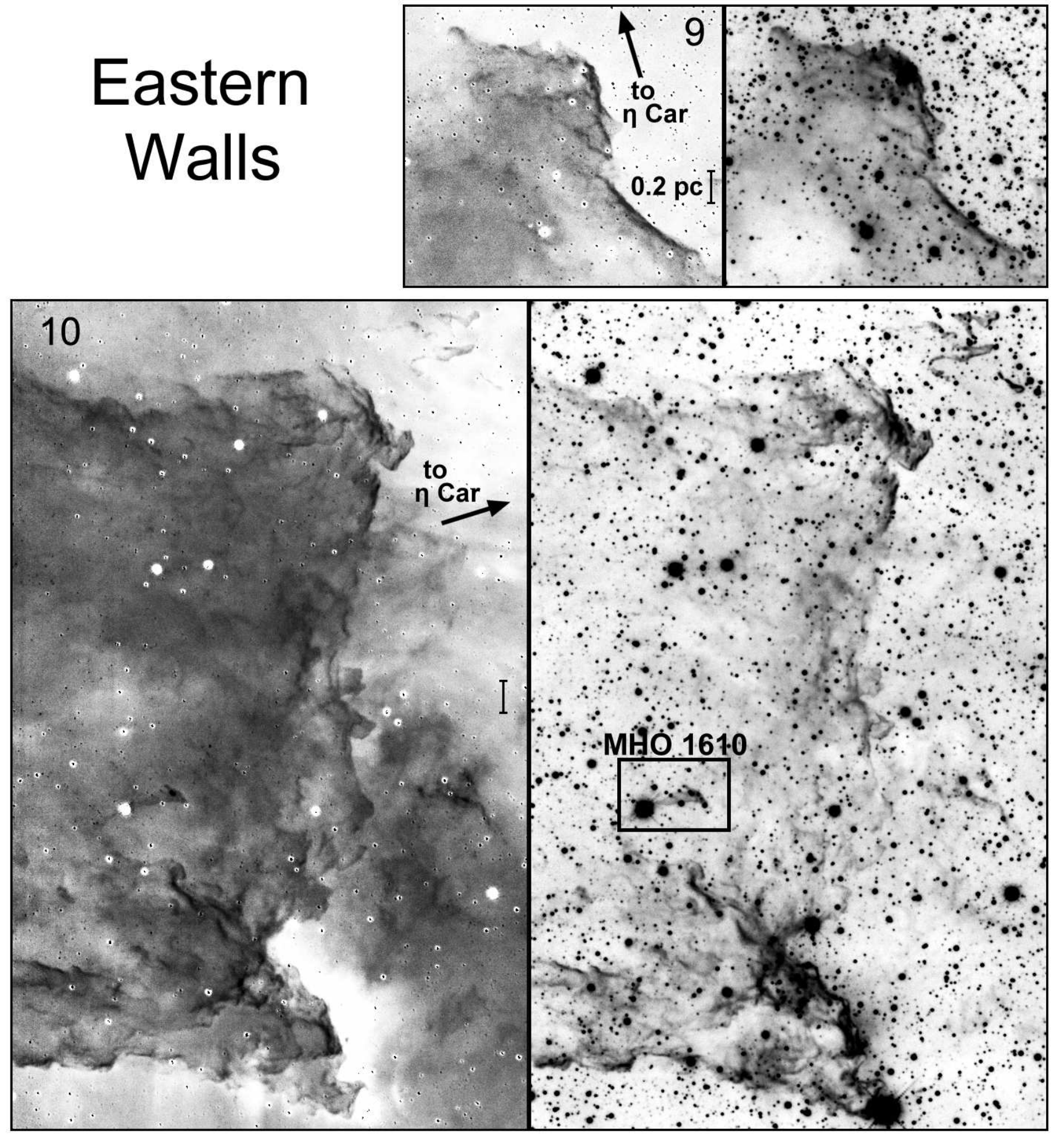}
\caption{Same as Fig.~\ref{fig:spillars} but for Areas 9 and 10 of the Eastern
Walls in Fig~\ref{fig:h2_summary}. The boxed region surrounding MHO 1610 is enlarged in
Fig.~\ref{fig:MHO1610}.} \label{fig:sewall}
\end{figure}

\begin{figure}  
\centering
\includegraphics[angle=0,scale=1.00]{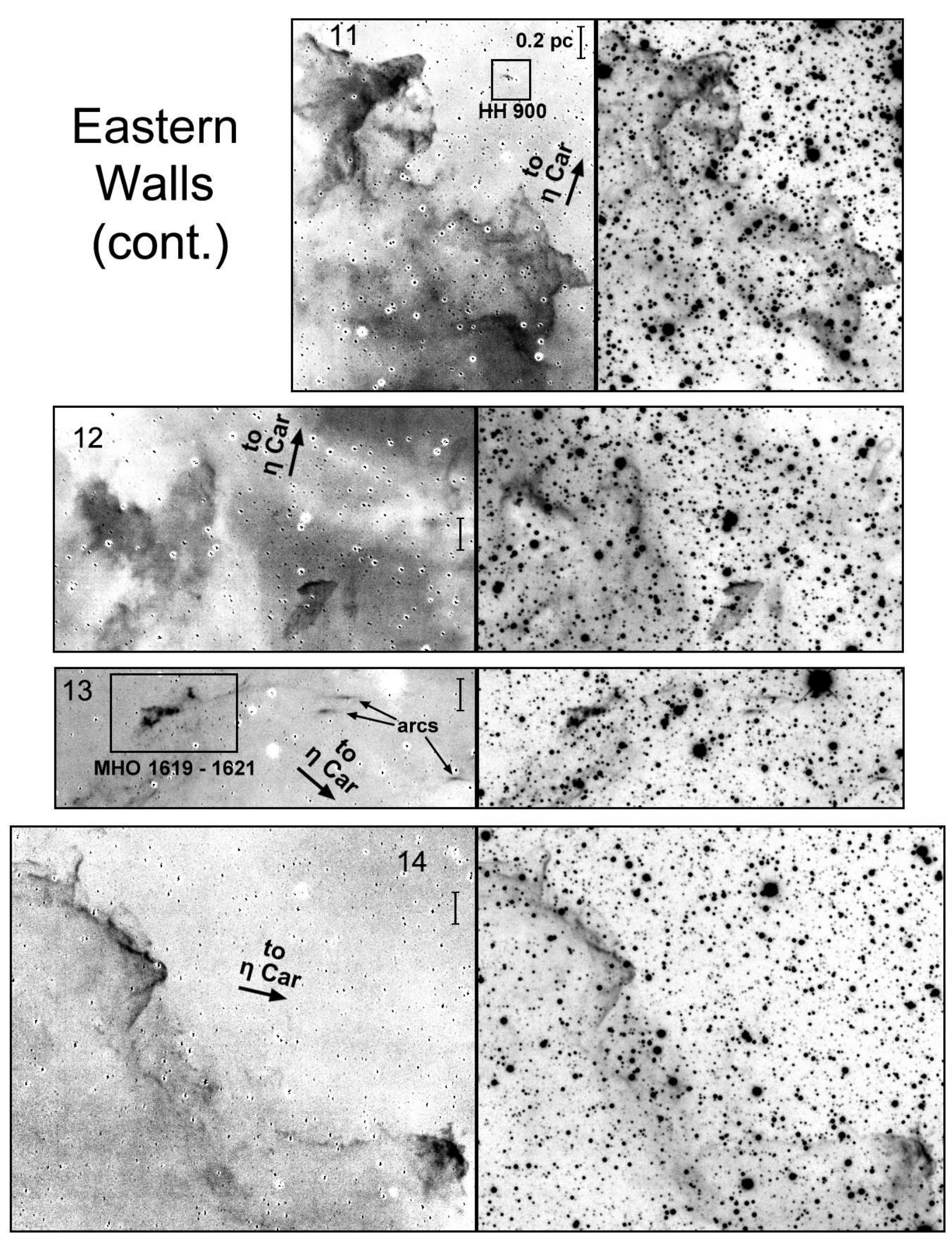}
\caption{Same as Fig.~\ref{fig:spillars} but for Areas 11 through 14 of the
Eastern Walls in Fig~\ref{fig:h2_summary}. The small boxed region in Area 11 is
HH~900, reproduced in Fig.~\ref{fig:hh900}, and MHO 1619 $-$ MHO 1621 in Area 13
is expanded in Fig.~\ref{fig:MHO1619}. The direction to $\eta$ Car is shown.} \label{fig:sewall2}
\end{figure}

\begin{figure}  
\centering
\includegraphics[angle=0,scale=1.00]{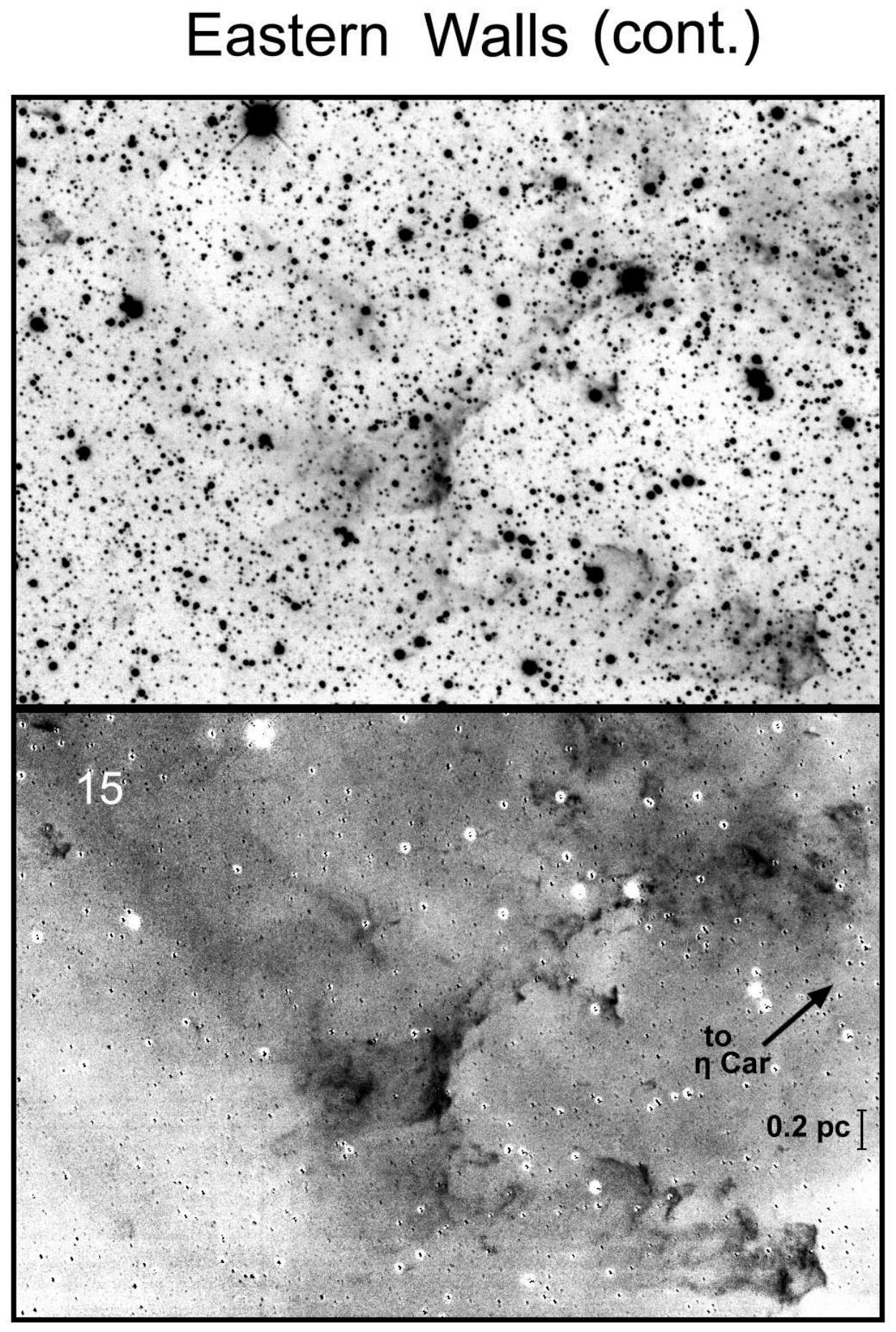}
\caption{Same as Fig.~\ref{fig:spillars} but for Area 15 of the Eastern
Walls in Fig~\ref{fig:h2_summary}. The H$_2$
image is at top with the difference image below it.} \label{fig:sewall3}
\end{figure}

\textit{HH 666}:
HH 666 is a straight bipolar jet that emerges from the head of a dust pillar
in the bright-rimmed globule G287.57-0.91 (\citealt{smith04a}; Area 3 in Fig.~\ref{fig:spillars};
Fig.~\ref{fig:hh666}).  The pillar itself points toward $\eta$ Carinae.
Optical imaging of this parsec-scale jet shows a chain of knots emitting in
H$\alpha$ and [S II] along the jet axis outside the globule, while narrow-band
near-IR [Fe~II] images trace the flow into the obscured parts of the globule and
back to the embedded driving source \citep{smith04a,rs13}. The composite image
in Fig.~\ref{fig:hh666} shows that knot E appears weakly in [O~III], which could arise
from a fast shock ($\gtrsim$ 100 km$\,$s$^{-1}$) into neutral gas, or from a slower shock into
ionized gas. 

NEWFIRM imaging uncovers some H$_2$ emission inside the pillar, but none that is
unambiguously associated with the jet.
The H$_2$ $-$ Br$\gamma$ image in Fig.~\ref{fig:hh666} shows H$_2$ emission along
what appears to be the edges of a V-shaped cavity that originates at the source and
opens to the northwest along the axis of the blueshifted jet \citep{smith04a}.  This feature 
has a similar morphology to the infrared reflection nebula seen in the
IR images of HH 666 \citep{rs13}, but it must have H$_2$ emission
because it persists strongly in the H$_2$ $-$ Br$\gamma$ image while reflected continuum would not.
Additional H$_2$ emission from the northern edge of the flow on both sides of the
driving source may trace the molecular material excited in the walls of an outflow
cavity cleared by the flow, although this is not certain given the complex filamentary background.

The driving source of the flow, HH666-IRS \citep[located at 10:43:51.3 -59:55:21.2;][]{smith04a},
does not appear in the X-ray catalog of \citet{broos11}. The object has a negative residual
(white) in the H$_2$ $-$ Br$\gamma$ image in Fig.~\ref{fig:hh666}. The 
ratio of the stellar fluxes in the Br$\gamma$ and H$_2$ narrowband images
is enhanced by $\sim$ 15\%\ relative to the median of stars in the field
and to that expected for a flat continuum. However, the very red SED
of HH666-IRS \citep[Fig.~9 of ][]{smith04a} will generate about 11\% more
flux in Br$\gamma$ than in H$_2$ owing to the 44~nm redder central
wavelengths of the Br$\gamma$ filter. If the remaining
4\%\ excess were caused solely by line emission at Br$\gamma$, it would imply an
emission equivalent width of about 4\% of the filter bandpass, or $\sim$ 9\AA ,
similar to what is observed for actively-accreting classical T Tauri stars and
Ae stars, both of which show correlations of Br$\gamma$ line luminosity with disk
accretion rates \citep{muzerolle98,donehew11}. Hence, the observed Br$\gamma$/H$_2$
ratio for HH666-IRS is in line with that expected given the extremely red
observed SED of the source, and Br$\gamma$ emission could explain the remaining
small excess. However this inference will need to be confirmed spectroscopically.

\textit{HH 900}: 
A dark globule (visible in Area 11 in Figure~\ref{fig:sewall2}, and in the [O~III] image in
Fig.~\ref{fig:hh900}) has a narrow ($\sim$ 0.2$^{\prime\prime}$) rope-like tail that extends
to the northeast that is visible only at HST resolution \citep[Fig.~4 of][]{smith10a}.
The globule lies within a wider $\sim$ 2$^{\prime\prime}$ streak of H$\alpha$ emission
that extends along the same direction on either side of the globule and points towards of a pair of
oppositely-directed bow shocks. \citet{smith10a} identified the H$\alpha$ streak as
a jet associated with the globule, but also noted the presence of a star near the western
edge of the streak that may drive a separate microjet.

Of the three brightest stars near the globule (marked with `+' in Fig.~\ref{fig:hh900}),
only the one labeled IRS (also known as 2MASS J10451881-5944238 or PCYC~838; \citealt{povich11}) 
at the western end of the H$\alpha$ streak appears unusual
in that it is significantly brighter in Br$\gamma$
than it is in H$_2$, indicative of a very red source and/or one with strong Br$\gamma$
emission (as described for HH666~IRS above). The spectral energy distribution of this
source based on 2MASS and Spitzer photometry exhibits a strong mid-IR excess
\citep{povich11}.  The best fit of the grid of YSO models of \citet{robo06} obtained
with the on-line SED fitting routine \citep{robo07} occurs for a nearly-edge-on
disk around a 2.3 M$_\odot$ star of age 8.9$\times 10^4$ yr, with a total luminosity
of 45 L$_\odot$ and a disk accretion rate of $1.1\times 10^{-9}$ M$_\odot\,$yr$^{-1}$.
However, uncertainties on such model fits can be large \citep{offner12}; for example,
\citet{povich11} report log L = 2.1 $\pm$ 1.4 L$_\odot$ and M = 2.5 $\pm$ 1.2 M$_\odot$ for
this source.

The H$_2$ emission in Fig.~\ref{fig:hh900} is brightest
on the northern side of the dark globule, and extends along the H$\alpha$
streak both to the northeast and to the southwest.  The H$_2$ is offset 
to the south of the H$\alpha$ as would occur if the globule were irradiated
by sources in Tr 16 to the north. However, further interpretation of this
unusual object is difficult to do from ground-based images because many of
the key features such as the dark rope-like tail are unresolved, and two
potential exciting sources are superposed upon the H$_2$, [S~II], and
H$\alpha$ nebular structures.  New HST images and spectroscopy of the object
will appear in a future work that will investigate this object more fully
(Reiter, Smith, \&\ Bally 2014; in preparation).

\textit{HH 901}: HH 901 is a jet that emerges from the head of a pillar 
that is irradiated by O stars in Tr 14 and Tr 16
(\citealt{smith10a}; Area 29 of Fig.~\ref{fig:tr14}).
Our images of this source in Fig.~\ref{fig:hh901} show the typical structure of an irradiated
pillar, with H$_2$ emission surrounded by limb-brightened H$\alpha$ and Br$\gamma$. A
diffuse arc which is bright in H$\alpha$ and [O~III] cuts across the middle of the
pillar. This feature lies along the extension of the bow shock, although
the high resolution HST images suggest the bow shock and the arc may be unrelated to one another.

The H$_2$ $-$ Br$\gamma$ panel in the upper right of Fig.~\ref{fig:hh901} shows that the
HH 901 jet emits in H$_2$ as it emerges from the pillar. An H$_2$ knot marks the location
where the jet emerges from the west side of the pillar, while it is possible to trace
the jet in H$_2$ for about 0.02~pc to the east of the pillar.
If H$_2$ emission came only from the irradiated edge of the pillar, we would
expect the emission to trace the ionization front and
peak at the apex of the arc facing Trumpler 14.  Instead, the H$_2$ emission is bright at the
edge of the globule on the west side, and extends along the optical jet to the east, arguing
that H$_2$ originates in the outflow itself.

A continuum source labeled IRS in Fig.~\ref{fig:hh901}
appears along the jet near the eastern edge of the pillar in
the K-band and Br-gamma images at 10:44:03.58 -59:31:01 (2000), and is also visible
in the I-band image. However, the stellar density is high enough
in the near-IR that chance superpositions are a concern, and in HST images
\citep{rs13} the source lies above the jet axis and just outside the edge of the pillar.
The object was not listed in the catalog of X-ray-selected members of Carina \citep{broos11}.

\textit{HH 902}:
Situated just above HH~901, the HH~902 jet is similar in that it originates from the tip
of an irradiated pillar, though in this case the pillar is much wider
(\citealt{smith10a}; Area 29 in Fig.~\ref{fig:tr14}; Fig.~\ref{fig:hh902}).
A jet appears to the west of the pillar in the H$\alpha$ images in Fig.~\ref{fig:hh902},
and also somewhat less-distinctly in Br$\gamma$. 
The situation to the east is more complex. The HST image reveals a
complex morphological structure in H$\alpha$ where the projection of the jet intersects
the base of the pillar that contains HH~901. \citet{rs14} show that the eastern 
and western knots both move away from the pillar along the axis of the jet.
This region has a strong continuum component in our K-band images. 

The H$_2$ emission near HH 902 is
rather unremarkable, and simply follows the outline of the pillar.
A small nebulous knot in the H$_2$ image along the jet labeled `cont' in Fig.~\ref{fig:hh902}
appears to be a continuum source, since its flux ratio of H$_2$/K is similar to that of
field stars. The object also appears in HST images in [Fe II] \citep{rs13}, and is likely
to be unrelated to HH~902 because knots on either side of it have
proper motions in the same direction \citep{rs14}.
A star in the K-band image exists near the head of the pillar at coordinate
J2000 10:44:01.61 -59:30:29 \citep{ohlendorf12}.  However, the star field here is fairly dense, and there are
no other indications (e.g. very red colors, reflected-light cavities, X-ray emission) that
this source is driving the jet. Its position is about 2 arcseconds away from where 
\citet{smith10a} suspected the driving source should be located based on morphologies in
their HST images, and is also about 2 arcseconds north of the axis defined by
the H$\alpha$ proper motions measured by \citet{rs14}. 
As for the other irradiated globules and pillars, the host pillar for HH~902 shows a spatial
offset between H$_2$ and Br$\gamma$ in its difference image (Figs.~\ref{fig:tr14} and \ref{fig:hh902})
in the sense that the Br$\gamma$ forms a sheath on the outside of the H$_2$ emission from the 
pillar.

\textit{HH 1066}: The newly-confirmed jet HH 1066 (formerly HH c-1) is driven from the
head of a bright C-shaped proplyd \citep{rs13}. The proplyd is superposed
upon a larger pillar in the background that itself has small-scale structure suggestive of
at least three more pillars (\citealt{smith10a}; Area 29 of Fig.~\ref{fig:tr14}). The illumination source is to the
south, in the general direction of Tr 14. The composite images in Fig.~\ref{fig:hh1066}
show the proplyd and the large pillar well in H$\alpha$, [S~II], and even [O~III] but
the jet is only visible weakly in H$\alpha$ and [S~II] in our ground-based images.

The H$_2$ $-$ Br$\gamma$ difference image reveals the typical outline of inner
H$_2$ emission enclosed by a shell of Br$\gamma$. However, towards the southwestern 
edge of the large pillar we find a spot of bright H$_2$ that appears
to separate more from the corresponding Br$\gamma$ emission to its south. The HST
images hint that several interfaces converge at this point which may account for
the additional H$_2$. It is also possible that a shocked knot exists at this location,
but high-resolution images of this region with HST or ground-based AO will be needed to
interpret the structure at this location. The exciting star located near the apex of
the proplyd is PCYC~429, which has a near-infrared excess and is an X-ray source \citep{broos11,povich11}.

\textit{HH c-9}: \citet{smith10a} drew attention to a parabolic arc in the Tr~14 region
that was bright in H$\alpha$ and [O~III], and present but fainter in [S~II]. It is located
one arcminute due east of the cluster in Area 30 (Figs.~\ref{fig:h2_summary} and \ref{fig:tr14}).
Our Br$\gamma$ images show a similar morphology to the H$\alpha$ and [O~III] images in
Fig.~\ref{fig:hhc-9}.  The object has no H$_2$ emission. The arc is a highly ionized
shell of some sort, probably either a bow shock or the edge of a
stellar wind bubble irradiated by stars in Trumpler 14. If it is a bow shock there is
no obvious source for the flow in either the H$_2$ or Br$\gamma$ images. Proper motion
measurements will be the best way to determine the nature of this object.

\textit{HH 903 and HH c-10}: The \citet{smith10a} HST ACS H$\alpha$ images of the southern pillar
G287.88-0.93 show two distinct outflows. HH~903 originates from the center of the 
pillar and drives an optical jet to the west, with bow shocks that extend about an 
arcminute to both the east and west of the pillar. A second system,
HH c-10, manifests as an H$\alpha$ shell that extends about 10 arcseconds
to the southwest of a small cluster of stars located at the head of the pillar.
Neither of these flows are easy to see in our NEWFIRM images (Area 2 of
Fig.~\ref{fig:spillars}), though it is possible to see faint traces of the HST
flows in our Br$\gamma$ images.  H$_2$ emission is not visible in the HH~903 jet and bow shocks
or in the HH c-10 shell.
However, at the location of the source of the HH~903 jet the pillar has several curved arcs
like we observe at the heads of other pillars. This morphology suggests the pillar is either
two pillars observed in projection, or that the pillar has a `shoulder' where the driving source
of the HH~903 jet lies.

\textit{HH 1008 and HH c-2}: The pillar G287.73-0.92 (Area 4 of Fig.~\ref{fig:spillars})
is one of the most remarkable of its kind in the sky. HH~1008 is situated near
the base of the pillar on the eastern side, and appears
as a southward-moving bow shock in HST H$\alpha$ images \citep{smith10a}.
This feature is visible in our Br$\gamma$ images as well, albeit with less
contrast owing to ground-based seeing. The candidate HH object c-2 located at the head of the
pillar is too small to be detected from the ground. Neither object has H$_2$ emission.

\subsection{New Outflows}\label{ss:new_flows} 

We identified new outflow candidates in the Carina Nebula region via their morphology
in our narrowband NEWFIRM and MOSAIC images. 
Optical outflow candidates all have bright [S~II] emission, typical of HH objects.
Several molecular jets and shocks appear in the H$_2$ images alone, but most
are easier to see in the H$_2$ $-$ Br$\gamma$ difference images because continuum is
present in similar amounts in both images and is largely removed in the subtraction.
However, H$_2$ emission along PDRs (section \ref{ss:pillars}) can be difficult to distinguish from
shocked molecular flows in Carina.  Following \citet{davis10}, we designate objects that have clumpy
or jet-like H$_2$ emission and do not define obvious PDR boundaries with the
acronymn MHO (molecular hydrogen object).  Section \ref{ss:others} summarizes
extended H$_2$ objects whose true nature remains uncertain from morphology alone.

\subsubsection{New Optically-Identified Outflows and Candidates} 

\textit{HH 1123}: HH 1123 consists of a group of four bright H$\alpha$ and
[S~II] knots arranged in a cross-shape (Fig.~\ref{fig:hh1123}). 
The knots have the typical clumpy morphology of HH objects, and
are not visible in [O~III], Br$\gamma$ or in H$_2$. 
A relatively faint star lies between knots A and C in the
infrared images, but this source is not clearly associated with the emission knots.
There is no indication of a jet, although a faint arc that 
extends from knot A to the east of knot C may mark the edge of a cavity.
HH~1123 is situated along the northern extension of the Southwestern Loop.
The Southwestern Loop is a region that hosts many filamentary structures and
knots that appear to result from impacts of winds upon a large, slowly-expanding shell
(Fig.~\ref{fig:swloop}), and it is possible that HH~1123 has a similar origin.

\begin{figure} 
\centering
\includegraphics[angle=0,scale=1.00]{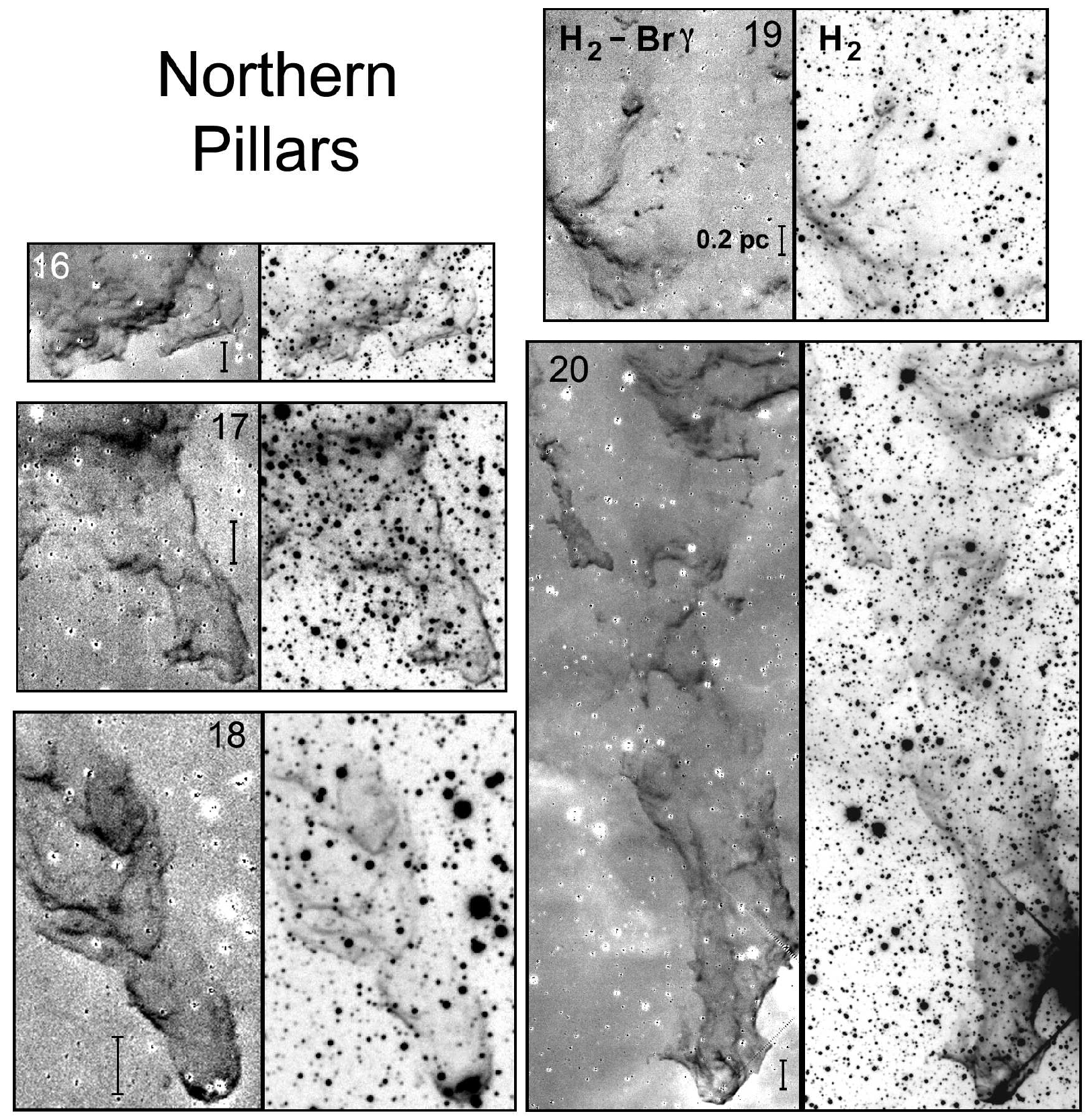}
\caption{Same as Fig.~\ref{fig:spillars} but for Areas 16 through 20 of the Northern
Pillars in Fig~\ref{fig:h2_summary}.  Fig.~\ref{fig:MHO1622} presents an expanded
version of the MHO objects in Area 19.} \label{fig:npillars}
\end{figure}

\begin{figure} 
\centering
\includegraphics[angle=0,scale=1.00]{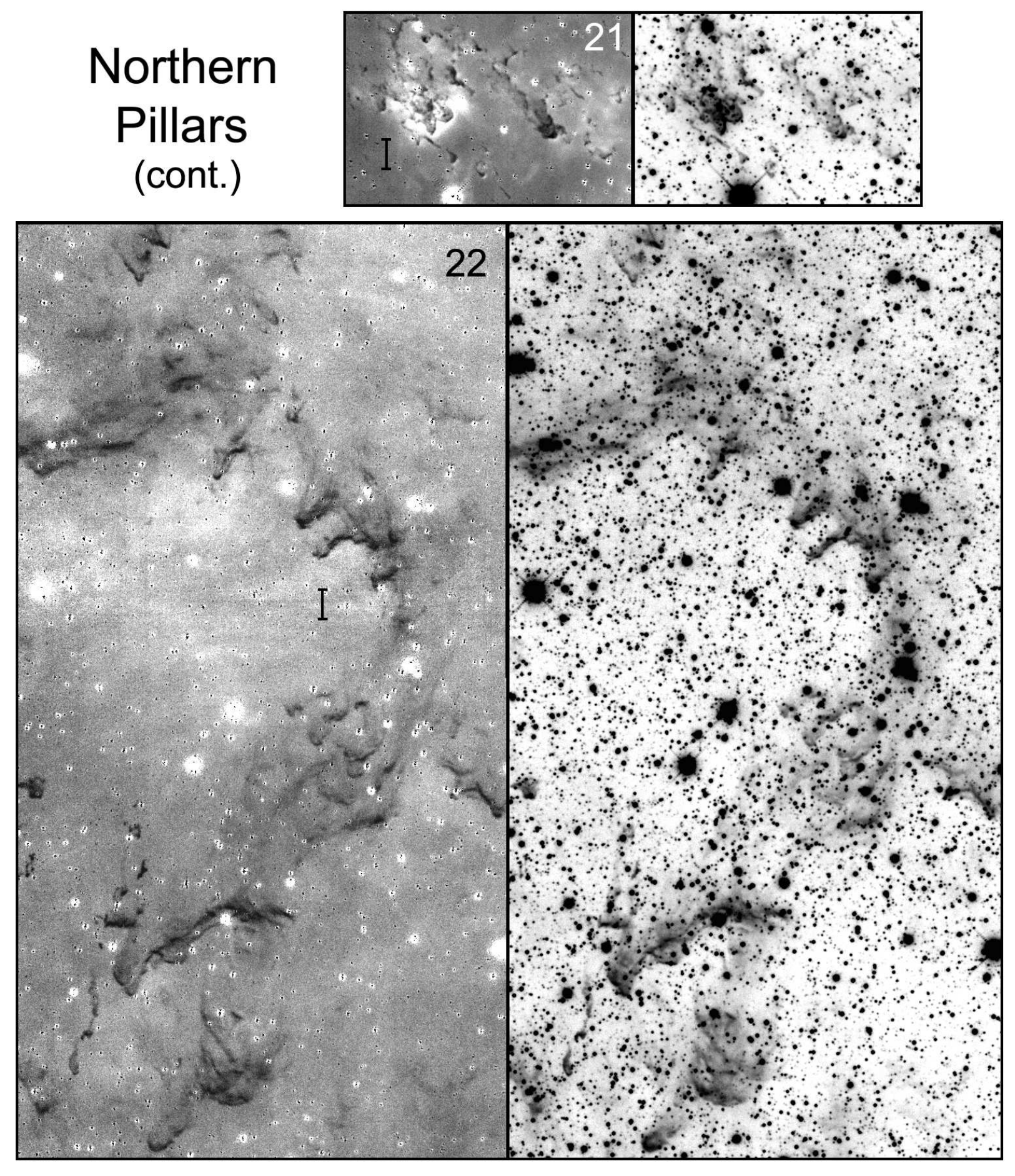}
\caption{Same as Fig.~\ref{fig:spillars} but for Areas 21 and 22 of the Northern
Pillars in Fig~\ref{fig:h2_summary}. } \label{fig:npillars2}
\end{figure}

\begin{figure} 
\centering
\includegraphics[angle=0,scale=1.00]{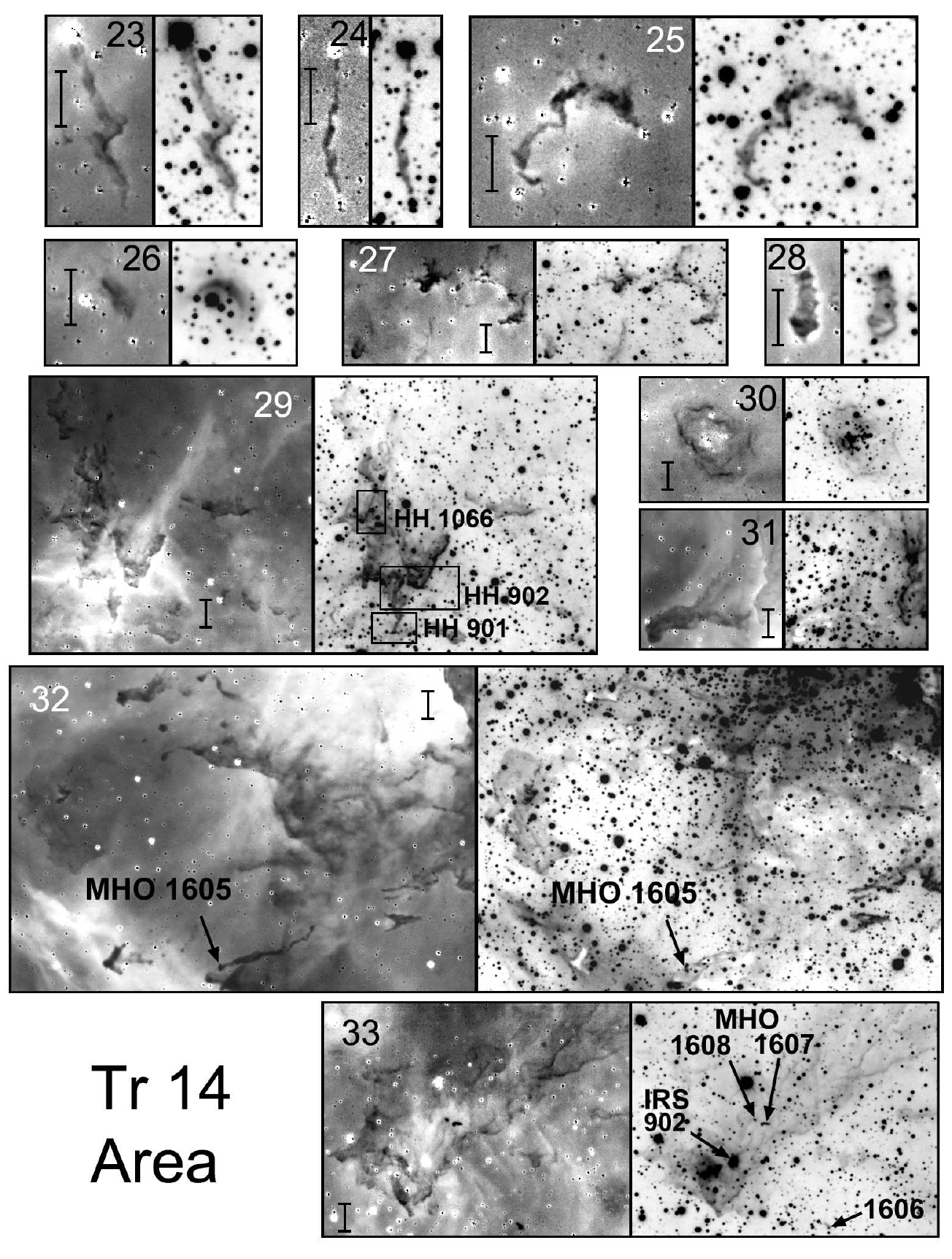}
\caption{Same as Fig.~\ref{fig:spillars} but for Areas 23 through 33 of the Trumpler 14 region in
Fig~\ref{fig:h2_summary}. Boxes around HH 901, HH 902, and HH~1066 are expanded in Figs.~\ref{fig:hh901}, 
\ref{fig:hh902}, and \ref{fig:hh1066}, respectively. The MHO objects in Areas 32 and 33 are
discussed in the text.} \label{fig:tr14}
\end{figure}

\begin{figure} 
\centering
\includegraphics[angle=0,scale=1.00]{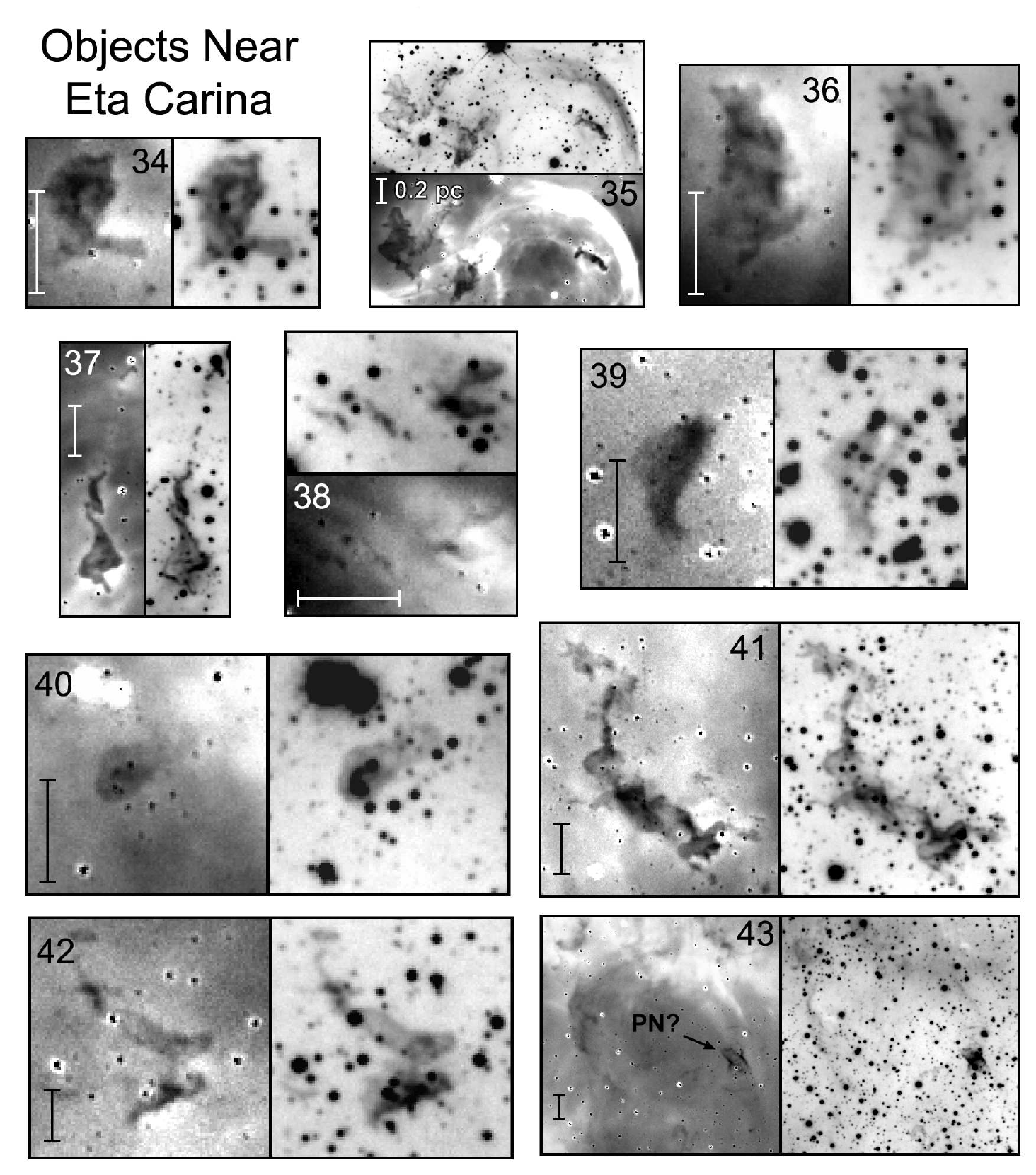}
\caption{Same as Fig.~\ref{fig:spillars} but for Areas 34 through 43 in the
vicinity of $\eta$ Car in Fig~\ref{fig:h2_summary}. The nebulous object 
on the right side of Area 43 is a candidate planetary nebula.} \label{fig:eta}
\end{figure}

\begin{figure} 
\centering
\includegraphics[angle=0,scale=1.00]{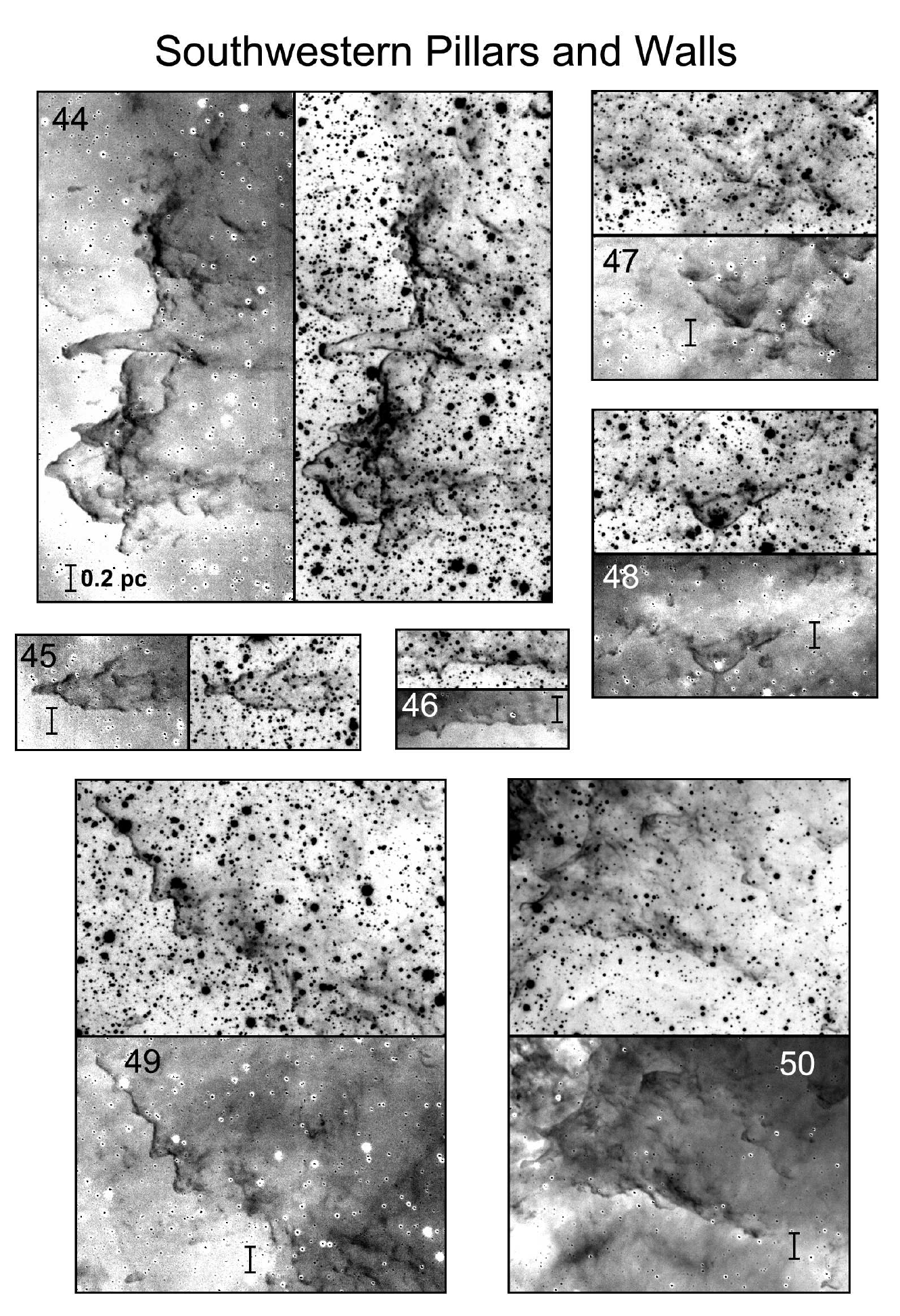}
\caption{Same as Fig.~\ref{fig:spillars} but for Areas 44 through 50 of the Southwestern Pillars and 
Walls in Fig~\ref{fig:h2_summary}.  The H$_2$ images are at the top or right in each panel, with the
H$_2$ $-$ Br$\gamma$ images at bottom or left.} \label{fig:swpillar}
\end{figure}

\textit{HH 1124}: Images of the head of the irradiated pillar in Area 18 of Fig.~\ref{fig:npillars}
in H$\alpha$, [S~II], I-band, and H$_2$ are shown in Fig.~\ref{fig:hh1124}. Three optical
emission-line knots (A, B, and C) lie to the west of the pillar, while one knot (D) and a
bow-shaped arc lie to the east. If we attribute all knots to a single source, then the
pattern is consistent with a jet that emerges at PA $\sim$ 53 degrees and is bent to the north
by winds from the same source or sources that shape the pillar.
The H$_2$ images show weak extensions along the axis of the putative jet that are not continuum because
they are also visible in the H$_2$ $-$ Br$\gamma$ images in Fig.~\ref{fig:npillars}.
However, this H$_2$ `jet' marked in Fig.~\ref{fig:hh1124}
could also simply define another irradiated surface along the pillar.
The axis of the outflow also has extended emission in IRAC Band 2 images, including
an extended green object (EGO) identified from a Band 2 $-$ Band 1 difference image.

There are several infrared sources located near the head of the main pillar, so confusion
between multiple outflows is always a concern. Two bright NIR sources positioned
near the center of the pillar, PCYC~884 and PCYC~889, are YSOs
as they emit strongly in the Spitzer bandpasses (\citealt{povich11}; Fig.~\ref{fig:hh1124})
PCYC~884 lies at the base of an arc-shaped 
feature in the I-bandpass and is the only IR source in the region associated with
such a feature. This arc has a morphology and orientation
consistent with it defining the edge of an evacuated cavity driven by the outflow that produces knots A, B, and C.
The source has a flat spectral energy distribution in the IRAC bands. 
Situated just to the west of PCYC~884,
PCYC~889 is fainter in the near-infrared, but becomes a magnitude brighter than PCYC~884 
at 24 $\mu$m \citep{povich11}.

A region outlined by a box in Fig.~\ref{fig:hh1124} and
located $\sim$ 0.1~pc to the west of a near-infrared source (IRS~1 in the Figure) 
has several faint knots and extended emission
in the H$\alpha$, [S~II] and H$_2$ images. The knots are not necessarily associated with IRS 1;
in fact, the H$_2$ $-$ Br$\gamma$ difference image in
Area 18 of Fig.~\ref{fig:npillars} has an arc-shape, so this region is most likely
another irradiated pillar.  The object labeled IRS~2 is another red object in the direction of
the globule. It lies along the axis of the main outflow and appears to be a
subarcsecond binary with a PA $\sim$ 105 degrees.  Finally, a bright star just outside the edge of
the pillar $\sim$ 0.06~pc south of PCYC~884 is probably a foreground object, as it is
bright in both the optical and infrared images and does not have any obvious influence on
the morphology of the pillar.

\textit{HH 1125}: HH 1125 is a thin, long, filamentary [S II] emission-line
object located near HH~666 in the region of the southern pillars to the west of the H$_2$
structures shown in Area 3 of Fig.~\ref{fig:spillars}.
The morphology of HH 1125 resembles other filamentary [S~II] structures 
that trace the remains of an outflow lobe from an unseen 
driving source, such as HH~400 in Orion \citep{bally01}.
The brightest feature in this object (marked `A' in Fig.~\ref{fig:hh1125}) has a morphology
similar to that of a bow shock in a jet that propagates at position angle 280$^\circ$.
HH 1125 is invisible in H$\alpha$ and [O~III],
and also has no infrared counterparts in H$_2$ or Br$\gamma$, so it is unlikely to mark the
edge of a dark cloud. 
The projected distance of the most distant observed knot B from the edge
of the nebulous region is 3.7~pc, so HH 1125 would represent a large-scale outflow if it traces a jet.
While any number of the bright stars in this densely-populated region
could be the exciting source for a jet, none of them particularly stand out as an
obvious driving source. Hence, without radial velocity and proper motion 
information we cannot rule out the possibility that HH 1125 
represents the edge of a shell driven by one of the many massive stars in Carina.

\textit{HHc-21}: Fig.~\ref{fig:MHO1622} shows a small but distinct optical knot located 
about 1.5 arcseconds at PA = 200$^\circ$ away from a star. The knot, also faintly visible in [S~II], but
not in [O~III], H$_2$, or Br$\gamma$, is a candidate HH object we label
HHc-21. The star is not listed as a member of Carina by \citet{povich11}.
The system lies to the north of the H$\alpha$ arc R2 in the region of MHO 1622 - MHO 1626.

\subsubsection{Candidate Outflows Identified in Narrowband H$_2$ Images} 

We have searched our images for evidence of shocked H$_2$ in outflows. Such objects typically appear
as distinct knots rather than extended H$_2$ that defines a PDR front
at the edge of a globule, wall or pillar. Morphologies of the PDRs in Carina can be quite complex, however,
and it is not always immediately obvious from images alone which objects are shocked flows and which
are PDRs. For this reason MHO objects in the Carina Nebula are best treated as candidate outflows pending verification
by proper motions, though most will likely turn out to be outflows. A few of these were 
discovered independently by \citet{preib11b}, and we discuss those objects together with the new
sources in this section.

\textit{MHO 1605}: MHO 1605 has the morphology of a barely-resolved bipolar outflow
emerging from a small pillar embedded in the dark dust lane $\sim 4$\arcmin\ south
of Trumpler 14 (\citealt{preib11b}; Area 32 of Fig.~\ref{fig:tr14}). 
The object consists of two bright H$_2$ knots located just outside the pillar,
along PA$\approx 3^{\circ}$.  The pillar is well-defined as an extinction feature 
in the Br$\gamma$, H$\alpha$, [S~II], and [O~III] images.
MHO 1605 is difficult to discern in the optical images, although there is
some diffuse [S~II] emission coincident with the H$_2$ knots.
Emission in IRAC bands 1 - 3 is evident near the pillar
head and is elongated along a north-south axis (not consistent with a point source),
but the feature is too small (only $\sim 4\arcsec$ long in our H$_2$ image) to
be clearly resolved with Spitzer.

\textit{Cr 232/N4 Area, MHO 1606 $-$ 1608}:
\citet{preib11b} and \citet{tapia11} discussed the cluster Cr 232
that is associated with the extended H$_2$ emission in 
Area 33 of Fig.~\ref{fig:tr14}. The bright K-band source labeled IRS in Fig.~\ref{fig:tr14} is
a massive star embedded within a nearly edge-on circumstellar disk that \citet{preib11d} refer
to as the `disk object' \citep[also known as source \#902,][]{tapia11}.
A separate foreground X-ray source, PCYC 556, is located about 2 arcseconds to the southeast of
the near-IR peak, and appears in our near-IR images and those of \citet{preib11d} as a distinct point source.
Fig.~\ref{fig:tr14} uncovers over a dozen localized bright regions of H$_2$ emission in this
region. Most of
these follow arcs that define PDR boundaries in the dark cloud that houses the cluster,
though several may turn out to result from shocks in outflows.  Three bright H$2$ features particularly
stand out in the H$_2$ $-$ Br$\gamma$ image as having a knotty morphology typical of shocked gas.
The brightest of these, MHO~1607, consists of three distinct knots aligned
approximately east to west over an extent of about 5 arcseconds. A few arcseconds to the northeast,
MHO~1608 is more point-like, with a slight elongation along PA 45$^\circ$. 
Both these MHO objects are prominent in the H$_2$ image but are not present in Br$\gamma$.
The H$_2$ $-$ Br$\gamma$ image reveals two faint arcs that trail to the southeast of MHO~1608 and 
may outline the edges of an irradiated pillar.  Neither the optical nor the IR images indicate
any obvious driving source associated with these two objects, though of course there are many
possibilities within the cluster, which has 72 members according to \citet{tapia11}. 
The IRAC Band 2 $-$ Band 1 image shows diffuse emission
in the area, with a modestly-bright feature located between MHO 1607 and 1608.
A third object, MHO~1606, is an isolated, bright, comma-shaped H$_2$ knot located away
from the cloud boundary $\sim$ 80 arcseconds to the southwest of MHO~1607 and MHO~1608. 
This object also has weak [S~II] emission.

\textit{MHO 1609}:
The detached pillar shown in Area 6 of Fig.~\ref{fig:spillars2} and in
Fig.~\ref{fig:MHO1609} is surrounded by several candidate HH objects (HH c-4 through HH c-8)
identified by \citet{smith10a}, but none of these are visible in our NEWFIRM images.
However, our H$_2$ images uncover two extended spots of bright emission near the 
middle of the irradiated pillar. These spots lie just inside the H$\alpha$
and [S~II] emission at the surface of the evaporating pillar.
A K-band point source lies just west of the H$_2$ knots, but is offset from the
probable flow axis enough that it is unlikely to be the driving source.
It is unclear if the two elongated features are related as they are bright in a
relatively narrow portion of the pillar, and at this resolution, are consistent
with emission from the irradiated pillar edge.  An alternative interpretation is
that bright H$_2$ emission comes from photoevaporation of the denser clumps within
a more diffuse pillar.  None of the H$_2$ features in the pillar appear to be
related to any of the candidate HH jets emerging from this globule that
\citet{smith10a} identified with HST H$\alpha$ imaging.  The five separate HH jet
candidates emerging from this elongated globule suggest ongoing star formation from
an object dense enough to survive despite harsh UV radiation that has largely
cleared the surrounding gas and dust.  In general, the H$\alpha$ features of the
candidate HH jets are too faint or too tenuous to be identified in the ground-based
images, although bright H$_2$ and Br$\gamma$ emission from the head of the globule
may reveal the protostar driving HH c-5.

\textit{MHO 1610}: MHO 1610 is a single bright, compact H$_2$ knot
embedded within more diffuse emission that probably outlines a globule
as does so much of the H$_2$ emission from this region of the Eastern Walls
(\citealt{preib11b}; Area 10 of Fig.~\ref{fig:sewall}; Fig.~\ref{fig:MHO1610}).
While the images show two other nebulous objects and several stars within
the boundaries of the diffuse emission, these sources also appear in the
Br$\gamma$ and K-band images and are dominated by continuum.
The faint continuum source labeled IRS
is coincident with a Herschel point source at 70 $\mu$m.
The H$_2$ knot is elongated by about
2.5 arcseconds along PA $\sim$ 45$^\circ$, and is slightly narrower on
its northeastern side.

\begin{figure} 
\centering
\includegraphics[angle=0,scale=1.00]{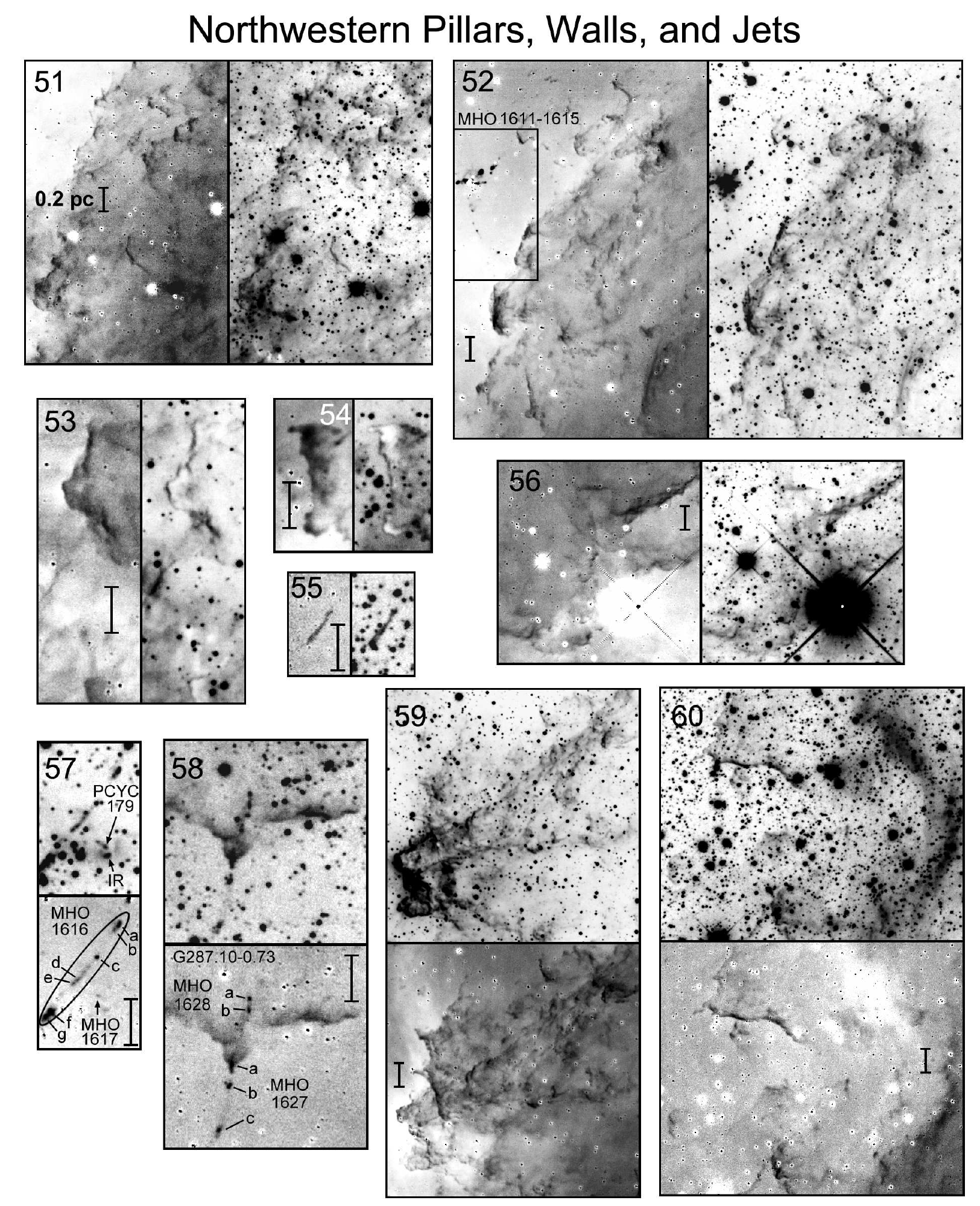}
\caption{Same as Fig.~\ref{fig:spillars} but for Areas 51 through 60 of the Northwestern Pillars,
Walls and Jets in Fig~\ref{fig:h2_summary}.  The H$_2$ images are at the top or right in each panel, with the
H$_2$ $-$ Br$\gamma$ images at bottom or left. The region marked in Area 52 is expanded in
Fig.~\ref{fig:MHO1611}.} \label{fig:nwpillar}
\end{figure}

\begin{figure} 
\centering
\includegraphics[angle=0,scale=1.00]{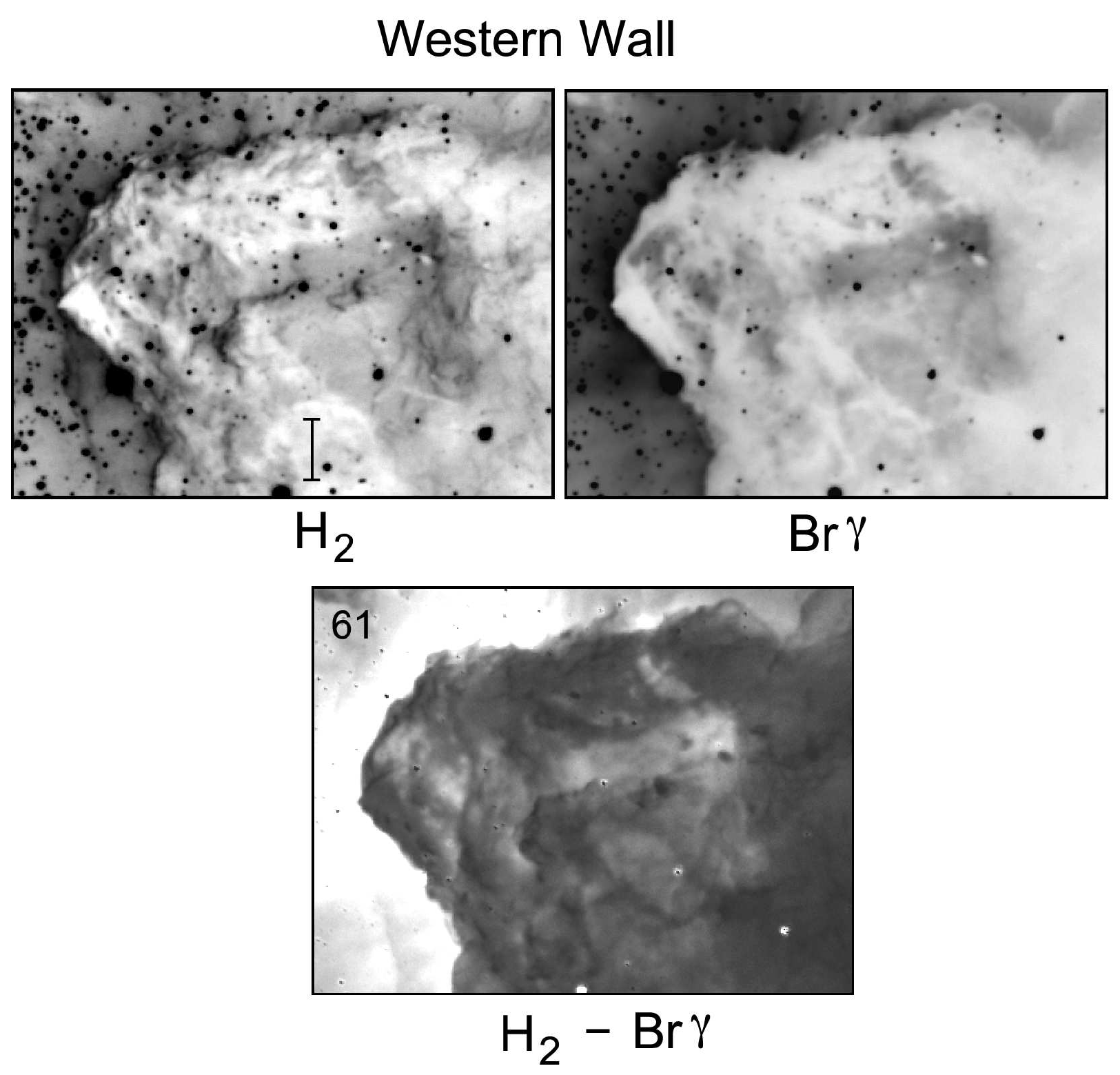}
\caption{Same as Fig.~\ref{fig:spillars} but for Area 61 (G287.38-0.62), the Western Wall
in Fig~\ref{fig:h2_summary}.  The H$_2$ and Br$\gamma$ images are at the top, with the
H$_2$ $-$ Br$\gamma$ images at bottom.} \label{fig:western_wall}
\end{figure}

\begin{figure} 
\centering
\includegraphics[angle=0,scale=1.00]{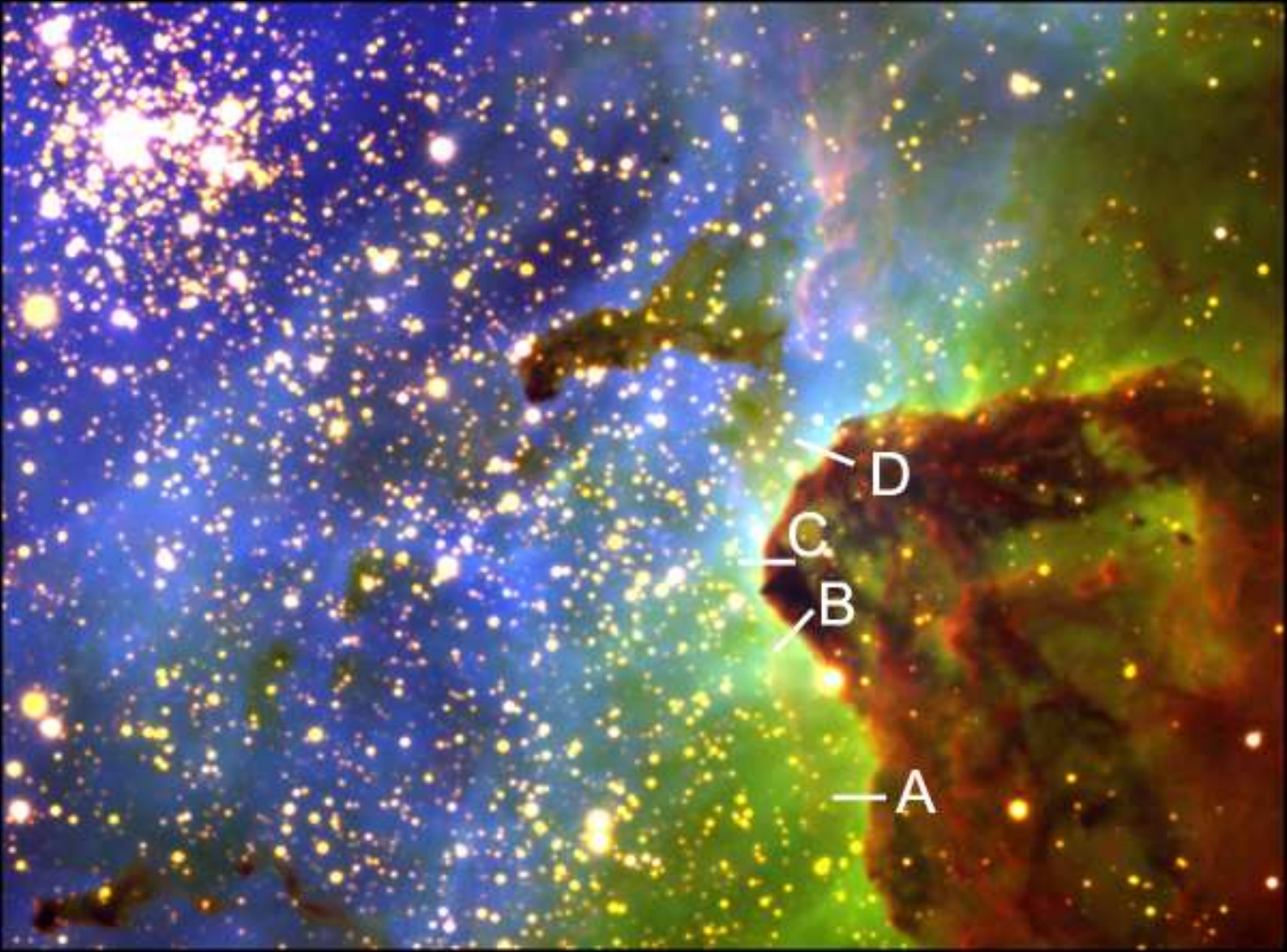}
\caption{Color composite of the Western Wall in Figure \ref{fig:western_wall}, with
H$_2$ in red, Br$\gamma$ in green, and [O III] 5007\AA\ in blue. Positions A, B, C, and D
mark the four slices along the PDR interface used in Figure \ref{fig:lineout}
to demonstrate positional offsets in the
emission line fluxes.}
\label{fig:western_color}
\end{figure}

\textit{MHO 1611 - 1615}: MHO 1611, MHO 1612, MHO 1613, MHO 1614, and MHO 1615 form a group of 
H$_2$ knots located $\sim 8$\arcmin\ northwest of Trumpler 14 (Area 52 of Fig.~\ref{fig:nwpillar};
Fig.~\ref{fig:MHO1611}).  These knots reside near the edge of the dark lane
that runs beneath Trumpler 14, and are centered around a loose cluster of at least
20 infrared-bright stars.  The brightest star at the center of the cluster is PCYC 139,
a class II infrared source with L$_{BOL}$ = 400 L$_\odot$ and an estimated mass of
4.9 M$_\odot$ \citep{povich11}.
MHO 1611 is a group of several faint H$_2$ knots that stretches from near PCYC 139 
to near the dark lane to the southwest. MHO 1612, MHO 1613, MHO 1614, and MHO 1615 lie within about
a half arcminute of PCYC 139, and could have different exciting sources. MHO 1615 connects
with a star situated 2.9 arcseconds at PA=250$^\circ$ from PCYC 139. This star appears
point-like in the optical images but has an extension along the direction of MHO 1615 in the
Br$\gamma$ and K-band images (Fig.~\ref{fig:MHO1611}), hence, it is the likely 
driving source for that portion of the H$_2$ emission.  A faint arc of H$_2$ extends to
the east of MHO 1613 towards MHO 1614.  Another bright young star and
infrared source, PCYC 147, is situated $\sim$ 20 arcseconds to the east of the infrared cluster. Some
6 arcseconds to the south of PCYC 147 lies the optically-bright B1.5V star
HD 303313, which does not appear heavily extincted and is probably situated in front of the cluster.

The PCYC 139 cluster is not associated with ionized pillars or globules, and is not 
surrounded by a distinct H$_2$ shell that would outline the photodissociation front
into a remnant cloud core.  The dark lane to the southwest appears as an H$_2$ wall
(Area 52 of Fig.~\ref{fig:nwpillar}), but this wall does not have any obvious connection
with the cluster. The lack of visual counterparts to the H$_2$ knots
suggests that the PCYC 139 cluster lies behind several magnitudes of extinction
on the far side of the Carina nebula. 

\textit{MHO 1616 - 1617}: MHO 1616  lies 3 arcminutes west of the PCYC 139 cluster, and resembles a
clumpy jet (Area 57 of Fig.~\ref{fig:nwpillar}).  None of the H$_2$ knots have any
optical counterparts. While there are several members of the Carina YSO association
in the vicinity, none are aligned with the knots of MHO 1616 . MHO 1617 is a faint jet of H$_2$
that extends for about 4 arcseconds from PCYC 179 at PA $\sim$ 75$^\circ$ and ends in 
a bright knot. Another source 3 arcseconds to the south labeled `IR' in Fig.~\ref{fig:nwpillar}
has an arcuate morphology that one would expect from a cavity created by an outflow
along PA $\sim$ 50$^\circ$.  This direction intersects with MHO 1617, so it is possible
that the IR source is responsible for some of the H$_2$ emission in this region. The IR
source falls within the positional errorbars of a Herschel point source at 70$\mu$m \citep{preib12}.

\textit{MHO 1618}: MHO 1618 is another shocked outflow candidate,
this one located in the Eastern Wall region just south of Area 11
(Fig.~\ref{fig:h2_summary}).  The H$_2$$-$Br$\gamma$ difference image uncovers a
faint H$_2$ jet that is aligned with a bright infrared source not
identified previously as a cluster member (Fig.~\ref{fig:MHO1618}). The source
is visible in the optical and the IR images and also appears
in all four IRAC Bands. We recorded the position of the IR source in 
Table~2, but it is also possible that the flow originates somewhere to the
northwest of knot a, where a dark cloud resides. Proper motions of the H$_2$ knots
would settle this question.  The putative jet is only visible in H$_2$.

\textit{MHO 1619 - 1621}:
A collection of bright H$_2$ clumps
labeled MHO 1619, MHO 1620, and MHO 1621 form part of a curve of faint extended
H$_2$ situated $\sim$ 11.5 arcminutes to the northeast of $\eta$ Car
(Area 13 of Fig.~\ref{fig:sewall2}; Fig.~\ref{fig:MHO1619}). MHO 1619 is arc-shaped
and has faint [S~II] emission located to the west-northwest of knot-c. MHO 1620
appears as a chain of knots oriented east-west. Knot-a is notable for its 
[S~II] emission, and knot-b has a complex internal structure down to the spatial
resolution of the images. A third string of H$_2$ knots to the south of MHO 1620
defines MHO 1621. Two stars located on either side of MHO 1621 (Fig.~\ref{fig:MHO1619})
have no clear association with the H$_2$. The H$\alpha$ image of the region shows
a filamentary structure in absorption, implying that foreground shells of material
exist in this region. None of these MHO objects has a clearly-defined exciting source.

\textit{G287.24-0.21, PCYC 438, and MHO 1622 - 1626}:
Area 19 of Fig.~\ref{fig:npillars} and Fig.~\ref{fig:MHO1622} reveal several potential outflow sources
and groups of interesting H$_2$ emission line objects.
Near the southeast corner of Area 19, a region of extended H$_2$ outlines the
boundaries of the molecular cloud G287.24-0.21.  The H$_2$ from this region forms a
backwards ``y'' shape in Fig.~\ref{fig:MHO1622} that is also bright in all four IRAC bands
in the Spitzer images.  The H$\alpha$ and [S~II] emission from G287.24-0.21 are filamentary
and more extensive than the H$_2$, but generally follows the H$_2$ shape, as expected for
an irradiated cloud boundary.

MHO 1622 - MHO 1626 define four groups of clumpy H$_2$ sources in Fig.~\ref{fig:MHO1622}.
None these show any optical counterparts or obvious driving sources, though the star field
in this region is rich and there are several cluster members in this region.
MHO 1623, MHO 1624, and MHO 1626 are jet-like in appearance, while MHO 1622 and MHO 1625
have less-collimated morphologies.  The bright NIR source PCYC 438
\citep[][Fig.~\ref{fig:MHO1622}]{povich11} has a bipolar nebula visible in the optical and NIR images
that extends $\sim$ 3 arcseconds on both sides of the star along PA = 225$^\circ$, though no jet is
visible along the axis of the nebula. The bipolar nebula shows hints of what could be H$_2$ in
the H$_2$ $-$ Br$\gamma$ image, but the nebula is primarily a continuum source. 

In addition to the MHO objects and bipolar nebula described above,
five bright arc-shaped H$_2$ and H$\alpha$ features (labeled R1 $-$ R5 in Fig.~\ref{fig:MHO1622})
define irradiated walls and pillars in this region. These all curve to the north, away from the dominant
sources of radiation from the Tr~14 cluster and $\eta$ Car. As in the other Carina PDR
interfaces, the H$\alpha$ emission from the R1 $-$ R5 arcs 
peaks closer to source of ionizing radiation than the H$_2$ emission does.

\textit{G287.10-0.73, MHO 1627 and MHO 1628}:
Of the four Extended Green Objects (EGOs) found in
Spitzer/IRAC images by \citet{smith10b}, only 
G287.10-0.73, an elongated IRAC Band 2 source located almost due west of Tr~14, is included
within the field of view that we imaged in H$_2$.
While we do not see H$_2$ 2.12$\mu$m emission at the location of G287.10-0.73,
probably owing to large extinction near the apex of the dusty pillar,
our H$_2$ image reveals a striking chain of knots that picks up where the Spitzer extension
leaves off and emerges at PA = 165$^\circ$ both to the north and to the south from the head
of the pillar (Area 58 of Fig.~\ref{fig:nwpillar}).
On the southern side, a bright H$_2$ jet followed by
two knots that are linked together by a stream of faint H$_2$ makes up MHO 1627, while on
the northern side, two other bright H$_2$ knots comprise MHO 1628.

Both MHO 1627 and MHO 1628 are visible faintly in the Spitzer images, albeit with poor spatial
resolution. No jet is visible in our Mosaic H$\alpha$, [S~II] or [O~III] images. A jet
also does not appear in Br$\gamma$, so that the lack of
H$\alpha$ is not simply due to foreground extinction.
A point source at 70$\mu$m is located between MHO~1627 and MHO~1628 in Herschel images \citep{preib12}.
The dark pillar associated with G287.10-0.73 is only visible in silhouette. Hence,
the ambient ionizing radiation field is not strong in this region, which makes it easier for
molecules to survive once the outflow leaves the confines of the dark cloud. 

Overall, the number of new candidate flows discovered in Carina with NEWFIRM
is smaller than would be expected based on other star-forming complexes at
comparable distances \citep[e.g. W5;][]{ginsburg11}.  Unlike 
W5, Carina hosts numerous uncovered O-stars 
that create a harsh radiative environment capable of destroying
molecules in much of the central portion of the nebula.
The South Pillars region, which
extends beyond the boundaries of our NEWFIRM images, has indications of
ongoing star formation \citep[e.g.][]{megeath96,smith00,rathborne04,smith10b}
that could produce additional flows.
The new H$_2$ flows are spatially intermixed with the irradiated jets discovered with HST
(\citealt{smith10a}; Fig.~\ref{fig:jet_distro}).
The outflows are generally most pronounced in either optical (H$\alpha$, [S~II])
or molecular (H$_2$) emission, but not both simultaneously.
None of the molecular jets have strong optical emission and only one of the
irradiated jets seen in H$\alpha$ has appreciable extended H$_2$ emission (see Sec.~\ref{ss:old_flows}).

\subsection{Other Extended Nebulous Objects}\label{ss:others} 

\textit{H$_2$ arcs}:
Three H$_2$ arcs to the west of the MHO 1619 $-$ 1621 group (Area 13 of Fig.~\ref{fig:sewall2})
do not clearly associate with a cloud boundary, and appear bright in H$_2$. These objects
look like shell fragments of some sort, but their nature is unclear.

\textit{Narrow [S II] filaments in the South Pillars region}:
The [S~II] image and the scaled continuum-subtracted version [S~II] $-$ I$_c$ in
Area 62 (Fig.~\ref{fig:shells}) reveal a remarkable pattern of five extremely thin, connected
filaments. The northeastern filament labeled `1' in the figure continues for several
parsecs and defines the western boundary of the large pillar that contains the two bright H$_2$ structures
in Area~3 (Figs.~\ref{fig:h2_summary}, \ref{fig:spillars}).  The curved shape of the cloud is
visible faintly to the west of filament 3. The southern portion of filament 3 also emits
weakly in H$\alpha$.  Filaments 2, 3, and 4 are most easily explained as constituting the leading edge of a bubble
that is impinging upon the cloud (see inset to Fig.~\ref{fig:shells}).
In this model filament 5 could define the southern edge of the cloud.
Proper motion measurements would be able to verify or disprove this scenario, which
predicts motion of filaments 2, 3, and 4 to be northward, eastward, and southward,
respectively.

The intersection points between shock fronts have generated a fair bit of interest 
recently because under some circumstances a normal shock known as a Mach stem may develop
at this location, leading to a hot spot that has anomalous pattern motion relative to the
larger bow shock structure.  Multiple-epoch HST images show
evidence for time-variable enhanced emission at intersection
points of bow shocks exists in HH~34S and possibly also HH~47A \citep{hartigan11}.
Laboratory experiments of this phenomena have quantified the critical angles where Mach stems 
can form \citep{foster10,yirak13}, and the experiments agree reasonably well with
the theoretical and numerical predictions.

In Fig.~\ref{fig:shells}, there is a small bright segment where filaments 3, 4,
and 5 intersect, but no Mach stem should form there because the angle between filaments 4 and 5 is 
too acute, and none exists in the object. The situation at the junction of filaments 1, 2, and 3
is more complex. One bright [S~II] knot, two fainter ones, and several foreground stars lie 
near the intersection point. The projected angle between filaments 1 and 2 is 
75 - 105 degrees, depending on how one interprets the spatial structure at that
point. An angle in the 75 - 90 degree range may produce a Mach stem, but the
shock is nearly normal to the interface in any case. Filament 3 is distinctly more
wavy in appearance than are the adjacent filaments, as expected if this filament
marks the location of the shock as it overtakes irregularities along
the surface of the cloud. The brightness variations along filament 3 seem uncorrelated with
whether or not the wiggles in the front are concave or convex.

\textit{Large [S II] shells in the South Pillars region}:
A region where several bright arcs of [S~II] overlap may define a fossil cavity
driven by the massive stars to the north (Area 63 of Fig.~\ref{fig:shells}). The
shells extend east-west for over a parsec and are visible faintly in the H$\alpha$
images though not in [O~III]

\textit{Shells and Filaments in the Southwest Loop}:
The southwestern boundary of the bright optical emission in the Carina Nebula
is defined by a large curved arc that spans several pc in size, and
within the arc lie a number of smaller knots and filaments.
This region lies to the southwest of our H$_2$ mosaic, but was covered by the optical
Mosaic images (see Fig.~\ref{fig:swloop}). The I$_c$-band image shows very little
continuum along the arc. The scaled H$\alpha$ $-$ I$_c$ images in
Fig.~\ref{fig:swloop} remove stars reasonably well, and show that
the arc has several bright knots situated along its inner part 
(labeled A $-$ G in Fig.~\ref{fig:swloop}),
as well as what appear to be smaller cavities (features H, I and J) embedded
in its outer portion. Knots C and F are bright and clumpy, and
emit strongly in [S~II].  

The arc has the morphology and line emission characteristics of an 
expanding bubble that is ionized by massive stars at the center of the 
Carina Nebula.  Features A $-$ G appear to mark locations where the
arc is impacted by winds that originate from the active central region
of the Nebula. The existence of distinct smaller shells within the arc
argues for multiple wind sources. \citet{mw86} discovered very similar
structures within the Rosette, so this phenomenon is not unique
to the Carina Nebula.

\textit{Candidate Background Planetary Nebula}:
Area 43 in Fig.~\ref{fig:eta} shows
limb-brightened, elliptical H$_2$ feature with some wisps extending beyond the ellipse along
a perpendicular axis. The object is unusual for Carina in that
Br$\gamma$ emission fills the H$_2$ ellipse, and weak H$\alpha$ is also present. A faint star is
located near the center of the object. 
An IRAC Band 2 $-$ Band 1 difference image reveals some extended 4.5 \micron\
(green) emission in the center of this object. IRAC emission matches the Br$\gamma$
emission, but not H$_2$, suggesting we are seeing Br$\alpha$ and not shocked H$_2$
in the 4.5$\mu$m image in this case.  The emission structure with H$_2$ surrounding Br$\gamma$ emission,
combined with a symmetrical elliptical morphology and a central star is best explained by a planetary
nebula situated behind the obscuring dark clouds of the Carina nebula.

\textit{Knots near the Carina `Defiant Finger'}:
A striking irradiated globule known as the Carina Defiant Finger \citep{smith04b}
exists in the vicinity of the Keyhole.  A chain of emission knots
extends from the globule's northern edge (from the ``wrist'')
that are bright in H$\alpha$, [S~II], and H$_2$ (Area 37 of Fig.~\ref{fig:eta}; 
Fig.~\ref{fig:finger}). However, the knots exhibit spatial offsets between [S~II]
and H$\alpha$, and between H$_2$ and Br$\gamma$ characteristic of PDRs. The [O~III]
image shows the knots in absorption against the nebular background, as expected for
dusty globules.  Faint emission from the knots in the IRAC bands is consistent with 
thermal emission from these dusty blobs.  This collection of objects most likely
define the remains of a pillar within Trumpler 16 that has been destroyed by 
the intense UV radiation in this region. The northern-most knot has a bright infrared
source on its northern boundary labeled as `IRS' in Fig.~\ref{fig:finger}.
This location is where a young star should be if it formed at the apex of the pillar.
\citet{smith04b} derived a photoablation rate of
$\dot{M} \approx 2 \times 10^{-5}$ M$_{\odot}$ yr$^{-1}$ in this region, so
small knots such as these should be transient objects.

\textit{The Runaway Star Trumpler 14 MJ 218}\label{ss:runaway} 

The Br$\gamma$ images reveal a curious arc-shaped feature situated about
6 arcseconds to the northwest of a bright star (Area 26). The arc stands out in
color composites owing to its relatively strong Br$\gamma$ emission as compared with
H$_2$ (Fig.~\ref{fig:tr14}).  The star, located at 
10:44:05.1 -59:33:41 (2000.0), was first observed photometrically by \citet{fein73}
and later spectroscopically by \citet{massey93}, and is listed in the \citet{reed03}
catalog of O and B stars. The above references named the star
Trumpler 14 29 and Trumpler 14 MJ 218, respectively. The stellar colors are
consistent with a reddened early B star, and the high-quality blue spectrum 
\citep{massey93} shows a spectral type of B1.5V.

The arc has recently been discussed in some detail by \citet{ngoumou}, who refer to
it as the `Sickle' \citep[see also Fig.~1 of][]{ascenso07}.
As shown by \citet{ngoumou}, the star appears to be moving past 
a dense clump observed in 870$\mu$m images, and the shape and offset of the bow shock
from the stellar image is consistent with a wind from the star impacting denser material
in the vicinity of the globule.  The star has a large proper motion of
8.7 mas yr$^{-1}$ at PA = 307 degrees (95 km$\,$s$^{-1}$ at a distance of 2.3 kpc),
consistent with producing the bow shock \citep{zach2013}, and
is most likely a runaway star ejected as its companion underwent a supernova explosion. 
However, errors in proper motion measurements at the distance of Carina are relatively
large, and if the X-ray radiation from this source arises from a companion, it may
be difficult to keep the binary system intact if a more massive primary became a supernova.
Our difference image (H$_2$ $-$ Br$\gamma$) in Fig.~\ref{fig:tr14} shows enhanced H$_2$
along a flattened structure that lies along the portion of the bow closest to the star,
as one would expect if molecular hydrogen in the shell were excited either by the
shock or by UV fluorescence from the star. 

Extrapolating forward, Tr~14~MJ~218 should move a degree on the sky in
about $4\times 10^5$ years, which amounts to $\sim$ 10\%\ of the star-forming epoch 
within a typical molecular cloud. Carina hosts $\sim$ 70 O stars, which typically become
supernovae after 5 $-$ 10 million years \citep{mm03}.  Many of these O stars are binaries,
with the companions being future runaway stars. 
Hence, we expect at most a few runaway stars to exist within the Carina nebula at any
given time.  We note that one neutron star has been discovered
in the Carina region, implying recent supernova activity \citep{hama09}, and abundance
variations of heavy metals across the nebula also point to recent supernovae \citep{hama07}.

Tr~14~MJ~218 represents the only clear candidate for a runaway star in our field $-$
many emission-line arcs appear in our images, but the bow shock ahead of
this star is unusual in that is has a smooth, diffuse shape,
a large Br$\gamma$-to-H$_2$ ratio, and a bright star contained within the bow.
This object, together with eight other early-type stars in Carina, were highlighted by
\citet{smith10b} as having extended emission in 8$\mu$m Spitzer images and classified as
extended red objects (EROs). Two of these EROs
lie to the south of our mosaics, while the other six appear 
as unremarkable point sources in the Br$\gamma$ and H$_2$ images.

\section{H$_2$ Within Embedded Clusters}\label{ss:embedded_clusters} 

The Carina star-formation region is sometimes referred to as a 'cluster of clusters';
as many as 24 separate clusters were identified in the recent deep CCCP X-ray survey 
\citet{feig11}. The Trumpler 16 region (near $\eta$ Car) alone consists of least six smaller
clusters, and as many as eight were identified from Spitzer data \citep{smith10b}. While H$_2$ walls
and pillars often exist in the vicinity of these clusters, most clusters visible in the
optical and near-IR are no longer embedded within their nascent molecular cloud cores.

However, two young clusters stand out in our images in that they have
what appear to be cavities defined by Br$\gamma$ emission at the cluster's center that is 
surrounded by circular arcs of H$_2$ that mark the boundary of the cavities evacuated by the cluster stars. 
The most notable of these from the standpoint of H$_2$ emission is undoubtedly the 
Treasure Chest (Area 8 in Fig.~\ref{fig:spillars2}). This region was described in some detail by
\citet{smith05}. The star at the center of the Br$\gamma$ emission is an O9.5V star named CPD -59$^\circ$2661, 
and this object excites the emission lines within the local H~II region that surrounds it.
A bright emission-line star \citep[][Hen 3-485 in Area 8]{ss73} lies to the upper left of the
cluster, and was identified as a likely Herbig Be star on the basis of its X-ray flux by \cite{gagne11}. 
The near-IR observations of \citet{preib11a} show that about one-third of the stars in the Treasure 
Chest have near-IR excesses indicative of circumstellar disks, a fraction about three times larger
than occurs in the rest of the Nebula.

The narrowband images in Fig.~3 of \citet{smith05} show that faint Pa$\beta$ extends
northward from the H$_2$ emission in the Treasure Chest, in the direction of $\eta$ Car
and Tr~16. We also observe this behavior in the reddening-independent
H$_2$ $-$ Br$\gamma$ difference image in Area 8 of Fig.~\ref{fig:spillars}.
Bright Br$\gamma$ emission defines the H~II region excited by the cluster, and
a sinuous arc of H$_2$ outlines the PDR from the cluster. Hence, the dark cloud in which
the Treasure Chest resides is undergoing photodissociation both internally from the
embedded cluster, and externally from $\eta$ Car and other young stars within Tr~16.

Another cluster of interest that has H$_2$ emission is shown in Area 30 (Fig.~\ref{fig:tr14}). 
Discovered independently by \citet{preib11b}, it was not identified by \citet{feig11}
as either a major cluster or as a smaller group. 
While less spectacular than the Treasure Chest, this cluster contains 30 - 40 stars and is
surrounded by filamentary H$_2$ emission. The brightest NIR source is G~287.51-0.49, located at
10:44:59.70 -59:31:19.5, is also a mid-IR source \citep{rathborne04,sanchawala07}, though
the lack of a prominent H~II region in Br$\gamma$ implies it is not a powerful source
of ionizing radiation.

\section{Discussion: Feedback, Triggering, and PDR Physics in the Carina Complex} 

\subsection{Feedback} 

Carina is among the most active regions of star formation within several
kpc of the Sun. The region contains more than 70 stars of spectral type O, has a total mass of
young stars of M$_{stars} \simeq 4 \times 10^4$ M$_\odot$ and a neutral
gas mass of at least M$_{gas} \simeq (0.5-1) \times 10^6$ M$_\odot$
\citep{smith06b,sb07,preib11b,preib11c}. The kinetic energy of
the expanding nebula is about 8$\times$10$^{51}$ ergs,
about 1/3 of the mechanical energy available from
stellar winds in the past $\sim$3 Myr \citep{sb07}.

Feedback from the many young massive stars in the Carina Nebula has had a
profound influence on the surrounding interstellar medium. 
These massive stars have cleared a cavity
and sculpted dozens of gas pillars that face the center of the
Nebula.  The detailed structure of these pillars and other cloud
formations stand out in the H$_2$ images presented in our paper
because the H$_2$ images trace the thin PDRs on the surfaces that
outline these irradiated clouds.  The relatively high sub-arcsecond
angular resolution provided by these images complements the longer
wavelength data that trace the warm dust and molecular gas inside the
clouds seen in previous surveys with {\it IRAS}, {\it MSX}, {\it
Spitzer}, and CO data \citep{yonekura05,sb07,smith10b}.

Recently \citet{rocca13} analyzed the impact of feedback in
Carina using ground-based APEX and space-based data \citep{preib11b}
and Herschel Space Observatory sub-mm observations of \citet{preib12}.
However, disentangling the emission from the thin PDRs and protostars embedded
within them that emit at long wavelengths remains challenging
because emission from the PDR and the protostar are not
resolved from one-another at long wavelengths.  Perhaps this is a task
for ALMA, and to this end our new H$_2$ images will provide a reliable
reference for the structure of the PDRs over the entire region.
 
Some authors have proposed that the initial formation of the massive
stars in Carina was itself triggered by external influences.  Maps of
the molecular gas surrounding the Carina Nebula reveal a CO supershell
containing about $8 \times 10^2$ M$_\odot$ of H$_2$, the Carina Flare,
which appears to be blowing out of the Galactic Plane
\citep{fukui99,dawson08a,dawson08b,wunsch12}.  Emission in the 21 cm
HI line from the Carina Flare and 10 to 100 $\mu$m dust emission also
uncover a shell that extends up to 10\arcdeg\ above the Galactic plane
with a mass of at least $2 \times 10^5$ M$_\odot$.  \Citet{fukui99}
propose that energetic feedback inflated a 300 $\times$ 400~pc
suberbubble that expands with a mean velocity of about 10
km$\,$s$^{-1}$.  These authors propose that the Carina Flare was
produced by massive stars that formed about 120~pc above the Galactic
mid-plane, and that the impact of their energy release may have
triggered star formation in the current Carina Nebula.

\subsection{Triggering (or Not)?} 

There are three main conceptual scenarios that describe how young massive stars
affect star formation in their neighborhoods: (1) ``pre-existing
uncovered'' star formation (PRE-SF), where winds and radiation from
massive stars evacuate material in the region and reveal stars that
have already formed in multiple locations throughout the molecular
cloud, (2) ``external pressure-induced collapse'' of pre-existing
clouds that would not otherwise have formed stars in isolation
(e.g. induced by radiation, winds, or supernovae; EPIC-SF), and (3)
``collect and collapse'' star formation (CC-SF), where the external
agent is also responsible for forming the parent cloud 
\citep[see, e.g.][]{el77,bertoldi89}.

Although our new images of Carina provide a rich harvest of phenomena
associated with stars currently forming under the influence of
external feedback, the degree to which this star formation activity
was actually triggered by that feedback remains quite unclear.  This situation
echoes many discussions of morphological evidence for triggered star
formation in other regions; despite abundant wide-field imaging data,
the questions persist and have been debated for decades. 
Our images show that the Carina Nebula contains dozens of pillars with
cometary tails that trail away from Carina's massive stars, and in
nearly all cases there is a young star with an infrared excess
embedded at the head of the pillar, sometimes driving an outflow.
Although morphologies of this kind have been interpreted as evidence
for triggered star formation, this geometry could also arise 
as winds and radiation from massive stars uncover star
formation events that would have occurred anyway in the absence of any
triggering effects. 

To demonstrate that triggering occurs in any given
region is a difficult problem, requiring numerical
simulations with multi-dimensional radiation hydrodynamics, and perhaps
magnetic fields.  Currently, models have not provided decisive observational
tests to determine if star formation is triggered based on the
morphology of the gas and protostars, and in the absence of such
discerning tests, morphology alone as seen in our images cannot be
used to support or refute triggering.  If future theoretical studies can
clarify specific morphologies that are or are not due to triggering,
the H$_2$ images presented here may be quite valuable to address this
question. 

In the meantime, detailed investigations of age gradients
in the stars as compared to gas, as well as star and gas kinematics
may be more fruitful avenues to investigate triggering, although we
cannot address these with the present dataset.  Here too, though,
clear predictions of ages or age-gradients for triggered star
formation as opposed to uncovered (and posibly truncated) star
formation are not obvious.  The upcoming GAIA mission will have 10 to
100 micro-arcsecond proper motion sensitivity, which for Carina and other
nearby regions, has the potential to distinguish
pre-existing uncovered star formation from triggered star formation,
at least among the young stars already exposed at visual wavelengths.
Protostars still embedded in Carina's pillars, young clusters such as
the Treasure Chest \citep{smith05}, and the driving sources of active
outflows such as HH~666 exhibit large infrared excesses,
indicating ages less than about 1 Myr.  In contrast, the massive stars
that have evolved to the LBV or red supergiant phases, and the ``naked"
portions of clusters such as Tr 15 must be at least 3~Myrs
old.  But, as noted above, it remains possible that the younger stars
and clusters embedded in the surviving molecular
clouds of Carina would have formed anyway in the absence of the older
generation. 

In the case of the Treasure Chest, our H$_2$ images
(Sec.~\ref{ss:embedded_clusters}; Area 8 of Fig.~\ref{fig:spillars2})
show an external PDR produced by radiation from the massive stars in
the environment, as seen previously, but they also reveal an {\it internal}
PDR that marks the boundary of where the embedded cluster stars
dissociate molecules in their natal cloud core. Forming such a
structure by triggering is only viable if the lifetime of the core
after it is exposed by the triggering event - in this case radiation
and winds from the nearby massive stars - is long enough to allow for
an entire cluster to form and to create its own internal H~II region
and PDR.  This complex problem should be addressed
numerically in light of this observational constraint.

\subsection{The Carina Nebula as a Laboratory for PDR Physics} 

Models of PDRs typically specify the FUV radiation flux
G$_\circ$ (6 eV - 13.6 eV) and the particle density n, and then track how the molecular,
atomic, and ionic elemental abundances vary with depth into the molecular cloud.
Most theoretical formalisms use the visual extinction A$_V$ as a convenient
independent variable, with the boundary of the H~II region
at A$_V$ $\lesssim$ 1, followed by the PDR at larger A$_V$.
Specific zones exist within a PDR: models show a transition from
H to H$_2$ at A$_V$ $\sim$ 2,
C$^+$ to C and CO at A$_V$ $\sim$ 4, and more complex chemistry ongoing
at larger depths \citep{th85,htt91}.

Models of H~II regions require less chemistry, and are correspondingly
more sophisticated than their PDR counterparts when it comes to dynamics
such as R- and D-fronts \citep{henney05}, 3-D capability,
magnetic fields, and time-dependence \citep{arthur11},
turbulence \citep{medina14}, and photoevaporation from pillars \citep{ercolano11}.
These H~II region models also produce synthetic emission-line images that have a direct
connection to observations.  Merged models of H~II regions and PDRs
developed by assuming pressure equilibrium between the PDR and H~II region as
a first approximation to the dynamics are a relatively recent theoretical development
\citep{abel05,kaufman06}, and fractal models of PDRs are under development
\citep{andree13}. 

\begin{figure} 
\centering
\includegraphics[angle=0,scale=1.00]{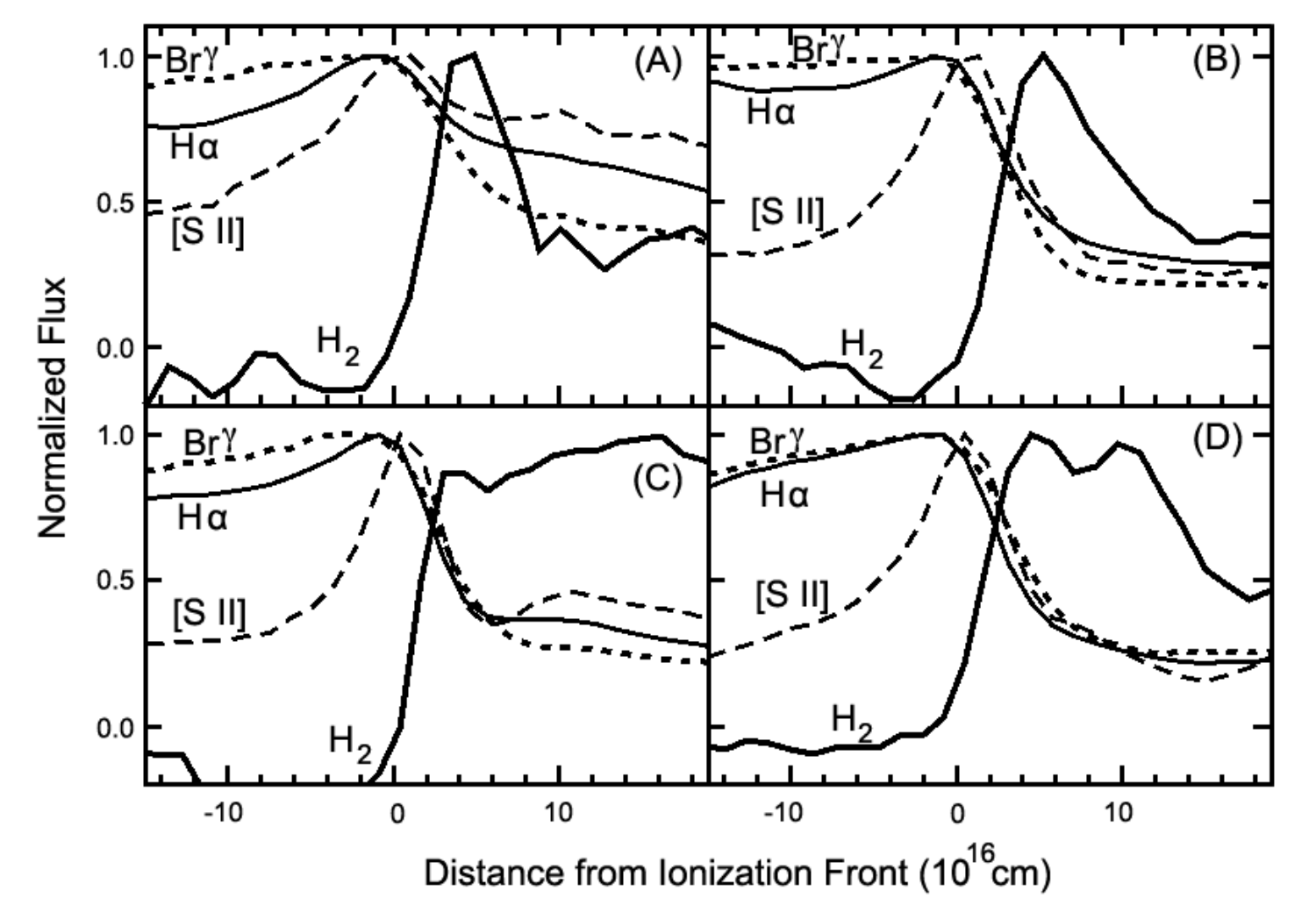}
\caption{Spatial offsets of Br$\gamma$ (short-dashed) and H$_2$ (bold) emission along the four positions
shown in Fig.~\ref{fig:western_color}. The fluxes are corrected for K-band continuum as described in the text. 
Note the distinct offset between H$_2$ and Br$\gamma$.} \label{fig:lineout}
\end{figure}

\begin{figure}
\centering
\includegraphics[angle=0,scale=1.00]{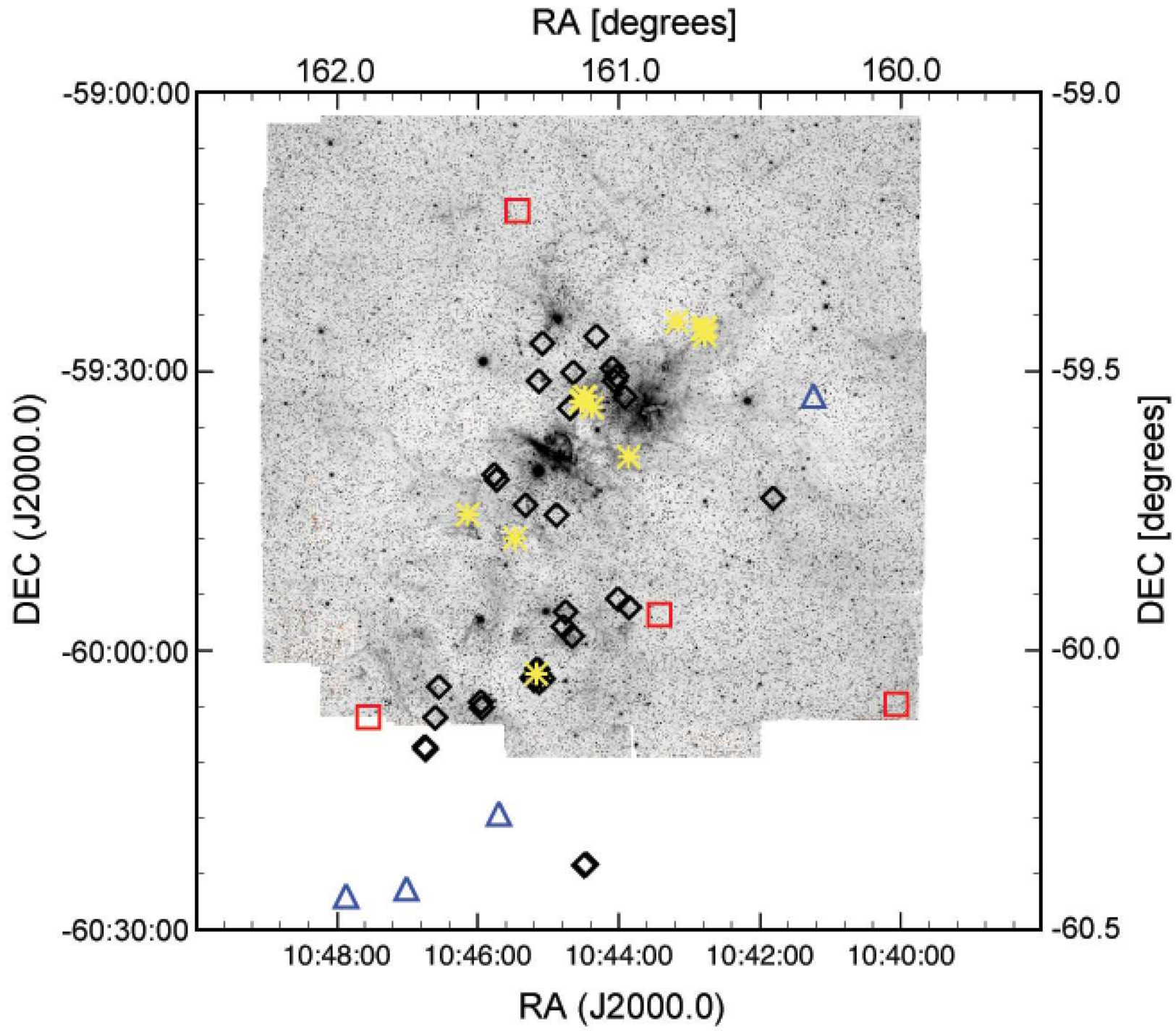}
\caption{Distribution of candidate jets and previously observed outflows in Carina plotted over the narrowband H$_2$ image.
Black diamonds indicate jets previously discovered with HST \citep{smith10a},
blue triangles indicate Spitzer EGOs from \citet{smith10b},
red squares are new candidate jets visible in [S~II],
and yellow stars are new H$_2$ flow candidates found in this work. }\label{fig:jet_distro}
\end{figure}

\begin{figure}
\centering
\includegraphics[angle=0,scale=1.00]{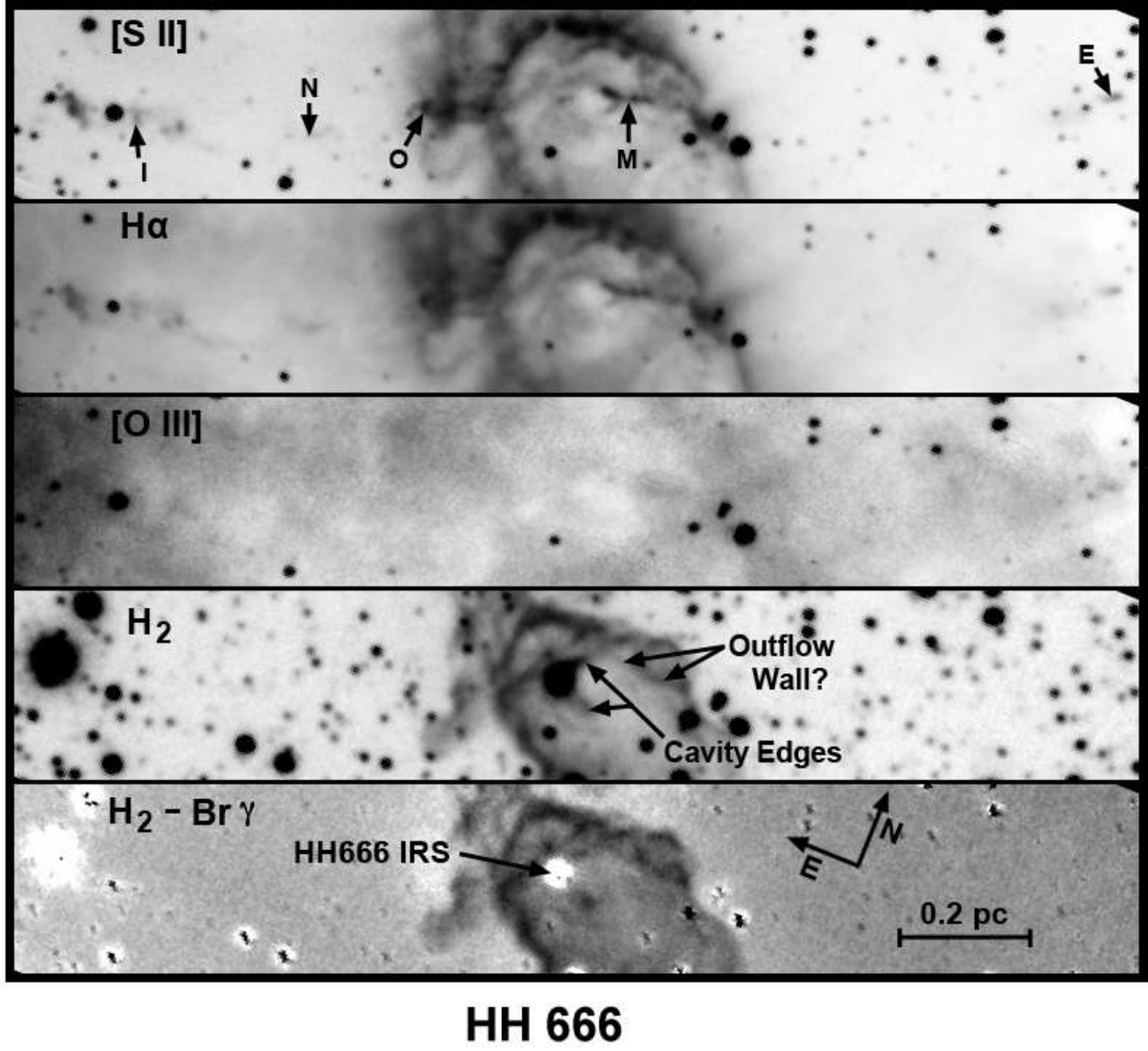}
\caption{Narrowband optical and IR images of HH 666. Labels of the jet features in the [S~II] image follow
the nomenclature of \citet{smith04a}. The H$_2$ emission outlines the edges of a cavity along the jet,
and may also trace the northern boundary of the outflow, but no H$_2$ is unambiguously associated with the jet.}
\label{fig:hh666}
\end{figure}

\begin{figure}
\centering
\includegraphics[angle=0,scale=1.00]{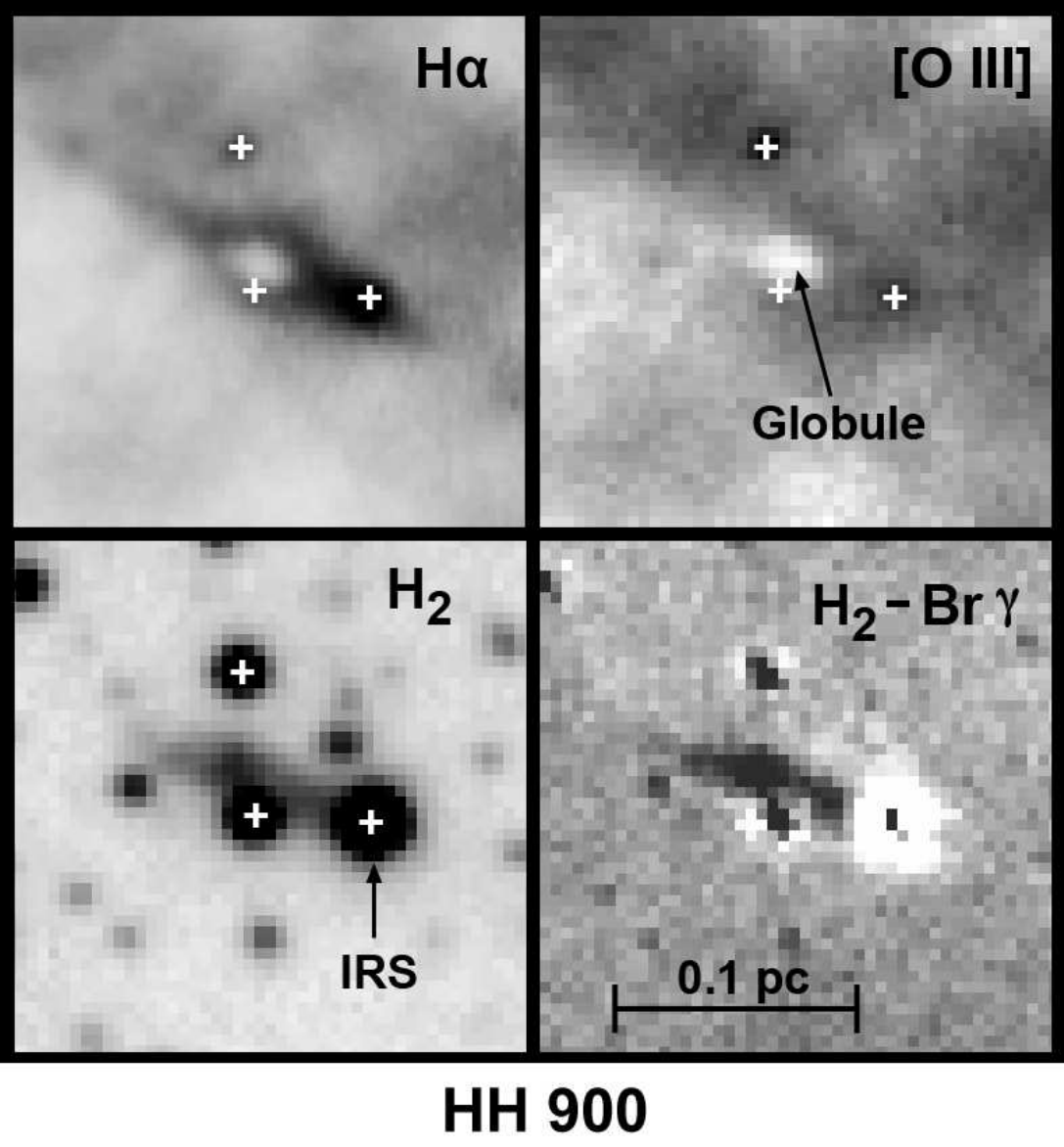}
\caption{Composite of HH 900. The dark globule and IRS (PCYC~838) discussed in the text
are marked. Plus-signs denote the location of three bright stars in the region.
}\label{fig:hh900}
\end{figure}

\begin{figure}
\centering
\includegraphics[angle=0,scale=0.90]{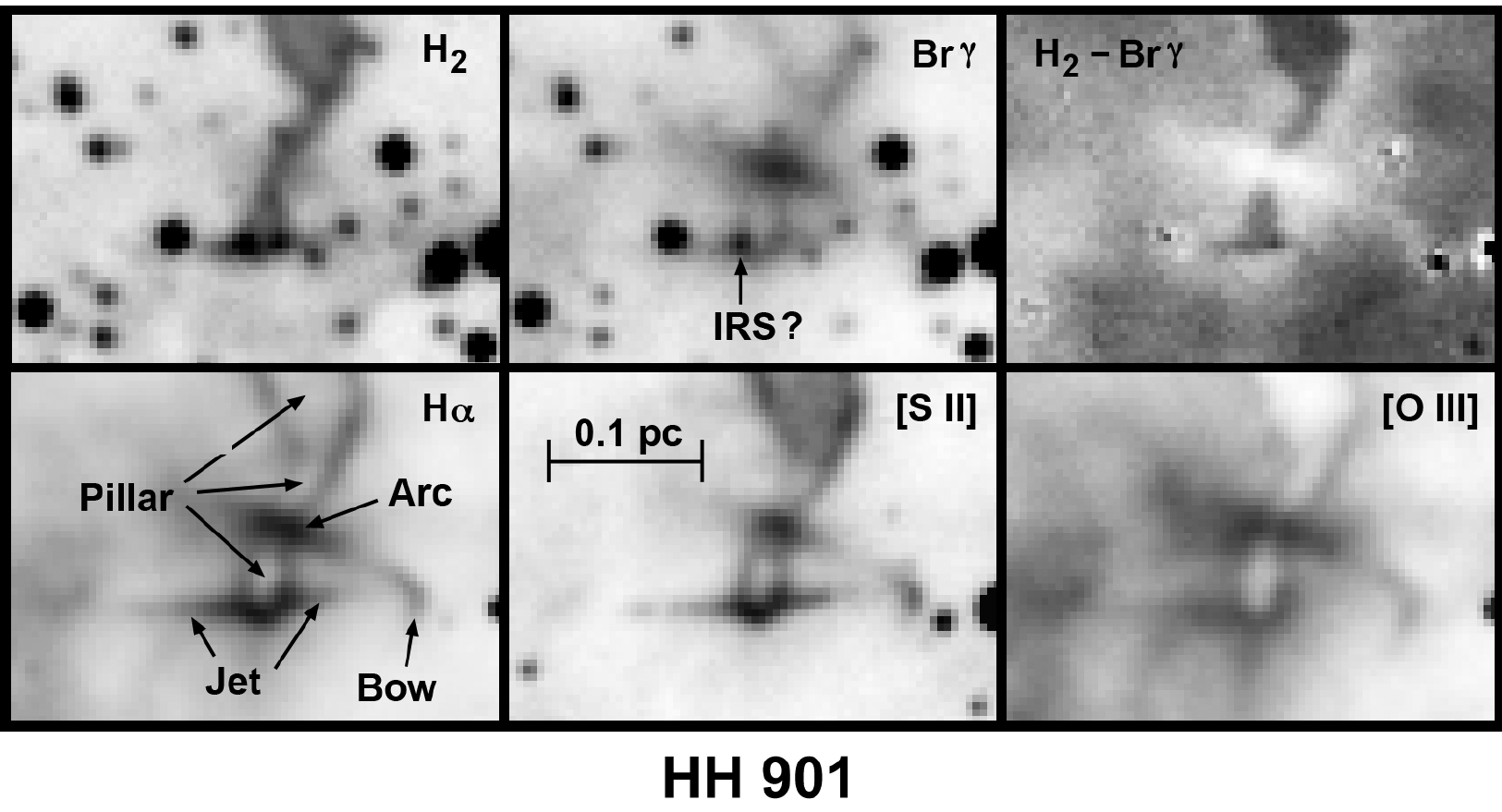}
\caption{Composite of HH 901. H$_2$ emission is evident as the jet emerges from the pillar. The
objects labeled arc, jet, pillar, and a potential driving source (IRS) are discussed in the text.}\label{fig:hh901}
\end{figure}

\begin{figure}
\centering
\includegraphics[angle=0,scale=0.95]{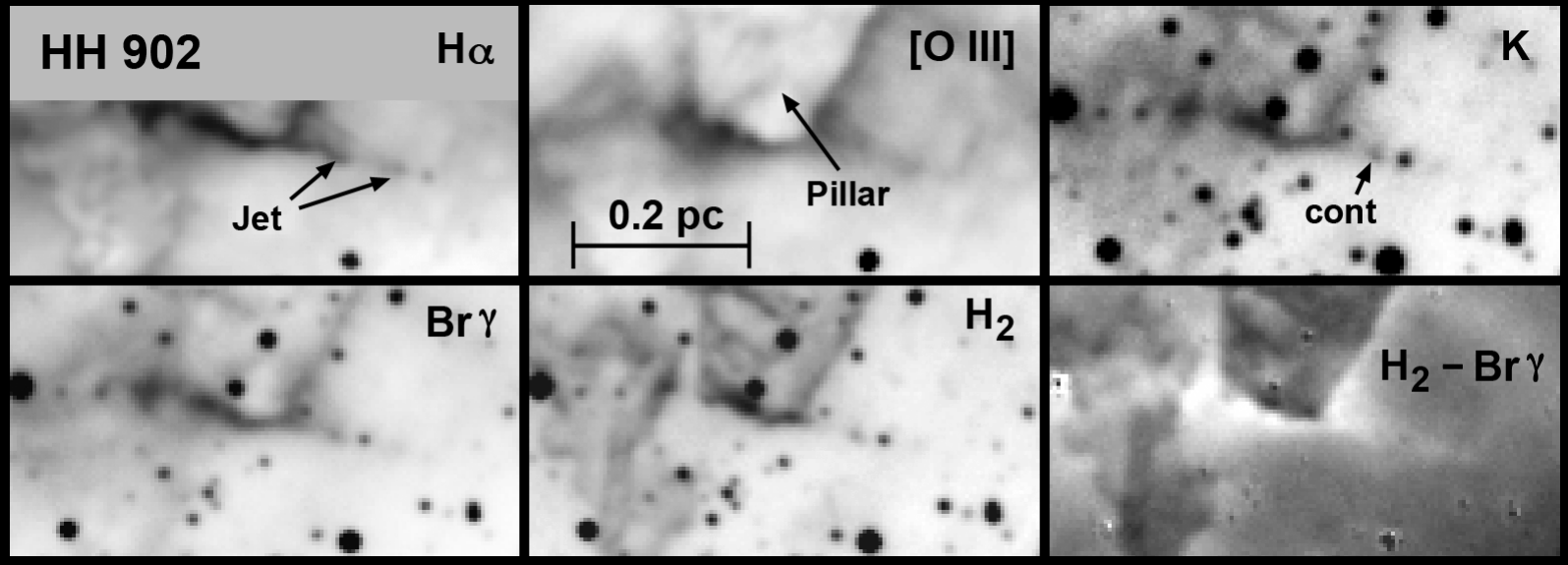}
\caption{Composite images of the HH~902 region. Labeled objects are described in the text.
}\label{fig:hh902}
\end{figure}

\begin{figure}
\centering
\includegraphics[angle=0,scale=0.95]{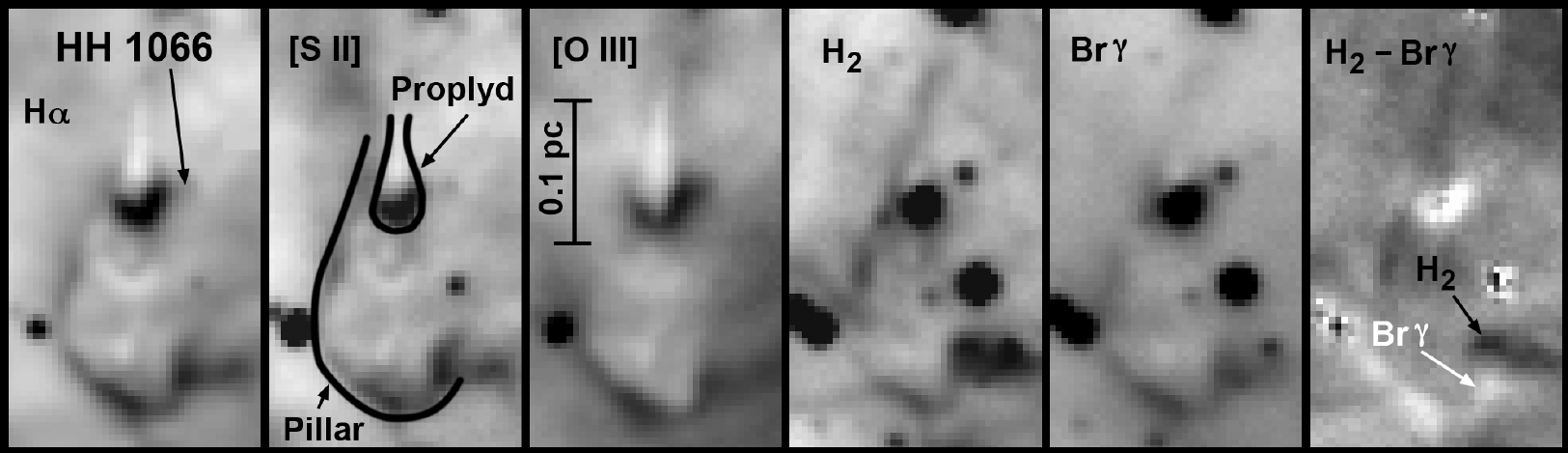}
\caption{Composite of the proplyd and pillars near HH~1066 (formerly HHc-1). The head of
the proplyd shows a distinct C-shape as it is illuminated by stars to the south. Regions marked
as bright Br$\gamma$ and H$_2$ emission are discussed in the text. The star at the head of the
proplyd is PCYC~429.
}\label{fig:hh1066}
\end{figure}

\begin{figure}
\centering
\includegraphics[angle=0,scale=1.00]{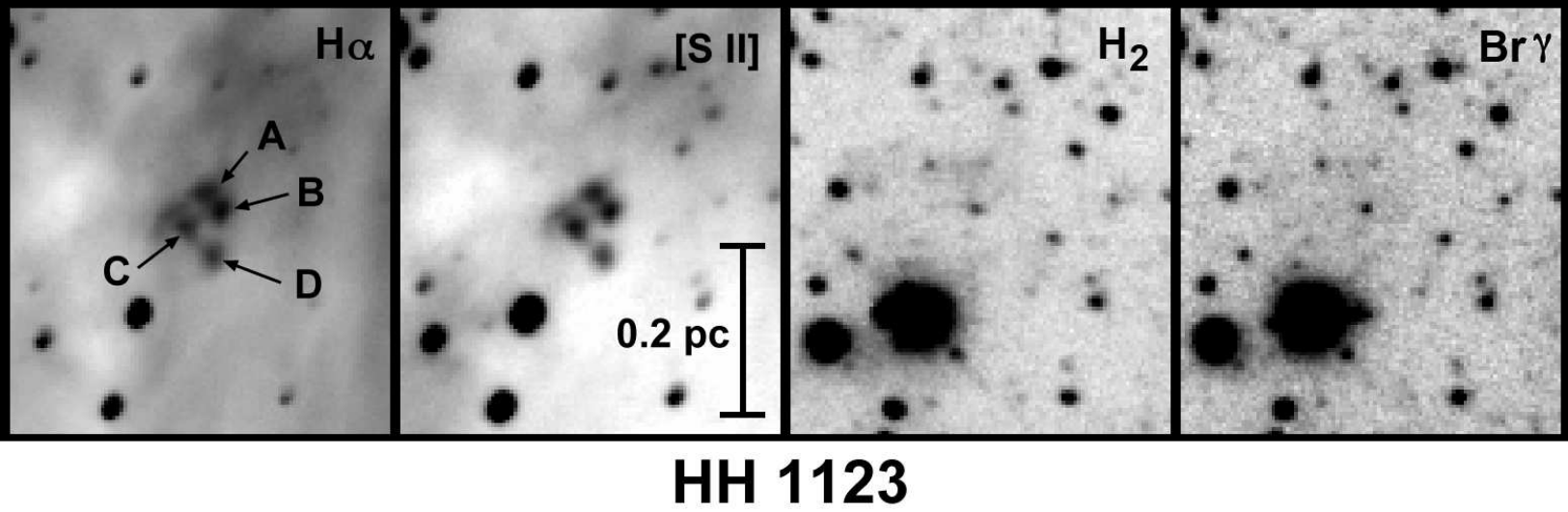}
\caption{Knots of H$\alpha$ and [S~II] in HH 1123.
No clear jet structure exists in any of the observed wavelengths.}\label{fig:hh1123}
\end{figure}

\begin{figure}
\centering
\includegraphics[angle=0,scale=1.00]{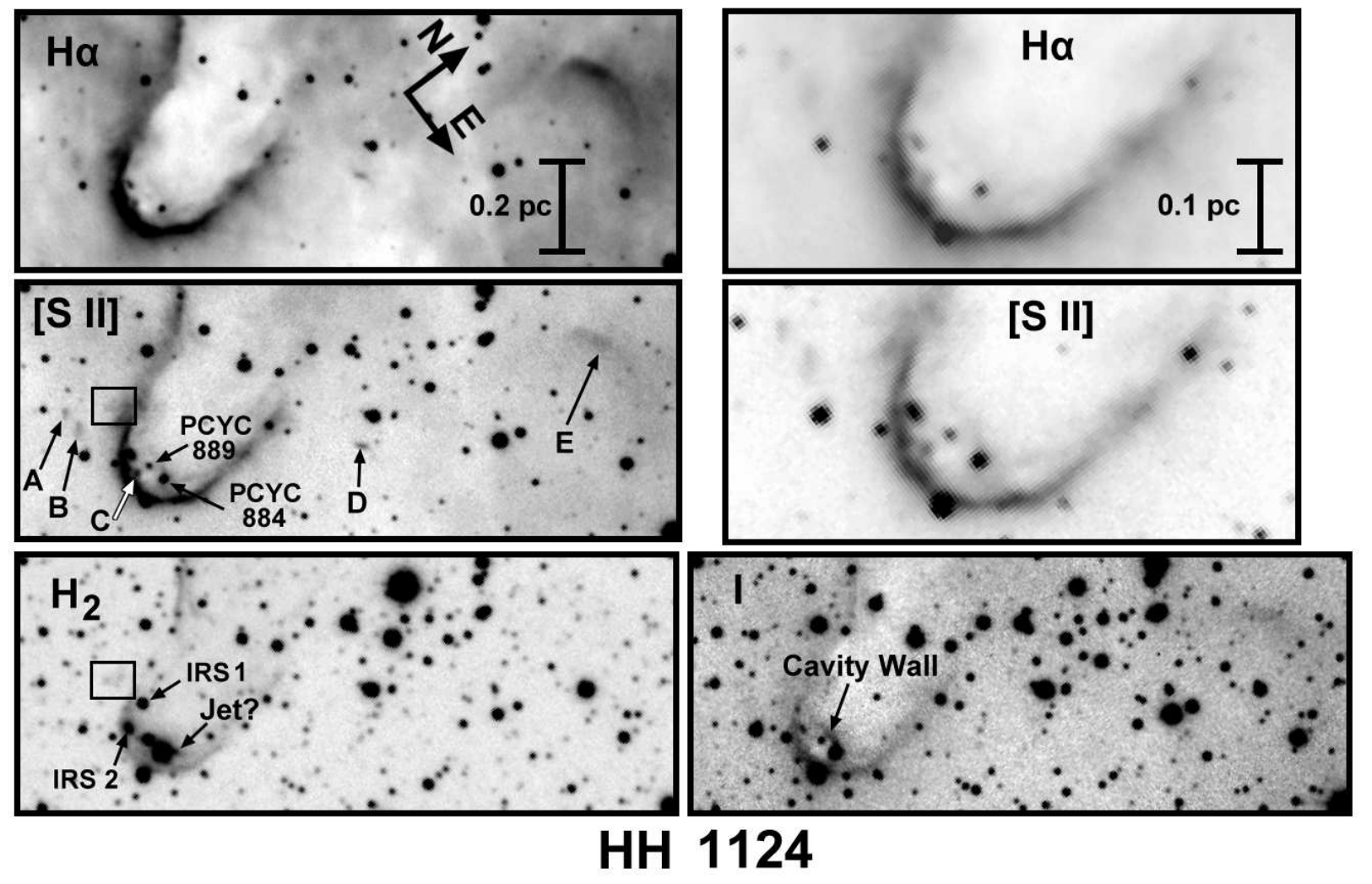}
\caption{Optical emission-line knots A$-$C outline a bent flow that emerges to the west of the
head of the pillar shown in Area 18 of Fig.~\ref{fig:npillars}. Knot D and the bow-shaped object E
lie to the east. The source labeled IRS (PCYC 884) shows what appears to be a cavity wall in reflected continuum.  
An extension in the H$_2$ image (also visible in the H$_2$ $-$ Br$\gamma$ image in Fig.~\ref{fig:npillars})
may outline a jet, but it could also define a photodissociation surface on the irradiated pillar.
Emission in the boxed region is discussed in the text.
}\label{fig:hh1124}
\end{figure}

\begin{figure}
\centering
\includegraphics[angle=0,scale=1.00]{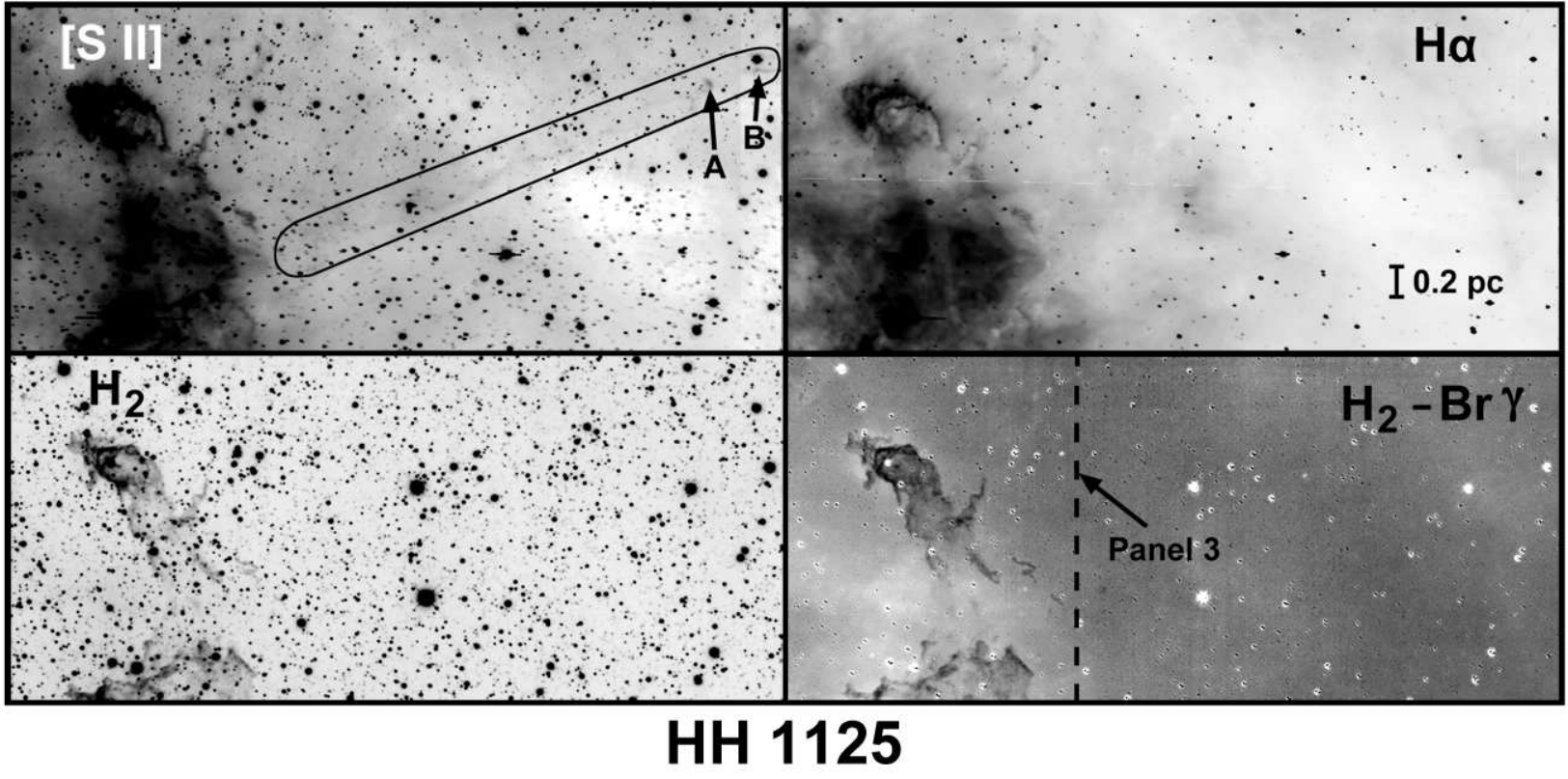}
\caption{The elongated region marked in the upper left panel encloses a
thin filament of [S~II] emission that comprises HH 1125.
There is no clear driving source if the object is a jet, though the
morphology of knot A resembles that of a bow shock. The filament is 
not visible in H$_2$, Br$\gamma$, [O~III] or H$\alpha$. A dashed line
marks the edge of Area 3 in Fig.~\ref{fig:spillars}.}
\label{fig:hh1125}
\end{figure}

\begin{figure}
\centering
\includegraphics[angle=0,scale=1.10]{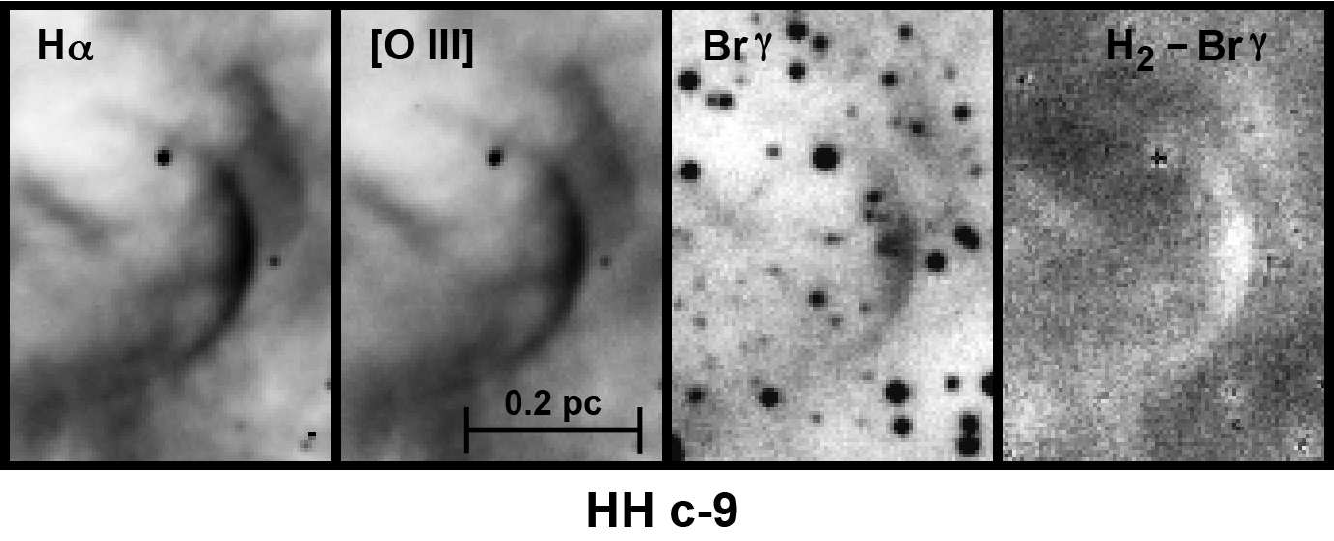}
\caption{The highly-ionized arc-shaped HH c-9 is bright in H$\alpha$ and [O~III], and has a similar
morphology in Br$\gamma$. No H$_2$ is present in this object. If the object is a bow shock, its
driving source remains unidentified.}\label{fig:hhc-9}
\end{figure}

\begin{figure}
\centering
\includegraphics[angle=0,scale=1.00]{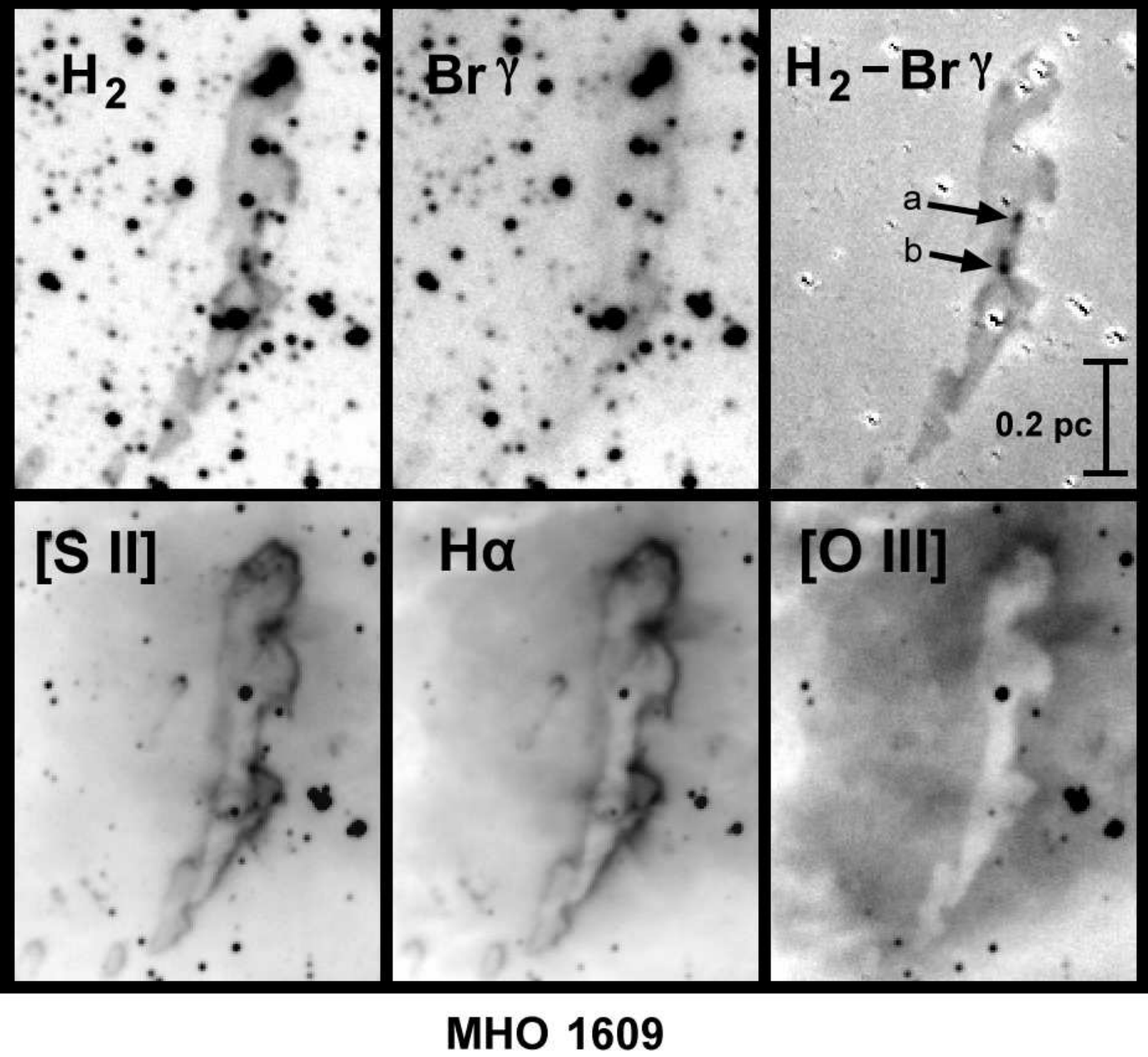}
\caption{Two bright, extended H$_2$ features within an irradiated pillar comprise MHO~1609.
Several optical outflow candidates were identified from this detached globule in the HST
data.
}\label{fig:MHO1609}
\end{figure}

\begin{figure}
\centering
\includegraphics[angle=0,scale=1.00]{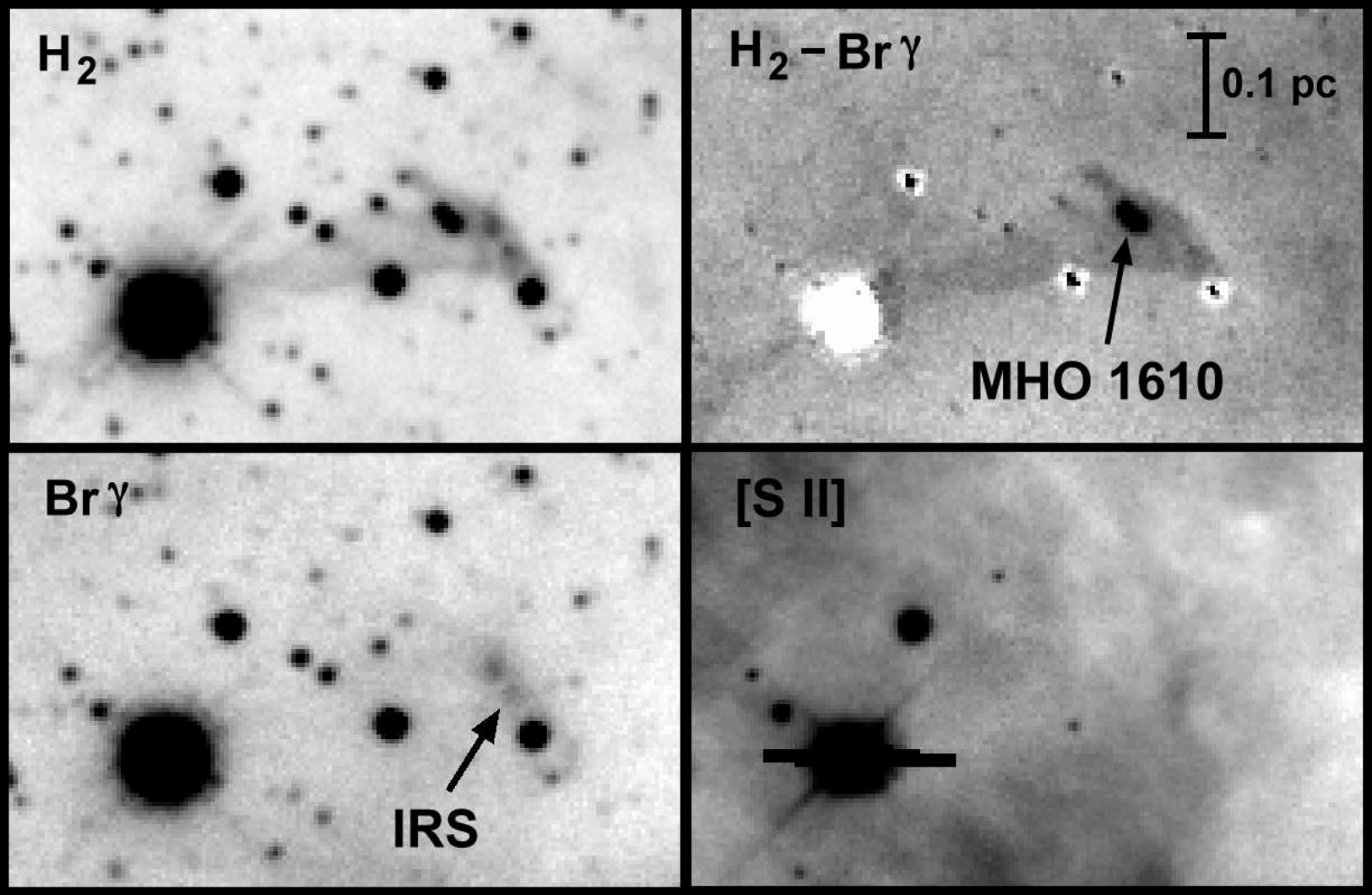}
\caption{Composite images of MHO~1610 (see also Area 10 of Fig.~\ref{fig:sewall})
reveal a single bright H$_2$ knot with no optical counterpart. A Herschel point source
at 70$\mu$m is coincident with the faint continuum source labeled IRS.}\label{fig:MHO1610}
\end{figure}

\begin{figure}
\centering
\includegraphics[angle=0,scale=1.00]{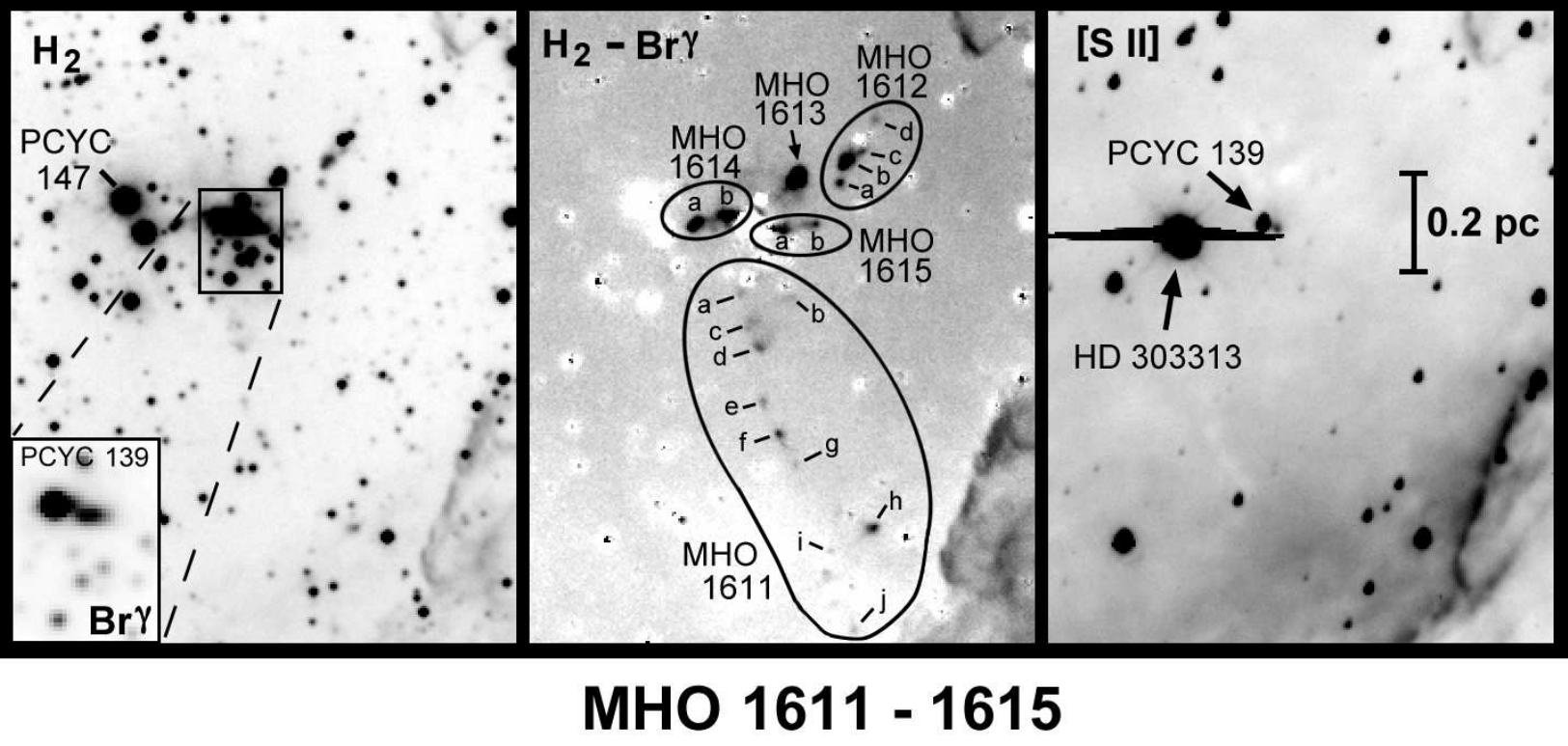}
\caption{Several bright H$_2$ knots in the vicinity of a small
cluster of stars centered on PCYC 139 comprise the candidate flows MHO 1611 - MHO 1615. 
These objects are identified by their H$_2$ emission, and lie outside a very
dark dust lane to the southwest. There are no obvious optical counterparts
to any of these outflow candidates. }\label{fig:MHO1611}
\end{figure}

\begin{figure}
\centering
\includegraphics[angle=0,scale=1.00]{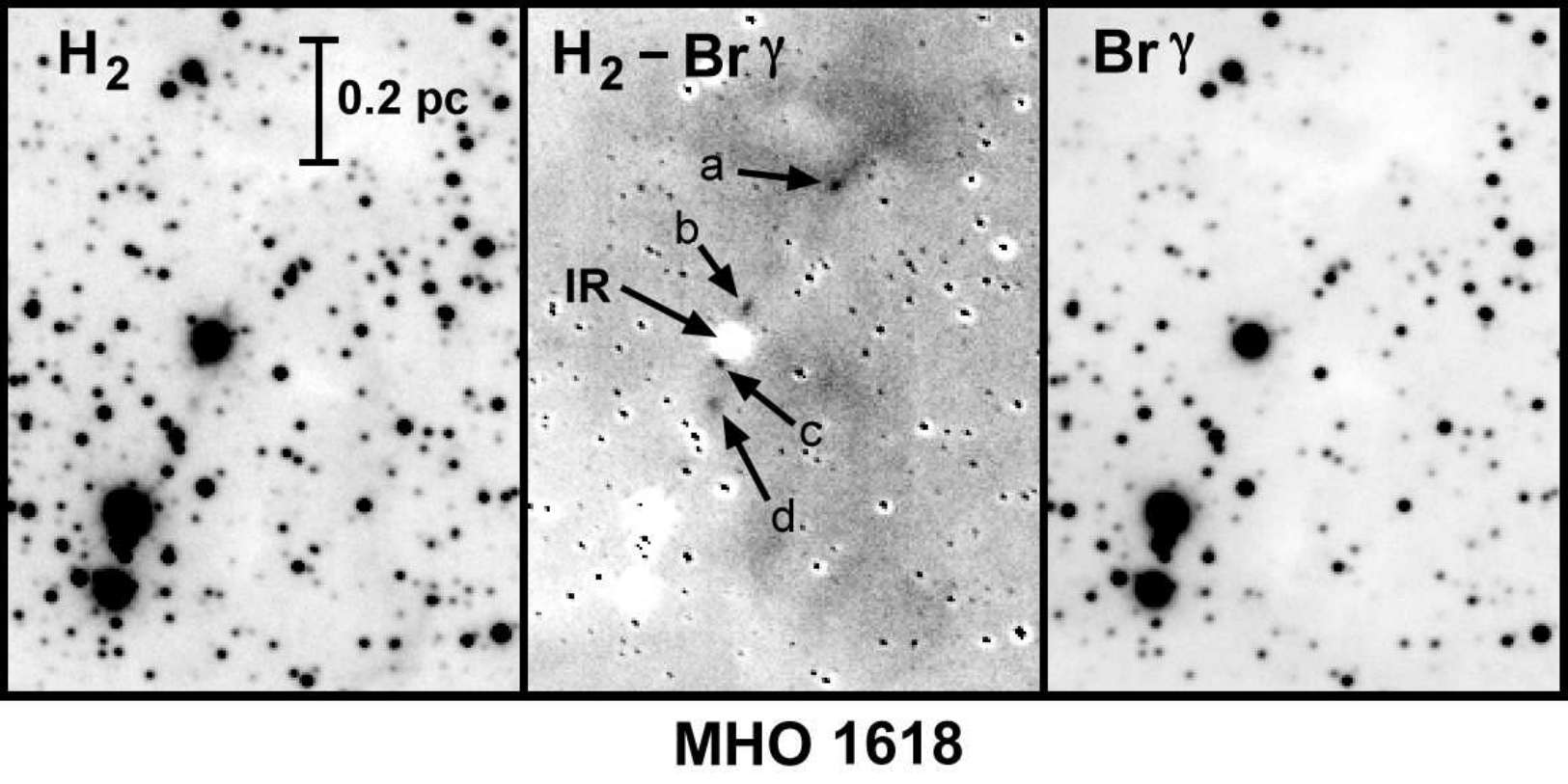}
\caption{Situated in the Eastern Walls,
the candidate outflow MHO 1618 appears as a curving string of H$_2$ knots on either
side of an infrared source.
}\label{fig:MHO1618}
\end{figure}

\begin{figure}
\centering
\includegraphics[angle=0,scale=1.00]{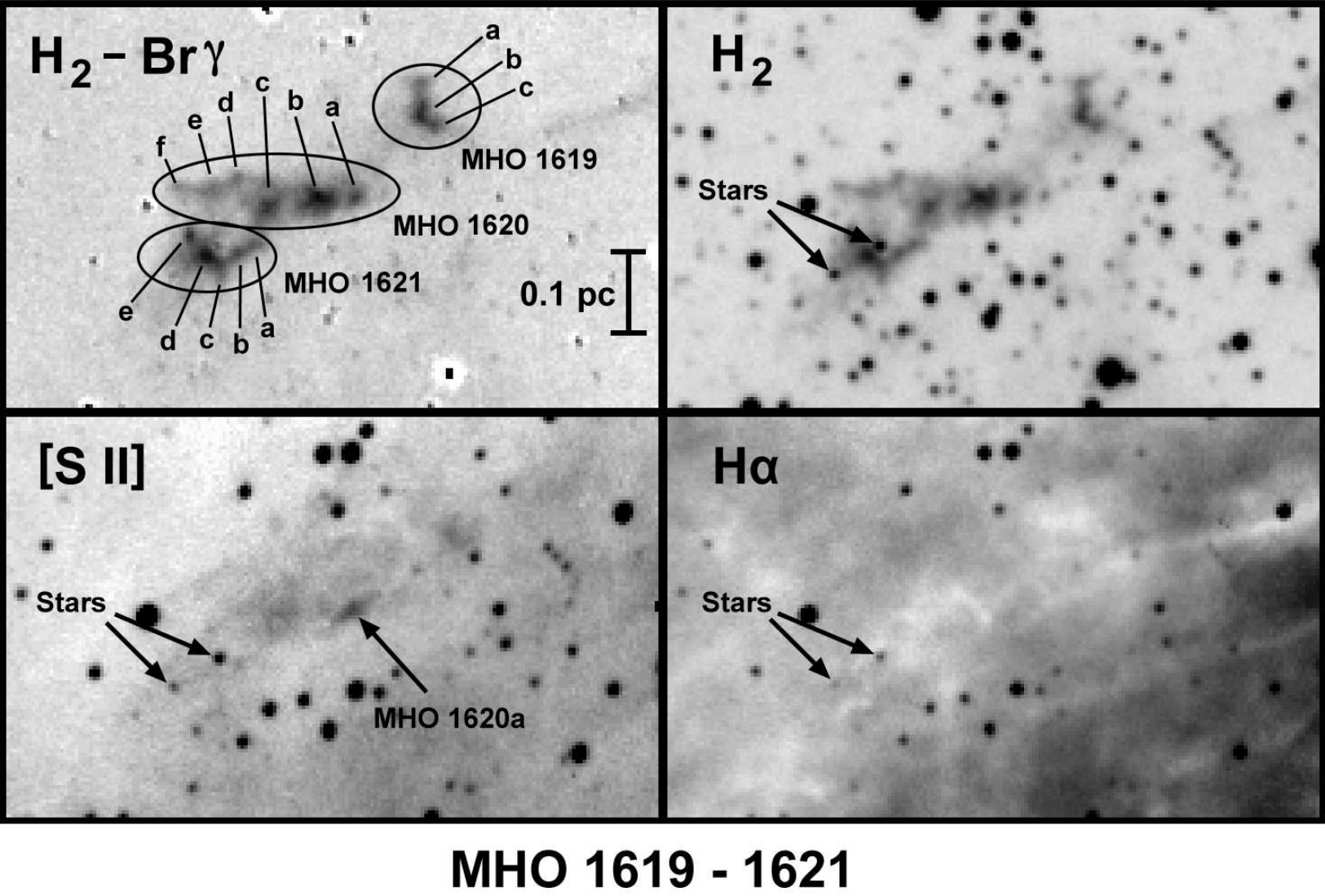}
\caption{Composite images of MHO 1619 $-$ MHO 1621 (see also Area 13 of Fig.~\ref{fig:sewall2}).
MHO 1620a is visible in the [S~II] image. No driving source is present.
}\label{fig:MHO1619}
\end{figure}

\begin{figure}
\centering
\includegraphics[angle=0,scale=1.00]{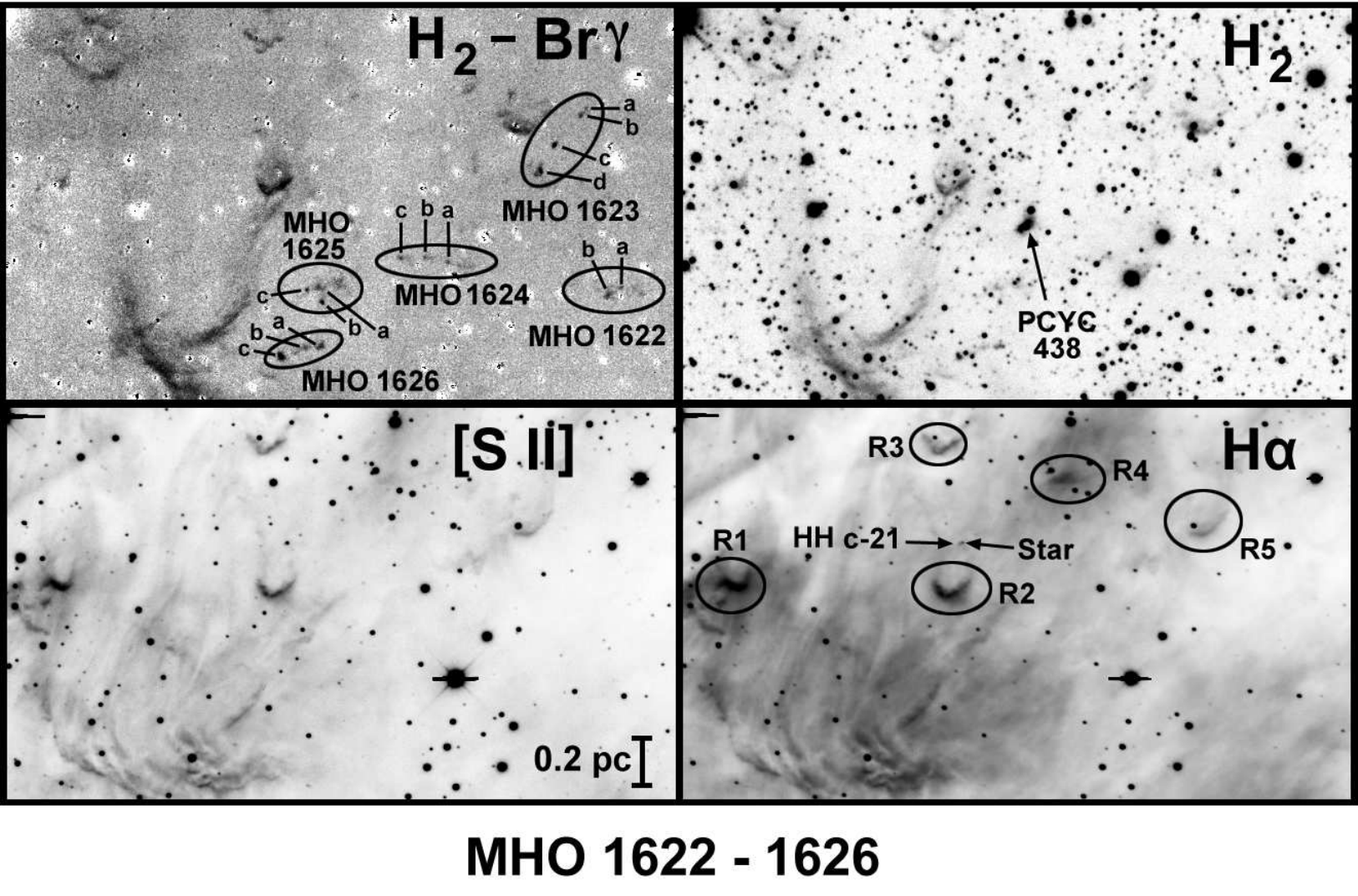}
\caption{Composite images in the G287.24-0.21 cloud, showing several strings of H$_2$ knots
groups as MHO 1622 $-$ MHO 1626 (see also Area 19 of Fig.~\ref{fig:npillars}).
The arc-shaped sources R1 $-$ R5 mark locations of irradiated pillars.
}\label{fig:MHO1622}
\end{figure}

\begin{figure}
\centering
\includegraphics[angle=0,scale=1.00]{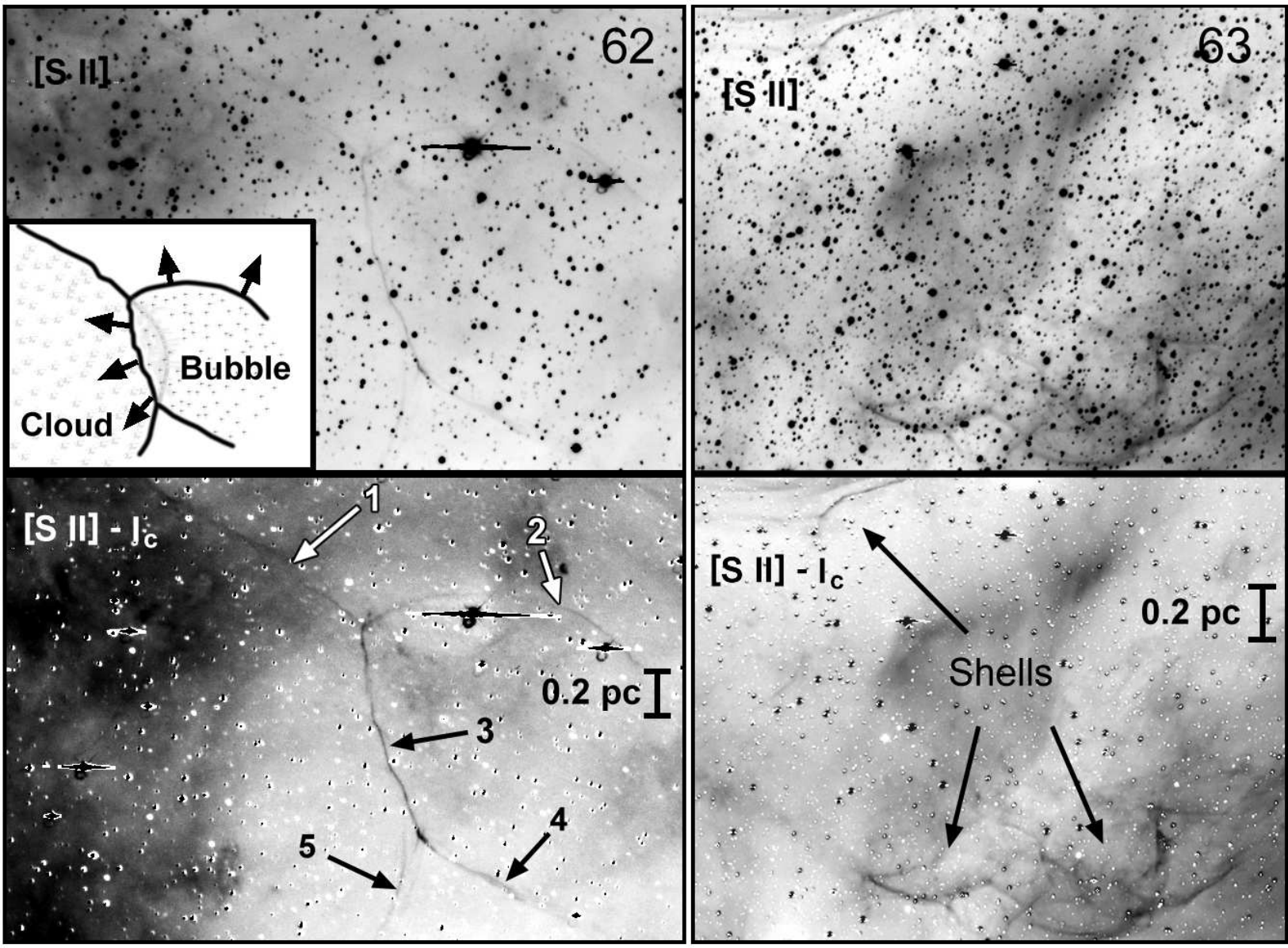}
\caption{Left: A network of five narrow filaments is visible in the
[S~II] image (top), and the continuum-subtracted image (bottom). The model described by the
inset is discussed in the text.  Right: Area 63 shows an extensive network of overlapping
shells in the [S~II] images. They are oriented as if driven by the massive stars to the
north of this region.
}\label{fig:shells}
\end{figure}

\begin{figure}
\centering
\includegraphics[angle=0,scale=1.00]{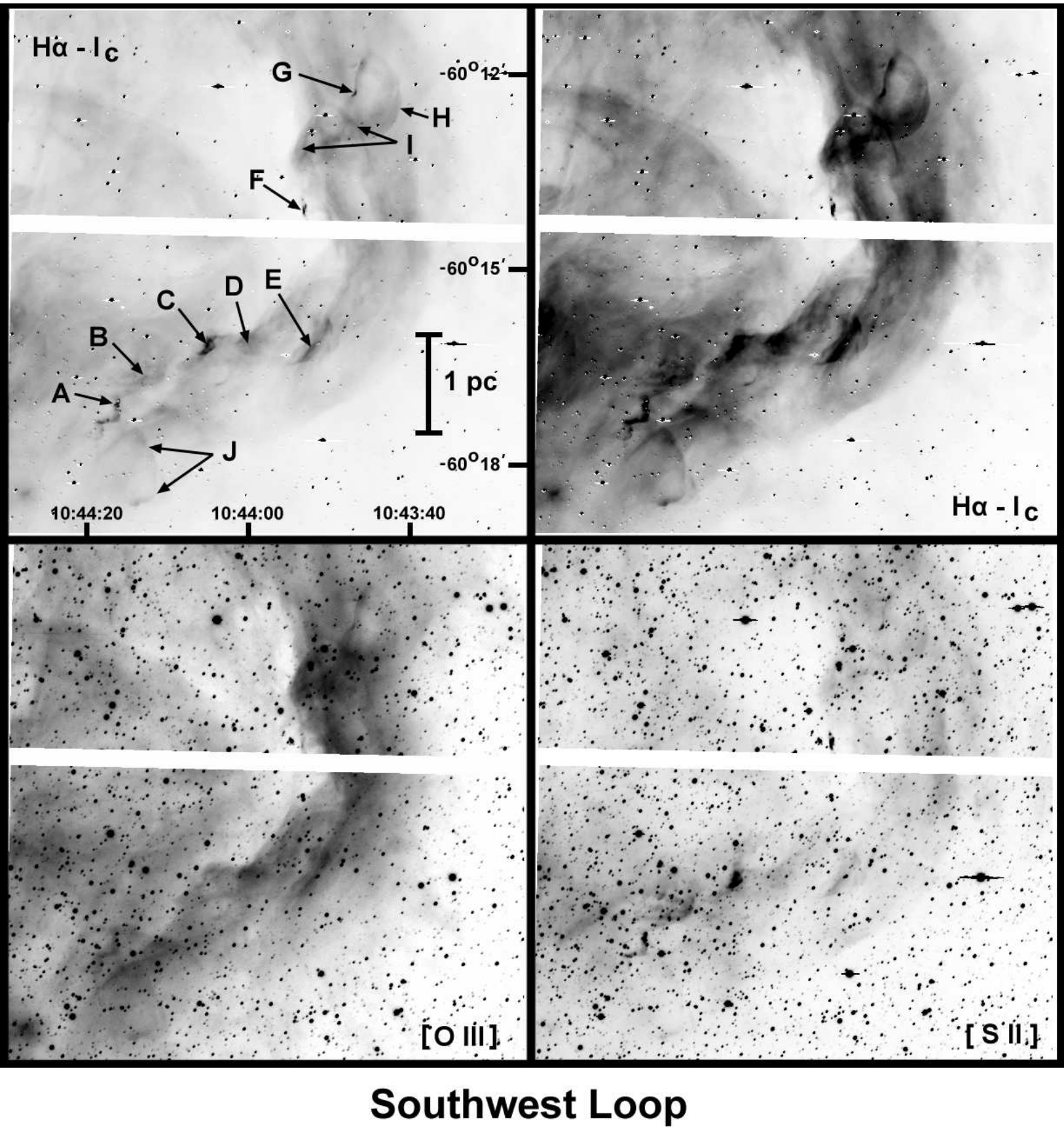}
\caption{Cavities, filaments and knots in the Southwestern Loop, located at the
edge of the bright optical emission in Carina. 
Top: H$\alpha$ images with scaled I$_c$ continnum subtracted, presented at two 
different greyscales. Bottom: A narrowband [O~III] image (left) and [S~II] image (right).
}\label{fig:swloop}
\end{figure}

\begin{figure}
\centering
\includegraphics[angle=0,scale=1.00]{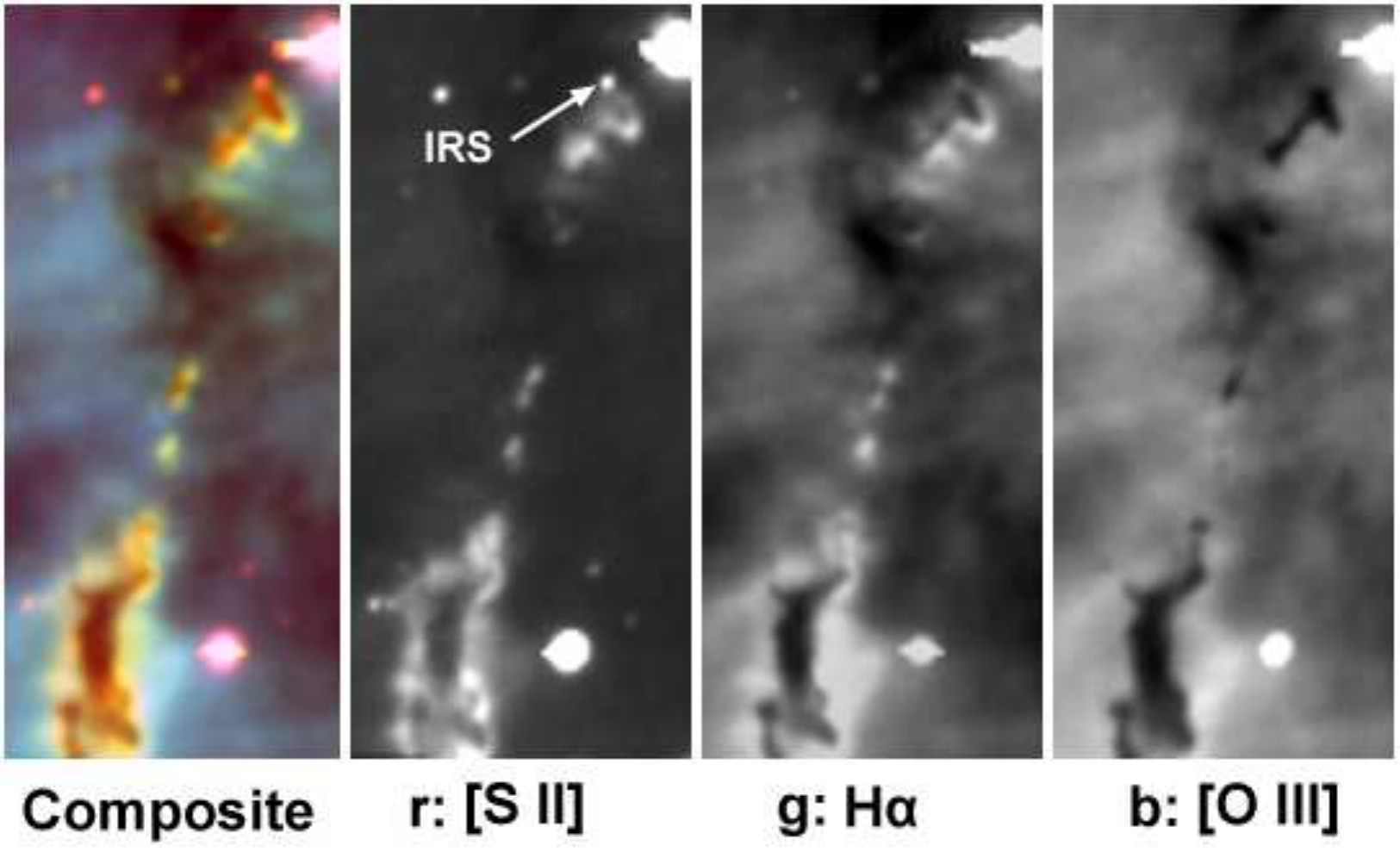}
\caption{An optical color composite (red = [S II], green = H$\alpha$ and blue = [O III]) of the remnants
of a pillar to the north of Carina's finger, shown in Area 37 of Fig.~\ref{fig:eta}.  The 
bright near-IR source labeled IRS is discussed in the text.  The emission 
lines arise from PDRs at the surface of the knots.
}\label{fig:finger}
\end{figure}

Our observations of H$_2$ 1-0 S(1), [S~II], Br$\gamma$ and H$\alpha$ probe
the boundary between the H~II region and the PDR and show clear offsets in the
spatial locations of these lines (Fig.~\ref{fig:lineout}).
While modern PDR models include H$_2$ (1-0) S1 emission from fluorescent processes,
and H~II region calculations include Br$\gamma$, H$\alpha$, [S~II], and [O~III] line
cooling, we are not aware of any simulations that predict synthetic emission maps for all
these lines in a single simulation, even for a 1-D model.
It should be possible to produce such maps for both a planar geometry
and pillars with existing codes. For example, motivated by new Spitzer mid-IR spectra,
\citet{kaufman06} made predictions for rotational H$_2$ 0-0 lines along with
forbidden lines of Si~II, Fe~II, and C~II. Relative to the Spitzer data,
our observations give an
order-of-magnitude improvement in spatial resolution and reliable photocenter
offsets between the H$_2$, H, and S~II lines. Simply predicting photocenter offsets
of all the bright emission lines in an H~II region/PDR interface in a 1-D model 
as a function of parameters such as densities, radiation fields, ages, and even
magnetic field strengths would be the easiest and most direct comparison with the
data in Fig.~\ref{fig:lineout}.

\section{Summary} 

We have presented a survey of the H$_2$ and Br$\gamma$ emission line sources over a square
degree in size centered on the Carina Nebula using the NEWFIRM camera at the CTIO 4-m telescope
with total integration times on-source of 12.5 and 8.7 ksec, respectively, and subarcsecond seeing.
Our near-IR images are complemented by new narrowband optical
images at [O~III], H$\alpha$, and [S~II], as well as 
a suite of broad-band images. The new data clarify the morphologies of irradiated interfaces
such as pillars, walls, and globules, identify jets, outflows and flared disks, and provide
insights into the star formation activity present within embedded clusters. We also quantified
spatial offsets that are present between various emission lines along the irradiated interfaces,
and compared each region with existing far-IR and X-ray images.

In many ways, the Carina Nebula is an ideal region to study
how massive stars affect their surroundings.
The large number of O-stars, combined with relatively low extinction and moderate
distance makes it possible to study star formation on a variety of scales. Surveys such
as ours provide an overview of the entire region, and also identify specific objects that
are worthy of followup with higher-resolution imaging and spectroscopy with HST and
ground-based AO systems. The high radiation environment in a region like Carina in some
ways makes it easier to interpret outflows, pillars, and walls because all irradiated dense regions become
visible in emission-line images. This advantage is particularly evident in fluorescent
lines such as H$_2$. However, in some regions projection of multiple irradiated surfaces
makes interpretation difficult. Future work centered on time-domain observations such
as high-precision proper motions should help to clarify the dynamics of the stars and gas
in this remarkable region.

\acknowledgements
We thank Jacob Palmer for his assistance with the NEWFIRM data acquisition at the 4-m,
and the CTIO and NOAO staff for their efforts in getting the instrument operational.
Megan Reiter would like to thank Andy Marble for his help generating the jet distribution plot.
We thank Bo Reipurth and Chris Davis for quickly providing new HH and MHO object numbers,
and an anonymous referee for helpful suggestions regarding the presentation and organization
of the paper. The PI was funded by DOE through the NLUF program during the course of this project.

\null

\begin{center}
\begin{deluxetable}{rrccc}\label{t:exptime}
\tabletypesize{\scriptsize}
\singlespace
\tablewidth{0pt}
\tablecolumns{5}
\tabcolsep = 0.08in
\parindent=0em
\tablecaption{Summary of Carina Mosaics}
\tablehead{
\colhead{Date} & \colhead{Filter} & \colhead{Exposure Time (ks)$^a$} & \colhead{FWHM (arcsec)$^b$} & \colhead{Camera}}
\startdata
9-10 Mar 2003  & [O III] 5007\AA       & 0.60 & 0.8 & Mosaic\\
9-10 Mar 2003  & H$\alpha$ 6563\AA     & 0.60 & 0.8 & Mosaic\\
9-10 Mar 2003  & [S II] 6720\AA        & 0.60 & 0.8 & Mosaic\\
9-10 Mar 2003  & I 0.9$\mu$m           & 0.06 & 0.8 & Mosaic\\
13-16 Mar 2011 & H$2$ 2.12$\mu$m       & 12.5 & 0.7 & Newfirm\\
14-17 Mar 2011 & Br$\gamma$ 2.16$\mu$m & 8.7  & 0.8 & Newfirm\\
 9-18 Mar 2011 & K 2.2$\mu$m           &      & 0.7 $-$ 2.2& Newfirm\\
21-22 Jan, 13 Jun, 17 Jul 2005 &  3.6$\mu$m   & 0.032 & $\sim 2$ & Spitzer\\
21-22 Jan, 13 Jun, 17 Jul 2005 &  4.5$\mu$m   & 0.032 & $\sim 2$ & Spitzer\\
21-22 Jan, 13 Jun, 17 Jul 2005 &  5.8$\mu$m   & 0.032 & $\sim 2$ & Spitzer\\
21-22 Jan, 13 Jun, 17 Jul 2005 &  8.0$\mu$m   & 0.032 & $\sim 2$ & Spitzer\\
\enddata
\tablenotetext{a} {Typical exposure time at center of mosaic}
\tablenotetext{b} {Image quality of final coadded mosaic}
\end{deluxetable}
\end{center}

\begin{center}
\begin{deluxetable}{lcccl}\label{t:coords}
\tabletypesize{\scriptsize}
\singlespace
\tablewidth{0pt}
\tablecolumns{4}
\tabcolsep = 0.08in
\parindent=0em
\tablecaption{Coordinates for images}
\tablehead{
\colhead{Object} & \colhead{$\alpha$ (2000)} & \colhead{$\delta$ (2000)} & \colhead{Figure} & \colhead{Notes}
}
\startdata
Area  1 center& 10:45:22.07& -59:58:48 &\ref{fig:spillars}&\\
Area  2 center& 10:45:58.27& -60:06:30 &\ref{fig:spillars}&\\
Area  3 center& 10:43:50.16& -59:57:38 &\ref{fig:spillars}&\\
Area  4 center& 10:44:41.04& -59:57:37 &\ref{fig:spillars}&\\
Area  5 center& 10:47:07.56& -60:02:15 &\ref{fig:spillars}&\\
Area  6 center& 10:45:13.57& -60:02:51 &\ref{fig:spillars2}&\\
Area  7 center& 10:44:03.21& -60:07:00 &\ref{fig:spillars2}&\\
Area  8 center& 10:45:52.71& -59:58:40 &\ref{fig:spillars2}&\\
Area  8 inset& 10:45:53.73& -59:57:02 &\ref{fig:spillars2}&\\
Area  9 center& 10:45:01.54& -59:47:43 &\ref{fig:sewall}&\\
Area 10 center& 10:46:03.19& -59:44:42 &\ref{fig:sewall}&\\
Area 11 center& 10:45:24.30& -59:45:37 &\ref{fig:sewall2}&\\
Area 12 center& 10:45:28.92& -59:53:37 &\ref{fig:sewall2}&\\
Area 13 center& 10:46:05.22& -59:33:46 &\ref{fig:sewall2}&\\
Area 14 center& 10:47:00.47& -59:37:49 &\ref{fig:sewall2}&\\
Area 15 center& 10:47:04.49& -59:51:43 &\ref{fig:sewall3}&\\
Area 16 center& 10:45:58.82& -59:12:50 &\ref{fig:npillars}&\\
Area 17 center& 10:46:18.41& -59:14:27 &\ref{fig:npillars}&\\
Area 18 center& 10:45:31.98& -59:12:23 &\ref{fig:npillars}&\\
Area 19 center& 10:44:07.69& -59:08:58 &\ref{fig:npillars}&\\
Area 20 center& 10:45:00.73& -59:21:60 &\ref{fig:npillars}&\\
Area 21 center& 10:45:15.82& -59:27:31 &\ref{fig:npillars2}&\\
Area 22 center& 10:44:01.97& -59:18:00 &\ref{fig:npillars2}&\\
Area 23 center& 10:44:47.22& -59:27:21 &\ref{fig:tr14}&\\
Area 24 center& 10:44:19.45& -59:25:44 &\ref{fig:tr14}&\\
Area 25 center& 10:44:22.42& -59:27:53 &\ref{fig:tr14}&\\
Area 26 center& 10:44:04.15& -59:33:42 &\ref{fig:tr14}&\\
Area 27 center& 10:43:30.73& -59:27:45 &\ref{fig:tr14}&\\
Area 28 center& 10:44:33.88& -59:34:57 &\ref{fig:tr14}&\\
Area 29 center& 10:43:58.22& -59:29:44 &\ref{fig:tr14}&\\
Area 30 center& 10:44:59.42& -59:31:19 &\ref{fig:tr14}&\\
Area 31 center& 10:43:38.38& -59:33:18 &\ref{fig:tr14}&\\
Area 32 center& 10:43:49.54& -59:37:40 &\ref{fig:tr14}&\\
Area 33 center& 10:44:26.54& -59:32:43 &\ref{fig:tr14}&\\
Area 34 center& 10:45:07.60& -59:39:17 &\ref{fig:eta}&\\
Area 35 center& 10:44:48.08& -59:37:39 &\ref{fig:eta}&\\
Area 36 center& 10:44:50.97& -59:40:55 &\ref{fig:eta}&\\
Area 37 center& 10:44:31.44& -59:39:21 &\ref{fig:eta}&\\
Area 38 center& 10:45:12.50& -59:37:41 &\ref{fig:eta}&\\
Area 39 center& 10:44:43.02& -59:44:03 &\ref{fig:eta}&\\
Area 40 center& 10:44:03.17& -59:40:42 &\ref{fig:eta}&\\
Area 41 center& 10:44:21.10& -59:42:04 &\ref{fig:eta}&\\
Area 42 center& 10:45:32.65& -59:37:41 &\ref{fig:eta}&\\
Area 43 center& 10:43:41.47& -59:41:45 &\ref{fig:eta}&\\
Area 44 center& 10:41:39.21& -59:43:33 &\ref{fig:swpillar}&\\
Area 45 center& 10:41:20.69& -59:48:23 &\ref{fig:swpillar}&\\
Area 46 center& 10:42:43.72& -59:36:02 &\ref{fig:swpillar}&\\
Area 47 center& 10:41:35.31& -59:35:07 &\ref{fig:swpillar}&\\
Area 48 center& 10:42:12.15& -59:35:30 &\ref{fig:swpillar}&\\
Area 49 center& 10:41:54.01& -59:37:59 &\ref{fig:swpillar}&\\
Area 50 center& 10:43:13.16& -59:37:24 &\ref{fig:swpillar}&\\
Area 51 center& 10:42:50.33& -59:31:43 &\ref{fig:nwpillar}&\\
Area 52 center& 10:42:36.51& -59:26:22 &\ref{fig:nwpillar}&\\
Area 53 center& 10:43:05.53& -59:32:36 &\ref{fig:nwpillar}&\\
Area 54 center& 10:43:20.57& -59:31:05 &\ref{fig:nwpillar}&\\
Area 55 center& 10:42:58.29& -59:17:25 &\ref{fig:nwpillar}&\\
Area 56 center& 10:42:09.74& -59:32:52 &\ref{fig:nwpillar}&\\
Area 57 center& 10:43:10.69& -59:24:44 &\ref{fig:nwpillar}&\\
Area 58 center& 10:41:12.93& -59:32:43 &\ref{fig:nwpillar}&\\
Area 59 center& 10:43:07.04& -59:29:04 &\ref{fig:nwpillar}&\\
Area 60 center& 10:42:10.27& -59:20:24 &\ref{fig:nwpillar}&\\
Area 61 center& 10:43:26.00& -59:34:45 &\ref{fig:western_wall}&\\
Area 62 center& 10:43:17.13& -60:02:51 &\ref{fig:shells}&\\
Area 63 center& 10:41:20.90& -60:08:18 &\ref{fig:shells}&\\
\noalign{\vskip 0.1in}
\multispan5{\hfil New Candidate Flows and Associated Objects\hfil}\\
HH 1123 A &  10:40:03.82 &  $-$60:05:43.1  &\ref{fig:hh1123}&H$\alpha$, [S II]  \\
HH 1123 B &  10:40:03.57 &  $-$60:05:44.9  &\ref{fig:hh1123}&H$\alpha$, [S II]  \\
HH 1123 C &  10:40:04.07 &  $-$60:05:46.8  &\ref{fig:hh1123}&H$\alpha$, [S II]  \\
HH 1123 D &  10:40:03.66 &  $-$60:05:49.8  &\ref{fig:hh1123}&H$\alpha$, [S II]  \\
HH 1124 PCYC 884&  10:45:28.53 &  $-$59:13:18.8&\ref{fig:hh1124}&Star \\
HH 1124 PCYC 889&  10:45:29.01 &  $-$59:13:19.5&\ref{fig:hh1124}&Star \\
HH 1124 IRS1&  10:45:29.84 &  $-$59:13:16.1&\ref{fig:hh1124}&Star\\
HH 1124 IRS2&  10:45:29.54 &  $-$59:13:21.0&\ref{fig:hh1124}&Star\\
HH 1124 A&  10:45:31.39 &  $-$59:13:26.6&\ref{fig:hh1124}&H$\alpha$, [S II]  \\
HH 1124 B&  10:45:30.83 &  $-$59:13:26.4&\ref{fig:hh1124}&H$\alpha$, [S II]  \\
HH 1124 C&  10:45:29.07 &  $-$59:13:22.3&\ref{fig:hh1124}&H$\alpha$, [S II]  \\
HH 1124 D&  10:45:26.17 &  $-$59:12:44.0&\ref{fig:hh1124}&H$\alpha$, [S II]  \\
HH 1124 E&  10:45:24.60 &  $-$59:11:54.9&\ref{fig:hh1124}&H$\alpha$, [S II]  \\
HH 1125 A&  10:43:04.18 &  $-$59:55:01.2&\ref{fig:hh1125}&[S II]  \\
HH 1125 B&  10:43:00.44 &  $-$59:54:52.4&\ref{fig:hh1125}&[S II]  \\
HHc-21 Star& 10:44:10.20 &  $-$59:08:06.5&\ref{fig:MHO1622}&\\  
MHO 1605 center&  10:43:51.48 &  $-$59:39:11.0&\ref{fig:tr14}&H$_2$, [S II]  \\
IRS 902&  10:44:31.09 &  $-$59:33:09.6&\ref{fig:tr14}&`Disk Object' \\
MHO 1606&  10:44:22.67 &  $-$59:33:53.6&\ref{fig:tr14}&H$_2$, [S~II]  \\
MHO 1607&  10:44:28.48 &  $-$59:32:45.5&\ref{fig:tr14}&H$_2$, east knot\\
MHO 1608&  10:44:29.12 &  $-$59:32:42.6&\ref{fig:tr14}&H$_2$  \\
MHO 1609a& 10:45:09.89 &  $-$60:02:25.8&\ref{fig:MHO1609}, \ref{fig:spillars2}&H$_2$, H$\alpha$, [S~II] \\
MHO 1609b& 10:45:10.21 &  $-$60:02:34.1&\ref{fig:MHO1609}, \ref{fig:spillars2}&H$_2$, H$\alpha$, [S~II] \\
MHO 1610&  10:46:08.45 &  $-$59:45:23.8&\ref{fig:MHO1610}&H$_2$ \\
MHO 1610 IRS&10:46:07.72 &  $-$59:45:26.6&\ref{fig:MHO1610}&Br$\gamma$, H$_2$\\
MHO 1611 a & 10:42:48.31 &  $-$59:25:42.1&\ref{fig:MHO1611}&H$_2$  \\
MHO 1611 b & 10:42:47.10 &  $-$59:25:42.8&\ref{fig:MHO1611}&H$_2$  \\
MHO 1611 c & 10:42:48.04 &  $-$59:25:47.3&\ref{fig:MHO1611}&H$_2$  \\
MHO 1611 d & 10:42:47.86 &  $-$59:25:52.1&\ref{fig:MHO1611}&H$_2$  \\
MHO 1611 e & 10:42:47.73 &  $-$59:26:01.8&\ref{fig:MHO1611}&H$_2$  \\
MHO 1611 f & 10:42:47.36 &  $-$59:26:07.8&\ref{fig:MHO1611}&H$_2$  \\
MHO 1611 g & 10:42:46.94 &  $-$59:26:13.6&\ref{fig:MHO1611}&H$_2$  \\
MHO 1611 h & 10:42:45.09 &  $-$59:26:24.8&\ref{fig:MHO1611}&H$_2$  \\
MHO 1611 i & 10:42:46.11 &  $-$59:26:29.1&\ref{fig:MHO1611}&H$_2$  \\
MHO 1611 j & 10:42:45.54 &  $-$59:26:43.6&\ref{fig:MHO1611}&H$_2$  \\
MHO 1612 a & 10:42:45.95 &  $-$59:25:21.8&\ref{fig:MHO1611}&H$_2$  \\
MHO 1612 b & 10:42:45.79 &  $-$59:25:17.7&\ref{fig:MHO1611}&H$_2$  \\
MHO 1612 c & 10:42:45.43 &  $-$59:25:15.8&\ref{fig:MHO1611}&H$_2$  \\
MHO 1612 d & 10:42:45.11 &  $-$59:25:10.4&\ref{fig:MHO1611}&H$_2$  \\
MHO 1613   & 10:42:46.99 &  $-$59:25:21.1&\ref{fig:MHO1611}&H$_2$  \\
MHO 1614 a & 10:42:49.40 &  $-$59:25:29.6&\ref{fig:MHO1611}&H$_2$  \\
MHO 1614 b & 10:42:48.68 &  $-$59:25:27.9&\ref{fig:MHO1611}&H$_2$  \\
MHO 1615 a & 10:42:47.28 &  $-$59:25:30.4&\ref{fig:MHO1611}&H$_2$  \\
MHO 1615 b & 10:42:46.52 &  $-$59:25:29.4&\ref{fig:MHO1611}&H$_2$  \\
MHO 1616  a & 10:43:09.13 &  $-$59:24:24.7&\ref{fig:nwpillar}&H$_2$  \\
MHO 1616  b & 10:43:09.40 &  $-$59:24:26.9&\ref{fig:nwpillar}&H$_2$  \\
MHO 1616  c & 10:43:10.30 &  $-$59:24:37.9&\ref{fig:nwpillar}&H$_2$  \\
MHO 1616  d & 10:43:11.30 &  $-$59:24:46.9&\ref{fig:nwpillar}&H$_2$  \\
MHO 1616  e & 10:43:11.50 &  $-$59:24:48.2&\ref{fig:nwpillar}&H$_2$  \\
MHO 1616  f & 10:43:12.70 &  $-$59:24:59.5&\ref{fig:nwpillar}&H$_2$  \\
MHO 1616  g & 10:43:12.95 &  $-$59:25:01.5&\ref{fig:nwpillar}&H$_2$  \\
MHO 1617   & 10:43:10.22 &  $-$59:24:53.1&\ref{fig:nwpillar}&H$_2$  \\
MHO 1617 IR& 10:43:09.68 &  $-$59:24:58.8&\ref{fig:nwpillar}&H$_2$  \\
MHO 1618 IR& 10:45:28.63 &  $-$59:47:55.1&\ref{fig:MHO1618}&H$_2$  \\
MHO 1618 a & 10:45:26.59 &  $-$59:47:32.6&\ref{fig:MHO1618}&H$_2$  \\
MHO 1618 b & 10:45:28.31 &  $-$59:47:51.2&\ref{fig:MHO1618}&H$_2$  \\
MHO 1618 c & 10:45:28.81 &  $-$59:47:58.5&\ref{fig:MHO1618}&H$_2$  \\
MHO 1618 d & 10:45:28.98 &  $-$59:48:04.4&\ref{fig:MHO1618}&H$_2$  \\
MHO 1619 a & 10:46:10.50 &  $-$59:33:17.9&\ref{fig:MHO1619}&H$_2$  \\
MHO 1619 b & 10:46:10.49 &  $-$59:33:21.6&\ref{fig:MHO1619}&H$_2$  \\
MHO 1619 c & 10:46:10.29 &  $-$59:33:22.4&\ref{fig:MHO1619}&H$_2$  \\
MHO 1620 a & 10:46:11.47 &  $-$59:33:30.6&\ref{fig:MHO1619}&H$_2$, [S~II]\\
MHO 1620 b & 10:46:11.99 &  $-$59:33:30.3&\ref{fig:MHO1619}&H$_2$  \\
MHO 1620 c & 10:46:12.73 &  $-$59:33:31.4&\ref{fig:MHO1619}&H$_2$  \\
MHO 1620 d & 10:46:13.19 &  $-$59:33:28.4&\ref{fig:MHO1619}&H$_2$  \\
MHO 1620 e & 10:46:13.56 &  $-$59:33:29.6&\ref{fig:MHO1619}&H$_2$  \\
MHO 1620 f & 10:46:14.03 &  $-$59:33:29.2&\ref{fig:MHO1619}&H$_2$  \\
MHO 1621 a & 10:46:12.96 &  $-$59:33:35.5&\ref{fig:MHO1619}&H$_2$  \\
MHO 1621 b & 10:46:13.18 &  $-$59:33:36.6&\ref{fig:MHO1619}&H$_2$  \\
MHO 1621 c & 10:46:13.43 &  $-$59:33:39.3&\ref{fig:MHO1619}&H$_2$  \\
MHO 1621 d & 10:46:13.62 &  $-$59:33:36.9&\ref{fig:MHO1619}&H$_2$  \\
MHO 1621 e & 10:46:13.86 &  $-$59:33:34.5&\ref{fig:MHO1619}&H$_2$  \\
PCYC 438 & 10:44:07.03 &  $-$59:08:39.1&\ref{fig:MHO1622}&Star\\ 
MHO 1622 a & 10:43:53.18 &  $-$59:09:02.5&\ref{fig:MHO1622}&H$_2$  \\
MHO 1622 b & 10:43:53.84 &  $-$59:09:05.4&\ref{fig:MHO1622}&H$_2$  \\
MHO 1623 a & 10:43:54.96 &  $-$59:07:52.8&\ref{fig:MHO1622}&H$_2$  \\
MHO 1623 b & 10:43:55.24 &  $-$59:07:55.8&\ref{fig:MHO1622}&H$_2$  \\
MHO 1623 c & 10:43:56.58 &  $-$59:08:07.0&\ref{fig:MHO1622}&H$_2$  \\
MHO 1623 d & 10:43:57.28 &  $-$59:08:18.1&\ref{fig:MHO1622}&H$_2$  \\
MHO 1624 a & 10:44:01.93 &  $-$59:08:53.0&\ref{fig:MHO1622}&H$_2$  \\
MHO 1624 b & 10:44:03.13 &  $-$59:08:51.0&\ref{fig:MHO1622}&H$_2$  \\
MHO 1624 c & 10:44:04.31 &  $-$59:08:51.6&\ref{fig:MHO1622}&H$_2$  \\
MHO 1625 a & 10:44:08.49 &  $-$59:09:03.0&\ref{fig:MHO1622}&H$_2$  \\
MHO 1625 b & 10:44:08.31 &  $-$59:09:08.6&\ref{fig:MHO1622}&H$_2$  \\
MHO 1625 c & 10:44:09.07 &  $-$59:09:03.9&\ref{fig:MHO1622}&H$_2$  \\
MHO 1626 a & 10:44:08.35 &  $-$59:09:25.9&\ref{fig:MHO1622}&H$_2$  \\
MHO 1626 b & 10:44:09.14 &  $-$59:09:26.4&\ref{fig:MHO1622}&H$_2$  \\
MHO 1626 c & 10:44:10.36 &  $-$59:09:30.0&\ref{fig:MHO1622}&H$_2$  \\
R1       & 10:44:21.6  &  $-$59:08:24  &\ref{fig:MHO1622}&H$_2$, [S~II], H$\alpha$  \\
R2       & 10:44:11.0  &  $-$59:08:26  &\ref{fig:MHO1622}&H$_2$, [S~II], H$\alpha$  \\
R3       & 10:44:11.5  &  $-$59:07:31  &\ref{fig:MHO1622}&H$_2$, [S~II], H$\alpha$  \\
R4       & 10:44:05.1  &  $-$59:07:42  &\ref{fig:MHO1622}&H$\alpha$  \\
R5       & 10:43:58.0  &  $-$59:08:02  &\ref{fig:MHO1622}&H$_2$, [S~II], H$\alpha$  \\
MHO 1627 a & 10:41:14.63 &  $-$59:32:49.9&\ref{fig:nwpillar}&H$_2$  \\
MHO 1627 b & 10:41:14.77 &  $-$59:32:58.7&\ref{fig:nwpillar}&H$_2$  \\
MHO 1627 c & 10:41:15.29 &  $-$59:33:16.3&\ref{fig:nwpillar}&H$_2$  \\
MHO 1628 a & 10:41:14.15 &  $-$59:32:35.9&\ref{fig:nwpillar}&H$_2$  \\
MHO 1628 b & 10:41:14.16 &  $-$59:32:38.2&\ref{fig:nwpillar}&H$_2$  \\
PN Candidate& 10:43:34.77 &  $-$59:42:01.5&\ref{fig:eta}&H$_2$ \\
Finger IRS & 10:44:30.16 &  $-$59:38:40.4&\ref{fig:finger}&Star \\
SW Loop A& 10:40:16.26 &  $-$60:17:09.7&\ref{fig:swloop}&H$\alpha$, Northern knot of arc \\
SW Loop B& 10:40:12.91 &  $-$60:16:44.7&\ref{fig:swloop}&H$\alpha$, Center of diffuse area\\
SW Loop C& 10:40:05.30 &  $-$60:16:14.4&\ref{fig:swloop}&H$\alpha$,  Brightest knot\\
SW Loop D& 10:40:00.20 &  $-$60:16:12.9&\ref{fig:swloop}&H$\alpha$, Diffuse structure\\
SW Loop E& 10:39:52.62 &  $-$60:16:17.0&\ref{fig:swloop}&H$\alpha$, Middle of a bright arc \\
SW Loop F& 10:39:53.64 &  $-$60:14:09.9&\ref{fig:swloop}&H$\alpha$ \\
SW Loop G& 10:39:47.60 &  $-$60:12:20.3&\ref{fig:swloop}&H$\alpha$ \\
\enddata
\end{deluxetable}
\end{center}

\clearpage
  
\clearpage
 
\clearpage

\begin{thebibliography}{}
\bibitem[Abel et al.(2005)]{abel05}
Abel, N.P., Ferland, G.J., Shaw, G., \& van Hoof, P.A.M. 2005, ApJS 161, 65
\bibitem[Andree et al.(2013)]{andree13}
Andree, S., R\"ollig, M., \& Ossenkopf, V. 2013, in ``New Trends in Radio Astronomy
in the ALMA Era'', ASPC 476, 311
\bibitem[Arthur et al.(2011)]{arthur11}
Arthur, J. et al. 2011, MNRAS 414, 1747
\bibitem[Ascenso et al.(2007)]{ascenso07}
Ascenso, J., Alves, J., Vicente, S., \& Lago, M.T.V.T. 2007, A\&A 476, 199
\bibitem[Bally et al.(2001)]{bally01}
Bally, J., Johnstone, D., Joncas, G., Reipurth, B., \& Mall\'{e}n-Ornelas, G. 2001, AJ 122, 1508
\bibitem[Bally et al.(2002)]{bally02}
Bally, J., Reipurth, B., Walawender, J., \& Armond, T. 2002, AJ 124, 2152
\bibitem[Bertoldi (1989)]{bertoldi89}
Bertoldi, F. 1989, ApJ 346, 735
\bibitem[Brooks et al.(2000)]{brooks00}
Brooks K.J., Burton M.G., Rathborne J.M., Ashley M.C.B., Storey J.W.V. 2000, MNRAS 319, 95
\bibitem[Brooks et al.(2001)]{brooks01}
Brooks K.J., Storey J.W.V., Whiteoak J.B. 2001, MNRAS 327, 46
\bibitem[Brooks et al.(2003)]{brooks03}
Brooks, K.J, Cox, P., Schneider, N., Storey, J.W.V., Poglitsch, A.,
Geis, N., \& Bronfman, L. 2003, A\&A 412, 751
\bibitem[Broos et al.(2011)]{broos11}
Broos, P.S., et al. 2011, ApJS 194, 2
\bibitem[Davis et al.(2010)]{davis10}
Davis, C., Gell, R. Khanzadyan, T., Smith, M.D., \& Jenness, T. 2010, A\&A 511, A24
\bibitem[Dawson et~al.(2008a)]{dawson08a}
Dawson, J.R., Kawamura, A., Mizuno, N., Onishi, T., \& Fukui, Y. 2008a, PASJ 60, 1297 
\bibitem[Dawson et~al.(2008b)]{dawson08b}
Dawson, J.R., Mizuno, N., Onishi, T., McClure-Griffiths, N.M., \& Fukui, Y. 2008b, MNRAS 387, 31 
\bibitem[Donehew \& Brittain(2011)]{donehew11}
Donehew, B. \& Brittain, S. 2011, AJ 141, 46
\bibitem[Elmegreen \& Lada(1977)]{el77}
Elmegreen, B.G. \& Lada, C.J. 1977, ApJ 214, 725
\bibitem[Ercolano et al.(2011)]{ercolano11}
Ercolano, B., Dale, J.E., Gritschneder, M., \& Westmoquette, M. 2011, MNRAS 420, 141
\bibitem[Gagne et al.(2011)]{gagne11}
Gagne, M. et al. 2011, ApJS 194, 5
\bibitem[Ginsburg et al.(2011)]{ginsburg11}
Ginsburg, A., Bally, J., \& Williams, J.P. 2011, MNRAS 418, 2121
\bibitem[Feigelson et al.(2011)]{feig11}
Feigelson, E. et al. 2011, ApJS 194, 9
\bibitem[Feinstein et al.(1973)]{fein73}
Feinstein, A., Marraco, H.G., \& Muzzio, J.C. 1973, A\&AS 12, 331
\bibitem[Foster et al.(2010)]{foster10}
Foster, J.M., Rosen, P.A., Wilde, B.H., Hartigan, P., and Perry, T.S. 
2010, Phys. Plasmas 17, 112704
\bibitem[Fukui et al.(1999)]{fukui99}
Fukui, Y., et~al.\ 1999, PASJ 51, 751
\bibitem[Ghosh et al.(1988)]{ghosh88}
Ghosh S.K., Iyengar K.V.K., Rengarajan T.N., Tandon S.N., Verma R.P., \& Daniel R.P.
1988, ApJ 330, 928
\bibitem[Grenman \& Gahm(2014)]{grenman14}
Grenman, T., \& Gahm, G.F. 2014, A\&A 565, A107
\bibitem[Hamaguchi et al.(2007)]{hama07}
Hamaguchi, K. et al. 2007 PASJ 59, 151
\bibitem[Hamaguchi et al.(2009)]{hama09}
Hamaguchi, K. et al. 2009 ApJ 695, L4
\bibitem[Hartigan et al.(2011)]{hartigan11}
Hartigan, P., et al. 2011, ApJ 736, 29
\bibitem[Hartigan et al.(2012)]{hedla}
Hartigan, P., Palmer, J., \& Cleeves, L.I. 2012, High Energy Density Phys. 8, 313
\bibitem[Haro(1952)]{haro52}
Haro, G. 1952, ApJ 115, 572
\bibitem[Haro(1953)]{haro53}
Haro, G. 1953, ApJ 117, 73
\bibitem[Harvey et al.(1979)]{harvey79}
Harvey P. M., Hoffmann W. F., Campbell M. F., 1979, ApJ 227, 114
\bibitem[Henney et~al.(2005)]{henney05}
Henney, W.J., Arthur, S.J., Williams, R.J.R., \& Ferland, G.J. 2005, ApJ, 621, 328
\bibitem[Herbig(1950)]{herbig50}
Herbig, G.H. 1950, ApJ 111, 11
\bibitem[Hester et al.(1996)]{hester96}
Hester, J.J. et al. 1996, AJ 111, 2349
\bibitem[Hollenbach et~al.(1991)]{htt91}
Hollenbach, D.J., Takahashi, T., \& Tielens, A.G.G.M. 1991, ApJ 377, 192
\bibitem[Hollenbach \& Tielens(1997)]{ht97}
Hollenbach, D.J., \& Tielens, A.G.G.M. 1997, ARA\&A 35, 179
\bibitem[Hur et al.(2012)]{hur12}
Hur, H., Hwankyung, S., \& Bessell, M.S. 2012 AJ 143, 41
\bibitem[Kaufman et al.(2006)]{kaufman06}
Kaufman, M.J., Wolfire, M.G., \& Hollenbach, D.J. 2006, ApJ 644, 283
\bibitem[Lee et al.(2012)]{lee12}
Lee, H.-T., Takami, M., Duan, H.-Y., Karr, J., Su, Y.-N., Liu, S.-Y.,
Froebrich, D., \& Yeh, C.C. 2012, ApJS 200, 2 
\bibitem[Lee et al.(2013)]{lee13}
Lee, H.-T., Liao, W.-T., Froebrich, D., Karr, J.,
Ioannidis, G., Lee, Y.-H., Su, Y.-N., Liu, S.-Y., Duan, H.-Y.,
\& Takami, M., 2013, ApJS 208, 23 
\bibitem[Massey \& Johnson(1993)]{massey93}
Massey, P., \& Johnson, J. 1993, AJ 105, 980
\bibitem[Meaburn \& Walsh(1986)]{mw86}
Meaburn, J., \& Walsh, J.R. 1986, MNRAS 220, 745
\bibitem[Medina et~al.(2014)]{medina14}
Medina, S.-N.X., Arthur, S.J., Henney, W.J., Mellema, G., \& Gazol, A. 2014, MNRAS 445, 1797
\bibitem[Megeath et al.(1996)]{megeath96}
Megeath, S.T., Cox, P., Bronfman, L., \& Roelfsema, P.R. 1996, A\&A 305, 296
\bibitem[Meynet \& Maeder(2003)]{mm03}
Meynet, G., \& Maeder, A. 2003, A\&A 404, 975
\bibitem[Muzerolle et al.(1998)]{muzerolle98}
Muzerolle, J., Calvet, N., \& Hartmann, L. 1998, AJ 116, 2965
\bibitem[Ngoumou et al.(2013)]{ngoumou}
Ngoumou, J., Preibisch, T., Ratzka, T., \& Burkert, A. 2013, ApJ 769, 139
\bibitem[O'Dell et al.(1993)]{odell93}
O'Dell, C.R., Wen, Z., \& Hu, X. 1993, ApJ 410, 696
\bibitem[Offner et al.(2012)]{offner12}
Offner, S., Robitaille, T., Hansen, C.E., McKee, C.F., \& Klein, R.I.
2012, ApJ 753, 98
\bibitem[Ohlendorf et al.(2012)]{ohlendorf12}
Ohlendorf, H., Preibisch, T., Gaczkowski, B., Ratzka, T., Grellmann, R., \& McLeod, A.M.
2012, A\&A 540, A81
\bibitem[Pekruhl et al.(2013)]{pekruhl13}
Pekruhl, S., Preibisch, T., Schuller, F., \& Menten, K. 2013, A\&A 550, A29
\bibitem[Preibisch et al.(2011a)]{preib11a}
Preibisch, T. et al. 2011a, A\&A 530, A34
\bibitem[Preibisch et al.(2011b)]{preib11b}
Preibisch, T. et al. 2011b, ApJS 194, 10 
\bibitem[Preibisch et al.(2011c)]{preib11c}
Preibisch, T., Schuller, F., Ohlendorf, H., Pekruhl, S., Menten, K., \& Zinnecker, H.
2011c, A\&A 525, A92
\bibitem[Preibisch et al.(2011d)]{preib11d}
Preibisch, T. et al. 2011d, A\&A 530, A40
\bibitem[Preibisch et al.(2012)]{preib12}
Preibisch, T., Roccatagliata, V., Gaczkowski, B., \& Ratzka, T. 2012, A\&A 541, A132 
\bibitem[Probst et al.(2008)]{newfirm}
Probst, R.G., et al. 2008, Proc. SPIE 7014, 70142S
\bibitem[Povich et al.(2011a)]{povich11}
Povich, M.S., et al. 2011, ApJS 194, 14
\bibitem[Povich et al.(2011b)]{povich11b}
Povich, M.S., et al. 2011, ApJS 194, 6
\bibitem[Rathborne et al.(2002)]{rathborne02}
Rathborne J.M., Burton M.G., Brooks K.J., Cohen M., Ashley M.C.B., Storey J.W.V.
2002, MNRAS 331, 85
\bibitem[Rathborne et al.(2004)]{rathborne04}
Rathborne, J.M., Brooks, K.M., Burton, M.G., Cohen, M. \& Bontemps, S.
2004, A\&A 418, 563
\bibitem[Reed(2003)]{reed03}
Reed, B.C. 2003, AJ 125, 2531
\bibitem[Reiter \& Smith(2013)]{rs13}
Reiter, M., \& Smith, N. 2013, MNRAS 433, 2226
\bibitem[Reiter \& Smith(2014)]{rs14}
Reiter, M., \& Smith, N. 2014, submitted to MNRAS
\bibitem[Retallack(1983)]{retallack83}
Retallack, D.S. 1983, MNRAS 204, 669 
\bibitem[Robitaille et al.(2006)]{robo06}
Robitaille, T.P., Whitney, B.A., Indebetouw, R., Wood, K., \& Denzmore, P.
2006, ApJS 167, 256
\bibitem[Robitaille et al.(2007)]{robo07}
Robitaille, T.P., Whitney, B.A., Indebetouw, R., \& Wood, K. 2007, ApJS 169, 328
\bibitem[Roccatagliata et al.(2013)]{rocca13}
Roccatagliata, V., Preibisch, T., Ratzka, T., \& Gaczkowski, B. 2013, A\&A 554, A6
\bibitem[Sanchawala et~al.(2007)]{sanchawala07}
Sanchawala, K. et~al. 2007, ApJ 667, 963
\bibitem[Sanduleak \& Stephenson(1973)]{ss73}
Sanduleak, N. \& Stephenson, C.B. 1973, ApJ 185, 899
\bibitem[Smith et al.(2000)]{smith00}
Smith, N., Egan, M.P., Carey, S., Price, S.D., Morse, J.A., \& Price, P.A. 2000, ApJ 532, 145
\bibitem[Smith et al.(2004a)]{smith04a}
Smith, N., Bally, J., \& Brooks, K. J. 2004a, AJ 127, 2793
\bibitem[Smith et al.(2004b)]{smith04b}
Smith, N., Barb\'{a}, R.H., \& Walborn, N.R. 2004b, MNRAS 351, 1457. 
\bibitem[Smith et al.(2005)]{smith05}
Smith, N., Stassun, K., \& Bally, J. 2005, AJ 129, 888
\bibitem[Smith(2006a)]{smith06a}
Smith, N. 2006a, ApJ 644, 1151
\bibitem[Smith(2006b)]{smith06b}
Smith, N. 2006b, MNRAS 367, 763
\bibitem[Smith \& Brooks(2007)]{sb07}
Smith, N., \& Brooks, K.J. 2007, MNRAS 379, 1279
\bibitem[Smith \& Conti(2008)]{sc08}
Smith, N., Conti, P.S.\ 2008, ApJ 679, 1467
\bibitem[Smith et al.(2010a)]{smith10a}
Smith, N., Bally, J., \& Walborn, N. 2010a, MNRAS 405, 1153
\bibitem[Smith et al.(2010b)]{smith10b}
Smith, N., et al. 2010b, MNRAS 406, 952
\bibitem[Stanke et al.(2000)]{stanke00}
Stanke, T., McCaughrean, M.J., \& Zinnecker, H., 2002, A\&A 355, 639
\bibitem[Stanke et al.(2002)]{stanke02}
Stanke, T., McCaughrean, M.J., \& Zinnecker, H., 2002, A\&A 392, 239
\bibitem[Tapia et al.(2011)]{tapia11}
Tapia, M., Roth, M., Bohigas, J., \& Persi, P. 2011, MNRAS 416, 2163
\bibitem[Tielens \& Hollenbach(1985)]{th85}
Tielens, A.G.G.M., \& Hollenbach, D.J. 1985, ApJ 291, 722
\bibitem[Townsley et al.(2011)]{townsley11}
Townsley, L. et al. 2011, ApJS 194, 1
\bibitem[Walborn et al.(2002)]{walborn02}
Walborn, N.R., et al.\ 2002, AJ 123, 2754
\bibitem[Wang et al.(2011)]{wang11}
Wang, J. et al. 2011, ApJS 194, 11
\bibitem[Wolk et al.(2011)]{wolk11}
Wolk, S.J. et al. 2011, ApJS 194, 12
\bibitem[W{\"u}nsch et al.(2012)]{wunsch12}
W{\"u}nsch, R., et~al. 2012, A\&A 539, A116 
\bibitem[Yirak et al.(2013)]{yirak13}
Yirak, K. et al. 2013, High Energy Density Phys. 9, 251
\bibitem[Yonekura et al.(2005)]{yonekura05}
Yonekura, Y., Asayama, S., Kimura, K., Ogawa, H., Kanai, Y., Yamaguchi, N.,
Barnes, P.J., \& Fukui, Y.  2005, ApJ 634, 476
\bibitem[Zacharias et al.(2013)]{zach2013}
Zacharias, N., Finch, C.T., Girard, T.M., Henden, A., Bartlett, J.L., Monet, D.G., \&
Zacharias, M.I.  2013, AJ 145, 44

\end{thebibliography}
\end{document}